\documentclass[twoside]{article}

\usepackage{PRIMEarxiv}

\usepackage[utf8]{inputenc} % allow utf-8 input
\usepackage[T1]{fontenc}    % use 8-bit T1 fonts
\usepackage{url}            % simple URL typesetting
\usepackage{booktabs}       % professional-quality tables
\usepackage{amsfonts}       % blackboard math symbols
\usepackage{nicefrac}       % compact symbols for 1/2, etc.
\usepackage{microtype}      % microtypography
\usepackage{lipsum}
\usepackage{fancyhdr}       % header
\usepackage{graphicx}       % graphics
\usepackage[square, sort, comma, numbers]{natbib}
\usepackage{hyperref}       % hyperlinks
\usepackage{bm}
\usepackage{amsmath}
\usepackage{amssymb}
\usepackage[utf8]{inputenc}
\usepackage{svg}
\usepackage{graphicx}
\graphicspath{{media/}}     % organize your images and other figures under media/ folder

\usepackage[disable]{todonotes}
\newcommand{\bdv}[2][] {\todo[inline,backgroundcolor=blue!20!white, #1]{(bdv) #2}}

\title{AIDA: An Active Inference-based Design Agent for Audio Processing Algorithms}

\author{
  Albert Podusenko$^{*\dagger,1}$, Bart van Erp$^{\dagger,1}$, Magnus Koudahl$^{\dagger,1, 2}$, Bert de Vries$^{1, 3}$ \\
  $^{1}$\href{https://biaslab.github.io/}{BIASlab}, Department of Electrical Engineering, Eindhoven University of Technology, Eindhoven, The Netherlands \\
  $^{2}$Nested Minds Solutions, Liverpool, England \\
  $^{3}$GN Hearing, Eindhoven, The Netherlands \\
  $^{*}$\texttt{a.podusenko@tue.nl} \\
}

\usepackage{pgfplots}
\usetikzlibrary{intersections}
\usepackage{bm}
\usetikzlibrary{positioning}
\usepgfplotslibrary{fillbetween}
\pgfplotsset{compat=1.5}
\usetikzlibrary{pgfplots.groupplots}
\usetikzlibrary{shapes.geometric,backgrounds}

\usepackage{tikz}
\usepackage{xifthen}

\pgfdeclarelayer{bg}    % declare background layer
\pgfsetlayers{bg,main}  % set the order of the layers (main is the standard layer)
\definecolor{beige}{RGB}{245, 245, 220}

\definecolor{darkgrey}{RGB}{75, 75, 75}
\definecolor{lightgrey}{RGB}{250, 250, 250}

\def\d{{\mathrm d}}

\usetikzlibrary{calc, arrows, fit, positioning, patterns, decorations.pathreplacing, shapes}
\tikzstyle{dash} = [dashed, -latex,>=latex]
\tikzstyle{line} = [draw, -latex,>=latex]
\tikzstyle{smallbox} = [draw, minimum size=5.0mm]
\tikzstyle{box} = [draw, minimum size=7.0mm]
\tikzstyle{bigbox} = [draw, minimum size=10.0mm]
\tikzstyle{rectangle} = [draw, minimum width=10.0mm, minimum height=20.0mm]
\tikzstyle{switch} = [trapezium, trapezium angle=120, draw, rotate=90,  inner ysep=5pt, outer sep=5pt,
minimum height=7mm, minimum width=7mm]
\tikzstyle{roundbox} = [draw, circle, inner sep=0pt, minimum size=3mm]
\tikzstyle{clamped} = [draw, fill=darkgrey, minimum size=0.15cm]
\tikzstyle{msgcircle} = [shape=circle, draw, inner sep=0pt, minimum size=4mm, fill=white, font=\scriptsize]
\tikzstyle{darkmsgcircle} = [shape=circle, draw, inner sep=0pt, minimum size=4mm, fill=darkgrey, text=white, font=\scriptsize]
\tikzstyle{redmsgcircle} = [shape=circle, draw=red, inner sep=0pt, minimum size=4mm, text=red, font=\scriptsize]
\tikzstyle{reddarkmsgcircle} = [shape=circle, draw=red, inner sep=0pt, minimum size=4mm, fill=red, text=white, font=\scriptsize]
\tikzstyle{msgdoublecircle} = [shape=circle, double, double distance=1.5pt, draw, inner sep=0pt, minimum size=5mm, fill=white]
\tikzstyle{darkmsgdoublecircle} = [shape=circle, double, double distance=1.5pt, draw, inner sep=0pt, minimum size=5mm, fill=darkgrey, text=white, font=\bfseries]
\newcommand{\msg}[6]{
      \ifthenelse{\isin{#1}{left} \AND \isin{#2}{down}}{
            \coordinate (anchor) at ($({#3})!{#5}!({#4})$);
            \node[xshift=-6.0mm] at (anchor) {#6};
            \node[xshift=-1.0mm] at (anchor) {$\downarrow$};
      }{}
      \ifthenelse{\isin{#1}{right} \AND \isin{#2}{down}}{
            \coordinate (anchor) at ($({#3})!{#5}!({#4})$);
            \node[xshift=6.0mm] at (anchor) {#6};
            \node[xshift=1.0mm] at (anchor) {$\downarrow$};
      }{}

      \ifthenelse{\isin{#1}{down} \AND \isin{#2}{right}}{
            \coordinate (anchor) at ($({#3})!{#5}!({#4})$);
            \node[ yshift=-4.0mm] at (anchor) {#6};
            \node[yshift=-1.0mm] at (anchor) {$\rightarrow$};
      }{}
      \ifthenelse{\isin{#1}{up} \AND \isin{#2}{right}}{
            \coordinate (anchor) at ($({#3})!{#5}!({#4})$);
            \node[ yshift=4.0mm] at (anchor) {#6};
            \node[yshift=1.0mm] at (anchor) {$\rightarrow$};
      }{}

      \ifthenelse{\isin{#1}{down} \AND \isin{#2}{left}}{
            \coordinate (anchor) at ($({#3})!{#5}!({#4})$);
            \node[ yshift=-4.0mm] at (anchor) {#6};
            \node[yshift=-1.0mm] at (anchor) {$\leftarrow$};
      }{}
      \ifthenelse{\isin{#1}{up} \AND \isin{#2}{left}}{
            \coordinate (anchor) at ($({#3})!{#5}!({#4})$);
            \node[ yshift=4.0mm] at (anchor) {#6};
            \node[yshift=1.0mm] at (anchor) {$\leftarrow$};
      }{}

      \ifthenelse{\isin{#1}{left} \AND \isin{#2}{up}}{
            \coordinate (anchor) at ($({#3})!{#5}!({#4})$);
            \node[ xshift=-6.0mm] at (anchor) {#6};
            \node[xshift=-1.0mm] at (anchor) {$\uparrow$};
      }{}
      \ifthenelse{\isin{#1}{right} \AND \isin{#2}{up}}{
            \coordinate (anchor) at ($({#3})!{#5}!({#4})$);
            \node[ xshift=6.0mm] at (anchor) {#6};
            \node[xshift=1.0mm] at (anchor) {$\uparrow$};
      }{}
}

\newcommand{\msgcircle}[6]{
      \ifthenelse{\isin{#1}{left} \AND \isin{#2}{down}}{
            \coordinate (anchor) at ($({#3})!{#5}!({#4})$);
            \node[msgcircle,xshift=-5.0mm] at (anchor) {#6};
            \node[xshift=-1.5mm] at (anchor) {$\downarrow$};
      }{}
      \ifthenelse{\isin{#1}{right} \AND \isin{#2}{down}}{
            \coordinate (anchor) at ($({#3})!{#5}!({#4})$);
            \node[msgcircle,xshift=5.0mm] at (anchor) {#6};
            \node[xshift=1.5mm] at (anchor) {$\downarrow$};
      }{}

      \ifthenelse{\isin{#1}{down} \AND \isin{#2}{right}}{
            \coordinate (anchor) at ($({#3})!{#5}!({#4})$);
            \node[msgcircle, yshift=-5.0mm] at (anchor) {#6};
            \node[yshift=-2.0mm] at (anchor) {$\rightarrow$};
      }{}
      \ifthenelse{\isin{#1}{up} \AND \isin{#2}{right}}{
            \coordinate (anchor) at ($({#3})!{#5}!({#4})$);
            \node[msgcircle, yshift=5.0mm] at (anchor) {#6};
            \node[yshift=2.0mm] at (anchor) {$\rightarrow$};
      }{}

      \ifthenelse{\isin{#1}{down} \AND \isin{#2}{left}}{
            \coordinate (anchor) at ($({#3})!{#5}!({#4})$);
            \node[msgcircle, yshift=-5.0mm] at (anchor) {#6};
            \node[yshift=-2.0mm] at (anchor) {$\leftarrow$};
      }{}
      \ifthenelse{\isin{#1}{up} \AND \isin{#2}{left}}{
            \coordinate (anchor) at ($({#3})!{#5}!({#4})$);
            \node[msgcircle, yshift=5.0mm] at (anchor) {#6};
            \node[yshift=2.0mm] at (anchor) {$\leftarrow$};
      }{}

      \ifthenelse{\isin{#1}{left} \AND \isin{#2}{up}}{
            \coordinate (anchor) at ($({#3})!{#5}!({#4})$);
            \node[msgcircle, xshift=-5.0mm] at (anchor) {#6};
            \node[xshift=-1.5mm] at (anchor) {$\uparrow$};
      }{}
      \ifthenelse{\isin{#1}{right} \AND \isin{#2}{up}}{
            \coordinate (anchor) at ($({#3})!{#5}!({#4})$);
            \node[msgcircle, xshift=5.0mm] at (anchor) {#6};
            \node[xshift=1.5mm] at (anchor) {$\uparrow$};
      }{}
}

\newcommand{\darkmsg}[6]{
      \ifthenelse{\isin{#1}{left} \AND \isin{#2}{down}}{
            \coordinate (anchor) at ($({#3})!{#5}!({#4})$);
            \node[darkmsgcircle, xshift=-5mm] at (anchor) {#6};
            \node[xshift=-1.5mm] at (anchor) {$\downarrow$};
      }{}
      \ifthenelse{\isin{#1}{right} \AND \isin{#2}{down}}{
            \coordinate (anchor) at ($({#3})!{#5}!({#4})$);
            \node[darkmsgcircle, xshift=5mm] at (anchor) {#6};
            \node[xshift=1.5mm] at (anchor) {$\downarrow$};
      }{}

      \ifthenelse{\isin{#1}{down} \AND \isin{#2}{right}}{
            \coordinate (anchor) at ($({#3})!{#5}!({#4})$);
            \node[darkmsgcircle, yshift=-5.0mm] at (anchor) {#6};
            \node[yshift=-2.0mm] at (anchor) {$\rightarrow$};
      }{}
      \ifthenelse{\isin{#1}{up} \AND \isin{#2}{right}}{
            \coordinate (anchor) at ($({#3})!{#5}!({#4})$);
            \node[darkmsgcircle, yshift=5.0mm] at (anchor) {#6};
            \node[yshift=2.0mm] at (anchor) {$\rightarrow$};
      }{}

      \ifthenelse{\isin{#1}{down} \AND \isin{#2}{left}}{
            \coordinate (anchor) at ($({#3})!{#5}!({#4})$);
            \node[darkmsgcircle, yshift=-5.0mm] at (anchor) {#6};
            \node[yshift=-2.0mm] at (anchor) {$\leftarrow$};
      }{}
      \ifthenelse{\isin{#1}{up} \AND \isin{#2}{left}}{
            \coordinate (anchor) at ($({#3})!{#5}!({#4})$);
            \node[darkmsgcircle, yshift=5.0mm] at (anchor) {#6};
            \node[yshift=2.0mm] at (anchor) {$\leftarrow$};
      }{}

      \ifthenelse{\isin{#1}{left} \AND \isin{#2}{up}}{
            \coordinate (anchor) at ($({#3})!{#5}!({#4})$);
            \node[darkmsgcircle, xshift=-5.0mm] at (anchor) {#6};
            \node[xshift=-1.5mm] at (anchor) {$\uparrow$};
      }{}
      \ifthenelse{\isin{#1}{right} \AND \isin{#2}{up}}{
            \coordinate (anchor) at ($({#3})!{#5}!({#4})$);
            \node[darkmsgcircle, xshift=5.0mm] at (anchor) {#6};
            \node[xshift=1.5mm] at (anchor) {$\uparrow$};
      }{}
}

\newcommand{\redbackmsg}[6]{
      \ifthenelse{\isin{#1}{left} \AND \isin{#2}{down}}{
            \coordinate (anchor) at ($({#3})!{#5}!({#4})$);
            \node[reddarkmsgcircle, xshift=-5mm] at (anchor) {#6};
            \node[xshift=-1.5mm] at (anchor) {$\downarrow$};
      }{}
      \ifthenelse{\isin{#1}{right} \AND \isin{#2}{down}}{
            \coordinate (anchor) at ($({#3})!{#5}!({#4})$);
            \node[reddarkmsgcircle, xshift=5mm] at (anchor) {#6};
            \node[xshift=1.5mm] at (anchor) {$\downarrow$};
      }{}

      \ifthenelse{\isin{#1}{down} \AND \isin{#2}{right}}{
            \coordinate (anchor) at ($({#3})!{#5}!({#4})$);
            \node[reddarkmsgcircle, yshift=-5.0mm] at (anchor) {#6};
            \node[yshift=-2.0mm] at (anchor) {$\rightarrow$};
      }{}
      \ifthenelse{\isin{#1}{up} \AND \isin{#2}{right}}{
            \coordinate (anchor) at ($({#3})!{#5}!({#4})$);
            \node[reddarkmsgcircle, yshift=5.0mm] at (anchor) {#6};
            \node[yshift=2.0mm] at (anchor) {$\rightarrow$};
      }{}

      \ifthenelse{\isin{#1}{down} \AND \isin{#2}{left}}{
            \coordinate (anchor) at ($({#3})!{#5}!({#4})$);
            \node[reddarkmsgcircle, yshift=-5.0mm] at (anchor) {#6};
            \node[yshift=-2.0mm] at (anchor) {$\leftarrow$};
      }{}
      \ifthenelse{\isin{#1}{up} \AND \isin{#2}{left}}{
            \coordinate (anchor) at ($({#3})!{#5}!({#4})$);
            \node[reddarkmsgcircle, yshift=5.0mm] at (anchor) {#6};
            \node[yshift=2.0mm] at (anchor) {$\leftarrow$};
      }{}

      \ifthenelse{\isin{#1}{left} \AND \isin{#2}{up}}{
            \coordinate (anchor) at ($({#3})!{#5}!({#4})$);
            \node[reddarkmsgcircle, xshift=-5.0mm] at (anchor) {#6};
            \node[xshift=-1.5mm] at (anchor) {$\uparrow$};
      }{}
      \ifthenelse{\isin{#1}{right} \AND \isin{#2}{up}}{
            \coordinate (anchor) at ($({#3})!{#5}!({#4})$);
            \node[reddarkmsgcircle, xshift=5.0mm] at (anchor) {#6};
            \node[xshift=1.5mm] at (anchor) {$\uparrow$};
      }{}
}

\newcommand{\redmsg}[6]{
      \ifthenelse{\isin{#1}{left} \AND \isin{#2}{down}}{
            \coordinate (anchor) at ($({#3})!{#5}!({#4})$);
            \node[redmsgcircle, xshift=-5mm] at (anchor) {#6};
            \node[xshift=-1.5mm] at (anchor) {$\downarrow$};
      }{}
      \ifthenelse{\isin{#1}{right} \AND \isin{#2}{down}}{
            \coordinate (anchor) at ($({#3})!{#5}!({#4})$);
            \node[redmsgcircle, xshift=5mm] at (anchor) {#6};
            \node[xshift=1.5mm] at (anchor) {$\downarrow$};
      }{}

      \ifthenelse{\isin{#1}{down} \AND \isin{#2}{right}}{
            \coordinate (anchor) at ($({#3})!{#5}!({#4})$);
            \node[redmsgcircle, yshift=-5.0mm] at (anchor) {#6};
            \node[yshift=-2.0mm] at (anchor) {$\rightarrow$};
      }{}
      \ifthenelse{\isin{#1}{up} \AND \isin{#2}{right}}{
            \coordinate (anchor) at ($({#3})!{#5}!({#4})$);
            \node[redmsgcircle, yshift=5.0mm] at (anchor) {#6};
            \node[yshift=2.0mm] at (anchor) {$\rightarrow$};
      }{}

      \ifthenelse{\isin{#1}{down} \AND \isin{#2}{left}}{
            \coordinate (anchor) at ($({#3})!{#5}!({#4})$);
            \node[redmsgcircle, yshift=-5.0mm] at (anchor) {#6};
            \node[yshift=-2.0mm] at (anchor) {$\leftarrow$};
      }{}
      \ifthenelse{\isin{#1}{up} \AND \isin{#2}{left}}{
            \coordinate (anchor) at ($({#3})!{#5}!({#4})$);
            \node[redmsgcircle, yshift=5.0mm] at (anchor) {#6};
            \node[yshift=2.0mm] at (anchor) {$\leftarrow$};
      }{}

      \ifthenelse{\isin{#1}{left} \AND \isin{#2}{up}}{
            \coordinate (anchor) at ($({#3})!{#5}!({#4})$);
            \node[redmsgcircle, xshift=-5.0mm] at (anchor) {#6};
            \node[xshift=-1.5mm] at (anchor) {$\uparrow$};
      }{}
      \ifthenelse{\isin{#1}{right} \AND \isin{#2}{up}}{
            \coordinate (anchor) at ($({#3})!{#5}!({#4})$);
            \node[redmsgcircle, xshift=5.0mm] at (anchor) {#6};
            \node[xshift=1.5mm] at (anchor) {$\uparrow$};
      }{}
}

\newcommand{\bwmsg}[6]{
      \ifthenelse{\isin{#1}{left} \AND \isin{#2}{down}}{
            \coordinate (anchor) at ($({#3})!{#5}!({#4})$);
            \node[msgdoublecircle, xshift=-5.5mm] at (anchor) {#6};
            \node[xshift=-1.5mm] at (anchor) {$\downarrow$};
      }{}
      \ifthenelse{\isin{#1}{right} \AND \isin{#2}{down}}{
            \coordinate (anchor) at ($({#3})!{#5}!({#4})$);
            \node[msgdoublecircle, xshift=5.5mm] at (anchor) {#6};
            \node[xshift=1.5mm] at (anchor) {$\downarrow$};
      }{}

      \ifthenelse{\isin{#1}{down} \AND \isin{#2}{right}}{
            \coordinate (anchor) at ($({#3})!{#5}!({#4})$);
            \node[msgdoublecircle, yshift=-6.0mm] at (anchor) {#6};
            \node[yshift=-2.0mm] at (anchor) {$\rightarrow$};
      }{}
      \ifthenelse{\isin{#1}{up} \AND \isin{#2}{right}}{
            \coordinate (anchor) at ($({#3})!{#5}!({#4})$);
            \node[msgdoublecircle, yshift=6.0mm] at (anchor) {#6};
            \node[yshift=2.0mm] at (anchor) {$\rightarrow$};
      }{}

      \ifthenelse{\isin{#1}{down} \AND \isin{#2}{left}}{
            \coordinate (anchor) at ($({#3})!{#5}!({#4})$);
            \node[msgdoublecircle, yshift=-6.0mm] at (anchor) {#6};
            \node[yshift=-2.0mm] at (anchor) {$\leftarrow$};
      }{}
      \ifthenelse{\isin{#1}{up} \AND \isin{#2}{left}}{
            \coordinate (anchor) at ($({#3})!{#5}!({#4})$);
            \node[msgdoublecircle, yshift=6.0mm] at (anchor) {#6};
            \node[yshift=2.0mm] at (anchor) {$\leftarrow$};
      }{}

      \ifthenelse{\isin{#1}{left} \AND \isin{#2}{up}}{
            \coordinate (anchor) at ($({#3})!{#5}!({#4})$);
            \node[msgdoublecircle, xshift=-5.5mm] at (anchor) {#6};
            \node[xshift=-1.5mm] at (anchor) {$\uparrow$};
      }{}
      \ifthenelse{\isin{#1}{right} \AND \isin{#2}{up}}{
            \coordinate (anchor) at ($({#3})!{#5}!({#4})$);
            \node[msgdoublecircle, xshift=5.5mm] at (anchor) {#6};
            \node[xshift=1.5mm] at (anchor) {$\uparrow$};
      }{}
}

\newcommand{\bwdarkmsg}[6]{
      \ifthenelse{\isin{#1}{left} \AND \isin{#2}{down}}{
            \coordinate (anchor) at ($({#3})!{#5}!({#4})$);
            \node[darkmsgdoublecircle, xshift=-5.5mm] at (anchor) {#6};
            \node[xshift=-1.5mm] at (anchor) {$\downarrow$};
      }{}
      \ifthenelse{\isin{#1}{right} \AND \isin{#2}{down}}{
            \coordinate (anchor) at ($({#3})!{#5}!({#4})$);
            \node[darkmsgdoublecircle, xshift=5.5mm] at (anchor) {#6};
            \node[xshift=1.5mm] at (anchor) {$\downarrow$};
      }{}

      \ifthenelse{\isin{#1}{down} \AND \isin{#2}{right}}{
            \coordinate (anchor) at ($({#3})!{#5}!({#4})$);
            \node[darkmsgdoublecircle, yshift=-6.0mm] at (anchor) {#6};
            \node[yshift=-2.0mm] at (anchor) {$\rightarrow$};
      }{}
      \ifthenelse{\isin{#1}{up} \AND \isin{#2}{right}}{
            \coordinate (anchor) at ($({#3})!{#5}!({#4})$);
            \node[darkmsgdoublecircle, yshift=6.0mm] at (anchor) {#6};
            \node[yshift=2.0mm] at (anchor) {$\rightarrow$};
      }{}

      \ifthenelse{\isin{#1}{down} \AND \isin{#2}{left}}{
            \coordinate (anchor) at ($({#3})!{#5}!({#4})$);
            \node[darkmsgdoublecircle, yshift=-6.0mm] at (anchor) {#6};
            \node[yshift=-2.0mm] at (anchor) {$\leftarrow$};
      }{}
      \ifthenelse{\isin{#1}{up} \AND \isin{#2}{left}}{
            \coordinate (anchor) at ($({#3})!{#5}!({#4})$);
            \node[darkmsgdoublecircle, yshift=6.0mm] at (anchor) {#6};
            \node[yshift=2.0mm] at (anchor) {$\leftarrow$};
      }{}

      \ifthenelse{\isin{#1}{left} \AND \isin{#2}{up}}{
            \coordinate (anchor) at ($({#3})!{#5}!({#4})$);
            \node[darkmsgdoublecircle, xshift=-5.5mm] at (anchor) {#6};
            \node[xshift=-1.5mm] at (anchor) {$\uparrow$};
      }{}
      \ifthenelse{\isin{#1}{right} \AND \isin{#2}{up}}{
            \coordinate (anchor) at ($({#3})!{#5}!({#4})$);
            \node[darkmsgdoublecircle, xshift=5.5mm] at (anchor) {#6};
            \node[xshift=1.5mm] at (anchor) {$\uparrow$};
      }{}
}

\tikzset{mainstyle/.style={fill=white, draw=black, shape=rectangle, align=center}}

\tikzset{dstyle/.style={mainstyle, minimum size=4mm, inner sep=0pt, text width=4mm}}

\tikzset{sstyle/.style={mainstyle, minimum size=5mm, inner sep=0pt, text width=5mm}}

\tikzset{ostyle/.style={fill=darkgrey, draw=black, shape=rectangle, minimum size=0.2cm, inner sep=0pt, text width=2mm}}

\tikzstyle{observation}=[ostyle]
\tikzstyle{deterministic}=[dstyle]
\tikzstyle{stochastic}=[sstyle]

\tikzstyle{filter}=[mainstyle, minimum width=1cm, minimum height=0.5cm]
\tikzstyle{selector}=[fill=white, draw=black, shape=trapezium, rotate=180, minimum width=1cm, minimum height=0.5cm]

\makeatletter
\DeclareRobustCommand{\cev}[1]{%
  {\mathpalette\do@cev{#1}}%
}
\newcommand{\do@cev}[2]{%
  \vbox{\offinterlineskip
    \sbox\z@{$\m@th#1 x$}%
    \ialign{##\cr
      \hidewidth\reflectbox{$\m@th#1\vec{}\mkern4mu$}\hidewidth\cr
      \noalign{\kern-\ht\z@}
      $\m@th#1#2$\cr
    }%
  }%
}

\DeclareMathOperator*{\argmin}{arg\,min}
\DeclareMathOperator*{\argmax}{arg\,max}

\makeatletter
\let\ORIbbl@fixname\bbl@fixname
\def\bbl@fixname#1{%
  \@ifundefined{languagealias@\expandafter\string#1}
    {\ORIbbl@fixname#1}
    {\edef\languagename{\@nameuse{languagealias@#1}}}%
}
\newcommand{\definelanguagealias}[2]{%
  \@namedef{languagealias@#1}{#2}%
}
\makeatother

\definelanguagealias{en}{english}
\definelanguagealias{eng}{english}

\begin{document}
\maketitle

\begin{abstract}
In this paper we present AIDA, which is an active inference-based agent that iteratively designs a personalized audio processing algorithm through situated interactions with a human client. The target application of AIDA is to propose on-the-spot the most interesting alternative values for the tuning parameters of a hearing aid (HA) algorithm, whenever a HA client is not satisfied with their HA performance. AIDA interprets searching for the "most interesting alternative" as an issue of optimal (acoustic) context-aware Bayesian trial design. In computational terms, AIDA is realized as an active inference-based agent with an Expected Free Energy criterion for trial design. This type of architecture is inspired by neuro-economic models on efficient (Bayesian) trial design in brains and implies that AIDA comprises generative probabilistic models for acoustic signals and user responses. We propose a novel generative model for acoustic signals as a sum of time-varying auto-regressive filters and a user response model based on a Gaussian Process Classifier. The full AIDA agent has been implemented in a factor graph for the generative model and all tasks (parameter learning, acoustic context classification, trial design, etc.) are realized by variational message passing on the factor graph. All verification and validation experiments and demonstrations are freely accessible at our GitHub repository. 

\end{abstract}

\keywords{Active inference, Bayesian trial design, Hearing aids, Noise reduction, Probabilistic modeling, Source separation, Speech enhancement, Variational Message passing}
\section{Introduction}
% an simplified overview of the tuning cycle of hearing aids
Hearing aids (HA) are often equipped with specialized noise reduction algorithms.
These algorithms are developed by teams of engineers who aim to create a single optimal algorithm that suits any user in any situation.
Taking a one-size-fits-all approach to HA algorithm design leads to two problems that are prevalent throughout today's hearing aid industry.
First, modeling all possible acoustic environments is simply infeasible.
The daily lives of HA users are varied and the different environments they traverse even more so.
Given differing acoustic environments, a single static HA algorithm cannot possibly account for all eventualities - even without taking into account the particular constraints imposed by the HA itself, such as limited computational power and allowed processing delays \citep{kates_multichannel_2005}.
Secondly, hearing loss is highly personal and can differ significantly between users. 
Each HA user consequently requires their own, individually tuned HA algorithm that compensates for their unique hearing loss profile \citep{van_de_laar_probabilistic_2016, nielsen_perception-based_2015, alamdari_personalization_2020} and satisfies their personal preferences for parameter settings \citep{reddy_individualized_2017}.
Considering that HAs nowadays often consist of multiple interconnected digital signal processing units with many integrated parameters, the task of personalizing the algorithm requires exploring a high-dimensional search space of parameters, which often do not yield a clear physical interpretation.
The current most widespread approach to personalization requires the HA user to physically travel to an audiologist who manually tunes a subset of all HA parameters. This is a burdensome activity that is not guaranteed to yield an improved listening experience for the HA user.

% situated design
From these two problems, it becomes clear that we need to move towards a new approach for hearing aid algorithm design that empowers the user.
Ideally, users should be in control of their own HA algorithms and should be able to change and update them at will instead of having to rely on teams of engineers that operate with long design cycles, separated from the users' living experiences.

% active inference for parameter tuning
The question then becomes, how do we move the HA algorithm design away from engineers and into the hands of the user?
While a naive implementation that allows for tuning HA parameters with sliders on, for example, a smartphone is trivial to develop, even a small number of adjustable parameters gives rise to a large, high-dimensional search space that the HA user needs to learn to navigate. 
This puts a large burden on the user, essentially asking them to be their own trained audiologist.
Clearly, this is not a trivial task and this approach is only feasible for a small set of parameters, which carry a clear physical interpretation. 
Instead, we wish to support the user with an agent that intelligently proposes new parameter trials.
In this setting, the user is only tasked to cast (positive or negative) appraisals of the current HA settings .
Based on these appraisal, the agent will autonomously traverse the search space with the goal of proposing satisfying parameter values for that user under the current environmental conditions in as few trials as possible. 

% brief overview of our approach
Designing an intelligent agent that learns to efficiently navigate a parameter space is not trivial.
In the solution approach in this paper, we rely on a probabilistic modeling approach inspired by the free energy principle (FEP) \citep{friston_free_2006}.
The FEP is a framework originally designed to explain the kinds of computations that biological, intelligent agents (such as the human brain) might be performing.
Recent years have seen the FEP applied to the design of synthetic agents as well \citep{van_de_laar_simulating_2019, van_de_laar_application_2019,millidge_deep_2019,tschantz_scaling_2020}.
A hallmark feature of FEP-based agents is that they exhibit a dynamic trade-off between exploration and exploitation \citep{friston_active_2015,da_costa_active_2020,friston_sophisticated_2021}, which is a highly desirable property when learning to navigate an HA parameter space.
Concretely, the FEP proposes that intelligent agents should be modeled as probabilistic models.
These types of models do not only yield point estimates of variables, but also capture uncertainty through modeling full posterior probability distributions.
Furthermore, user appraisals and actions can be naturally incorporated by simply extending the probabilistic model.
Taking a model-based approach also allows for fewer parameters than alternative data-driven solutions, as we can incorporate field-specific knowledge, making it more suitable for computationally constrained hearing aid devices.
The novelty of our approach is rooted in the fact that the entire proposed systems is framed as a probabilistic generative model in which we can perform (active) inference through (expected) free energy minimization.

% contributions + overview
In this paper we present AIDA\footnote{Aida is a girl's name of Arabic origin, meaning ``happy''. We use this name as an abbreviation for an "Active Inference-based Design Agent" that aims to make an end user ``happy''.}, an active inference-based design agent for the situated development of context-dependent audio processing algorithms, which provides the user with her own controllable audio processing algorithm.
This approach embodies an FEP-based agent that operates in conjunction with an acoustic model and actively learns optimal context-dependent tuning parameter settings.
After formally specifying the problem and solution approach in Section~\ref{sec:problem} we make the following contributions:

\begin{enumerate}
    \item We develop a modular probabilistic model that embodies situated, (acoustic) scene-dependent, and personalized design of its corresponding hearing aid algorithm in Section~\ref{sec:model:acoustic}.
    \item We develop an expected free energy-based agent (AIDA) in Section~\ref{sec:model:EFE}, whose proposals for tuning parameter settings are well-balanced in terms of seeking more information about the user's preferences (explorative agent behavior) versus seeking to optimize the user's satisfaction levels by taking advantage of previously learned preferences (exploitative agent behavior). 
    \item Inference in the acoustic model and AIDA is elaborated upon in Section~\ref{sec:inference} and their operations are individually verified through representative experiments in Section~\ref{sec:experiments}. Furthermore, all elements are jointly validated through a demonstrator application in Section~
    \ref{sec:experiments:demo}.
\end{enumerate}
We have intentionally postponed a more thorough review of related work to Section~\ref{sec:related} as we deem it more relevant after the introduction of our solution approach.
Finally, Section~\ref{sec:discussion} discusses the novelty and limitations of our approach and Section~\ref{sec:conclusion} concludes this paper.
\section{Problem statement and proposed solution approach} \label{sec:problem}

\subsection{Automated hearing aid tuning by optimization}
\begin{figure}[t]
    \centering
    \includegraphics[width=0.8\textwidth]{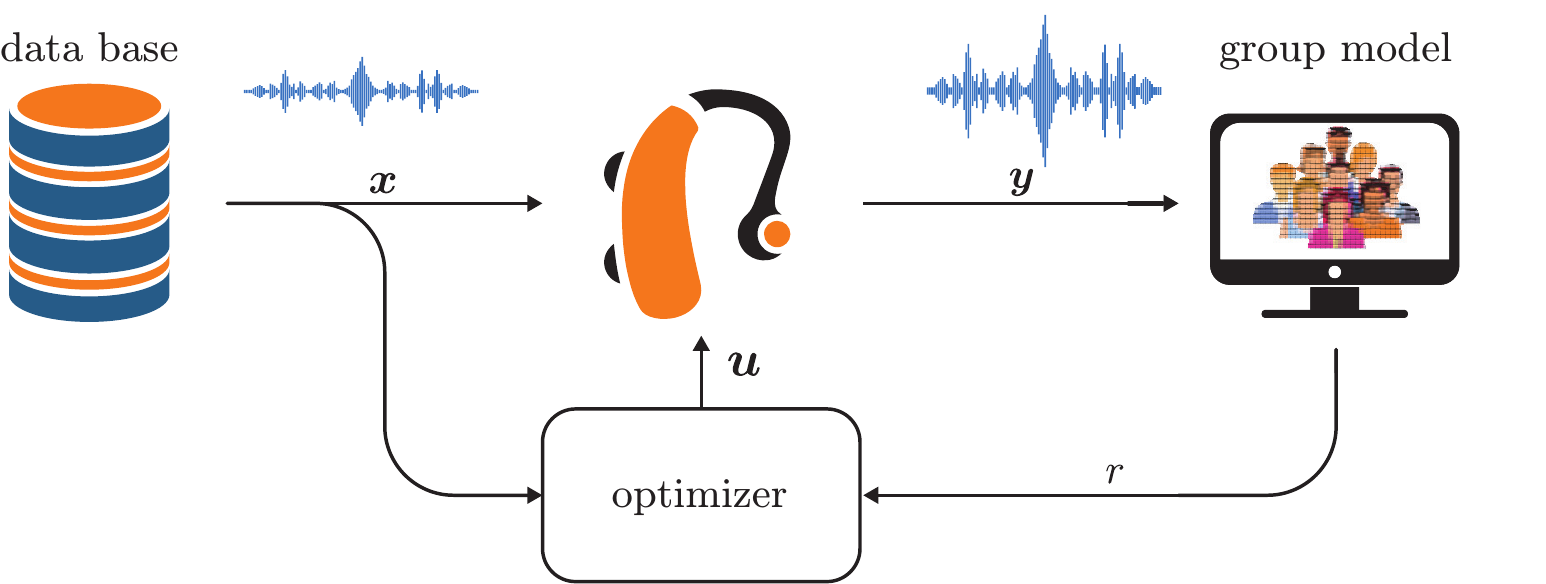}
    \caption{A schematic overview of the conventional approach to hearing aid algorithm tuning. Here the parameters of the hearing aid $\bm{u}$ are optimized with respect to some generic user rating model $r(\bm{y})$ for a large data base $\mathcal{X}$ of input data $\bm{x}$.
    }
    \label{fig:direct-tuning}
\end{figure}

In this paper we consider the problem of choosing values for the tuning parameters $\bm{u}$ of a hearing aid algorithm that processes an acoustic input signal $\bm{x}$ to output signal $\bm{y}$. In Figure~\ref{fig:direct-tuning}, we sketch an automated optimization-based approach to this problem. Assume that we have access to a generic ``signal quality'' model which rates the quality of a HA output signal $\bm{y} = f(\bm{x},\bm{u})$, as a function of the HA input $\bm{x}$ and parameters $\bm{u}$, by a rating $r(\bm{x},\bm{u}) \triangleq r(\bm{y})$. If we run this system on a representative set of input signals $\bm{x} \in \mathcal{X}$, then the tuning problem reduces to the optimization task
\begin{equation}
    \bm{u}^* = \argmax_{\bm{u}} \sum_{\bm{x}\in\mathcal{X}} r(\bm{x},\bm{u})\,.
\end{equation}
Unfortunately, in commercial practice, this optimization approach does not always result in satisfactory HA performance, because of two reasons. First, the signal quality models in the literature have been trained on large databases of preference ratings from many users and therefore only model the average HA client rather than any specific client \citep{rix_perceptual_2001, kates_hearing-aid_2010, taal_algorithm_2011, beerends_perceptual_2013, hines_visqol_2015, chinen_visqol_2020}. Secondly, the optimization approach averages over a large set of different input signals, so it will not deal with acoustic context-dependent client preferences. By acoustic context, we consider signal properties that depend on environmental conditions such as being inside, outside, in a car or at the mall. Generally, client preferences for HA tuning parameters are both highly \emph{personal} and \emph{context-dependent}. Therefore, there is a need to develop a \emph{personalized}, \emph{context-sensitive} controller for tuning HA parameters $\bm{u}$.

\subsection{Situated hearing aid tuning with the user in-the-loop} \label{sec:problem:AIDA}
\begin{figure}[t]
    \centering
    \includegraphics[width=0.8\textwidth]{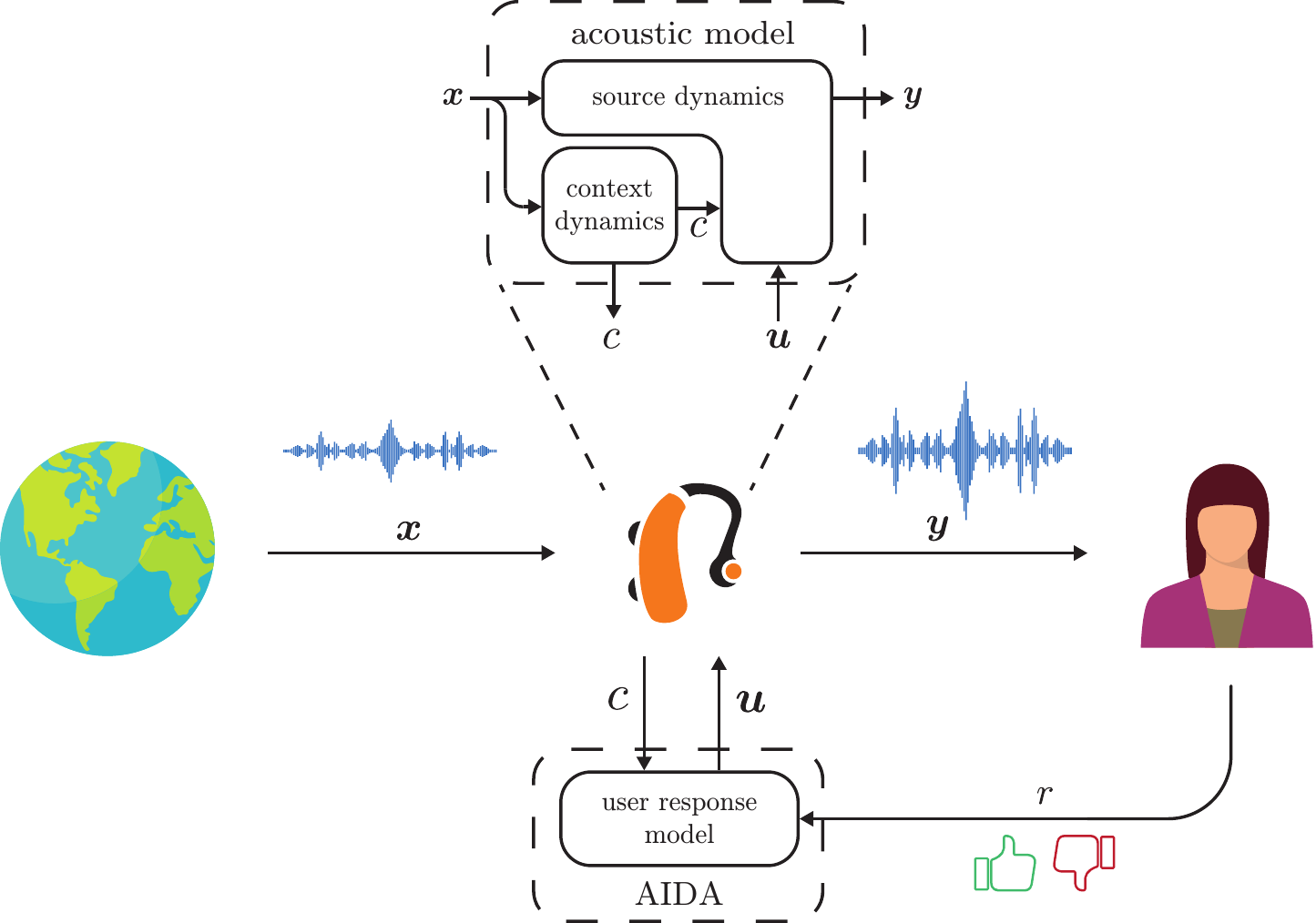}
    \caption{A schematic overview of the proposed situated HA design loop containing AIDA. An incoming signal $\bm{x}$ enters the hearing aid and is used to infer the context of the user $c$. Based on this context and previous user appraisals, AIDA proposes a new set of parameters $\bm{u}$ for the hearing aid algorithm. Based on the input signal, the proposed parameters and the current context, the output $\bm{y}$ of the hearing aid is determined, which are used together with the context in the hearing aid algorithm. The parameters $\bm{u}$ are actively optimized by AIDA, based on the inferred context $c$ from the input signal $\bm{x}$ and appraisals $r$ from the user in the loop. All individual subsystems represent parts of a probabilistic generative model as described in Section \ref{sec:model}, where the corresponding algorithms follows from performing probabilistic inference in these models as described in Section~\ref{sec:inference}.}
    \label{fig:AIDA-design-loop}
\end{figure}

In this paper, we will develop a personalized, context-aware design agent, based on the architecture shown in Figure~\ref{fig:AIDA-design-loop}. In contrast to Figure~\ref{fig:direct-tuning}, the outside world (rather than a database) produces an input signal $\bm{x}$ under situated conditions that is processed by a hearing aid algorithm to produce an output signal $\bm{y}$. A particular human hearing aid client listens to the signal $\bm{y}$ and is invited to cast at any time binary appraisals $r\in\{0,1\}$ about the current performance of the hearing aid algorithm, where $1$ and $0$ correspond to the user being satisfied and unsatisfied, respectively. Context-aware trials for HA tuning parameters are provided by AIDA. Rather than an offline design procedure, the whole system designs continually under \emph{situated} conditions. The HA device itself houses a custom hearing aid algorithm, based on state inference in a generative acoustic model. The acoustic model contains two sub-models: 1) a source dynamics model and 2) a context dynamics model.

Inference in the acoustic model is based on the observed signal $\bm{x}$ and yields the output $\bm{y}$ and context $c$. Based on this context signal $c$ and previous user appraisals $r$, AIDA will actively propose new parameters trials $\bm{u}$ with the goal of making the user happy. Technically, the objective is that AIDA expects to receive fewer negative appraisals in the future, relative to not making parameter adaptations, see Section~\ref{sec:model:EFE} for details.

The design of AIDA is non-trivial. For instance, since there is a priori no personalized model of HA ratings for any particular user, AIDA will have to build such a model on-the-fly from the context $c$ and user appraisals $r$. Since the system operates under situated conditions, we want to impose as little burden on the end user as possible. As a result, most users will only once in a while cast an appraisal and this complicates the learning of a personalized HA rating model. 

To make this desire for very light-weight interactions concrete, we now sketch how we envision a typical interaction between AIDA and a HA client. Assume that the HA client is in a conversation with a friend at a restaurant. The signal of interest, in this case, is the friend's speech signal while the interfering signal is an environmental babble noise signal. The HA algorithm tries to separate the input signal $\bm{x}$ into its constituent speech and noise source components, then applies gains $\bm{u}$ to each source component and sums these weighted source signals to produce output $\bm{y}$. If the HA client is happy with the performance of her HA, she will not cast any appraisals. After all, she is in the middle of a conversation and has no imperative to change the HA behavior. However, if she cannot understand her conversation partner, the client may covertly tap her watch or make another gesture to indicate that she is not happy with her current HA settings. In response, AIDA, which may be implemented as a smartwatch application, will reply instantaneously by sending a tuning parameter update $\bm{u}$ to the hearing aid algorithm in an effort to fix the client's current hearing problem.
Since the client's preferences are context-dependent, AIDA needs to incorporate information about the acoustic context from HA input $\bm{x}$. As an example, the HA user might leave the restaurant for a walk outside. Walking outside presents a different type of background noise and consequently requires different parameter settings.

Crucially, we would like HA clients to be able to tune their hearing aids without interruption of any ongoing activities. Therefore, we will not demand that the client has to focus visual attention on interacting with a smartphone app. At most, we want the client to apply a tap or make a simple gesture that does not draw any attention away from the ongoing conversation.
A second criterion is that we do not want the conversation partner to notice that the client is interacting with the agent. The client may actually be in a situation (e.g., a business meeting) where it is not appropriate to demonstrate that her priorities have shifted to tuning her hearing aids. In other words, the interactions must be very light-weight and covert.
A third criterion is that we want the agent to learn from as few appraisals as possible. Note that, if the HA has 10 tuning parameters and 5 interesting values (very low, low, middle, high, very high) per parameter, then there are $5^{10}$ (about 10 million) parameter settings. We do not want the client to get engaged in an endless loop of disapproving new HA proposals as this will lead to frustration and distraction from the ongoing conversation. Clearly, this means that each update of the HA parameters cannot be selected randomly: we want the agent to propose the most interesting values for the tuning parameters, based on all observed past information and certain goal criteria for future HA behavior. In Section~\ref{sec:inference:trial}, we will quantify what \emph{most interesting} means in this context.

In short, the goal of this paper is to design an intelligent agent that supports user-driven situated design of a personalized audio processing algorithm through a very light-weight interaction protocol.

In order to accomplish this task, we will draw inspiration from the way how human brains design algorithms (e.g., for speech and object recognition, riding a bike, etc.) solely through environmental interactions. Specifically, we base the design of AIDA on the Active Inference (AIF) framework. Originating from the field of computational neuroscience, AIF proposes to view the brain as a prediction engine that models sensory inputs. Formally, AIF accomplishes this through specifying a probabilistic generative model of incoming data. Performing approximate Bayesian inference in this model by minimizing free energy then constitutes a unified procedure for both data processing and learning. To select tuning parameter trials, an AIF agent predicts the \emph{expected} free energy in the near future, given a particular choice of parameter settings. AIF provides a single, unified method for designing all components of AIDA. 
The design of a HA system that is controlled by an AIF-based design agent involves solving the following tasks:

\begin{enumerate}
    \item Classification of acoustic context
    \item Selecting acoustic context-dependent trials for the HA tuning parameters.
    \item Execution of the HA signal processing algorithm (that is controlled by the trial parameters).
\end{enumerate}

Task 1 (context classification) involves determining the most probable current acoustic environment. Based on a dynamic context model (described in Section~\ref{sec:model:context}), we infer the most probable acoustic environment as described in Section~\ref{sec:inference:context}.

Task 2 (trial design) encompasses proposing alternative settings for the HA tuning parameters. Sections~\ref{sec:model:EFE} and \ref{sec:inference:trial} describe the user response model and execution of AIDA's trial selection procedure based on expected free energy minimization, respectively. 

Finally, task 3 (hearing aid algorithm execution) concerns performing variational free energy minimization with respect to the state variables in a generative probabilistic model for the acoustic signal. In Section~\ref{sec:model:acoustic} we describe the generative acoustic model underlying the HA algorithm and Section~\ref{sec:inference:acoustic} describes the inferred HA algorithm itself.

Crucially, in the AIF framework, all three tasks can be accomplished by variational free energy minimization in a generative probabilistic model for observations. Since we can automate variational free energy minimization by a probabilistic programming language, the only remaining task for the human designer is to specify the generative models. The next section describes the model specification.

\section{Model specification} \label{sec:model}

In this section, we present the generative model of the AIDA controlled HA system, as illustrated in Figure~\ref{fig:AIDA-design-loop}. %Following a good regulator theorem \citetemp{Conant and Ashby, 1970}, our agent should embrace a model of its environment to regulate the state of the environment. 
In Section~\ref{sec:model:acoustic}, we describe a generative model for the HA input and output signals $\bm{x}$ and $\bm{y}$ respectively.  
In this model, the hearing aid algorithm follows through performing probabilistic inference, as will be discussed in Section~\ref{sec:inference}. Part of the hearing aid algorithm is a mechanism for inferring the current acoustic context. In Section~\ref{sec:model:EFE} we introduce a model for agent AIDA that is used to infer new parameter trials. A concise summary of the generative model is also presented in Appendix~\ref{appendix:model} and an overview of the corresponding symbols is given in Table~\ref{tab:symbols-model}.

Throughout this section, we will make use of factor graphs for visualization of probabilistic models. In this paper we focus on Forney-style factor graphs (FFG), as introduced in \cite{forney_codes_2001} with notational conventions adopted from \cite{loeliger_introduction_2004}.
FFGs represent factorized functions by undirected graphs, whose nodes represent the individual factors of the global function.
The nodes are connected by edges representing the mutual arguments of the factors.
In an FFG, a node can be connected to an arbitrary number of edges, but edges are constrained to have a maximum degree of two.
A more detailed review of probabilistic modeling and factor graphs has been provided in Appendix~\ref{sec:appendix:background}.

\subsection{Acoustic model}\label{sec:model:acoustic}

Our acoustic model of the observed signal and hearing aid output consists of a model of the source dynamics of the underlying signals and a model for the context dynamics.

\subsubsection{Model of source dynamics}\label{sec:model:source}
\begin{figure}[!htb]
    \centering
    \begin{minipage}{.65\textwidth}
        \centering
        \resizebox{\textwidth}{!}{\begin{tikzpicture}

    % speech
    \node [style=filter] (AR_speech) {$AR$};
    \node [left of=AR_speech, node distance=20mm] (speech_prev) {$\scriptstyle{\bm{s}_{t-1}}$};
    \node [style=deterministic, right of=AR_speech, node distance=20mm] (constraint_speech) {$=$};
    \node [right of=constraint_speech, node distance=15mm] (speech) {$\scriptstyle{\bm{s}_{t}}$};
    \node [right of=speech, node distance=5mm] (next) {};
    \node [left of=speech_prev, node distance=6mm] (prev) {};

    % evolving theta for speech
    \node [style=deterministic, above left= 12mm and -5mm of AR_speech] (theta_speech) {$=$};
    \node [style=stochastic, left of=theta_speech, node distance=15mm] (theta_speech_prev) {$\mathcal{N}$};
    \node [style=observation, above of=theta_speech_prev, node distance=10mm] (omega) {};
    \node [above of=omega, node distance=3mm] (tau_clamped) {$\scriptstyle{\omega\bm{I}}$};
    \node [right of=theta_speech, node distance=15mm] (theta_speech_next) {$\scriptstyle{\bm{\theta}_t}$};
    \node [left of=theta_speech_prev, node distance=15mm] (theta_speech_half) {$\scriptstyle{\bm{\theta}_{t-1}}$};
    \draw[<-] (theta_speech) -- (theta_speech_prev);
    \draw[->] (theta_speech) -- node[above] {$\scriptstyle{}$} (theta_speech_next);

    \node [style=deterministic, above right = 5mm and -3mm of AR_speech] (gamma_speech) {$=$};
    \node [left of=gamma_speech, node distance=25mm] (gamma_speech_prev) {$\scriptstyle{\gamma}$};
    \node [right of=gamma_speech, node distance=15mm] (gamma_speech_next) {};
    \draw[<-] (gamma_speech) -- (gamma_speech_prev);
    \draw[->] (gamma_speech) -- node[above] {} (gamma_speech_next);

    % connections
    \draw [->] (theta_speech) -- ([xshift=-2mm]AR_speech.north);
    \draw [->] (AR_speech) -- (constraint_speech);
    \draw [->] (gamma_speech) -- ([xshift=4mm]AR_speech.north);
    \draw[->] (speech_prev) -- (AR_speech);
    \draw[<-] (theta_speech_prev) -- (theta_speech_half);
    \draw[->] (constraint_speech) -- (speech);

    % noise
    \node [style=filter, above left = 30mm and 50mm of AR_speech] (AR_noise) {$AR$};
    \node [left of=AR_noise, node distance=20mm] (noise_prev) {$\scriptstyle{\bm{n}_{t-1}}$};
    \node [style=deterministic, right of=AR_noise, node distance=20mm] (constraint_noise) {$=$};
    \node [right of=constraint_noise, node distance=15mm] (noise) {$\scriptstyle{\bm{n}_{t}}$};
    \node [right of=noise, node distance=5mm] (noise_next) {};
    % \node [left of=noise_prev, node distance=6mm] (prev) {};

    % observation noise
    \node [style=deterministic, below of=constraint_noise, node distance=45mm] (e_noise) {$\scriptstyle \bm{e}_1^\top $};
    \node [style=deterministic, below of=constraint_speech, node distance=10mm] (e_speech) {$\scriptstyle \bm{e}_1^\top$};
    \node [style=deterministic, below of = e_speech, node distance=10mm] (plus) {$+$};
    \node [style=observation, below of=plus, node distance=10mm] (yt) {};
    \node [below of=yt, node distance=4mm] (y_n) {$\scriptstyle{x}_t$};

    \draw [->] (constraint_speech) -- (e_speech);
    \draw [->] (constraint_noise) -- (e_noise);
    \draw [->] (e_speech) -- (plus);
    \draw [->] (e_noise) |- (plus);
    \draw [->] (omega) -- (theta_speech_prev);

    \draw[->] (plus) -- node[left] {} (yt);

    \node [style=smallbox, above left= 15mm and -6mm of AR_noise, node distance=10mm] (GMM) {$\scriptstyle{\text{GMM}}$};
    \node [style=smallbox, above right = 8mm and -6mm of AR_noise, node distance=10mm] (GammaMM) {$\scriptstyle{\Gamma\text{MM}}$};
    \node[style=deterministic, above = 12mm of GammaMM] (ckm_eq_1) {$=$};
    \node[style=deterministic, above = 5mm of GMM] (ckm_eq_2) {$=$};
    \node[style=stochastic, left = 10mm of ckm_eq_2] (cat) {$\scriptscriptstyle{\mathrm{Cat}}$};
    \node[left = 10mm of cat] (ckm_prev) {$\scriptstyle{\bm{c}_{k-1}}$};
    \node[right = 10mm of ckm_eq_1] (ckm_next) {$\scriptstyle{\bm{c}_{k}}$};

    \node[style=deterministic, above = 5mm of cat] (dir) {$=$};
    \node[left = 5mm of dir] (dir_prev) {$\scriptstyle{\mathrm{T}}$};
    \node[right = 5mm of dir] (dir_next) {};

    \draw [->] (noise_prev) -- (AR_noise);
    \draw [->] (AR_noise) -- (constraint_noise);
    \draw [->] (constraint_noise) -- (noise);
    \draw [->] (GMM) -- node[pos=0.7, left] {$\scriptstyle{\bm{\zeta}}_k$} ([xshift=-3.5mm]AR_noise.north);
    \draw [->] (GammaMM) -- node[pos=0.5, right] {$\scriptstyle{\tau}_k$} ([xshift=3.4mm]AR_noise.north);
    \draw [->] (ckm_eq_2) -- (GMM);
    \draw [->] (ckm_eq_1) -- (GammaMM);
    \draw[->] (ckm_eq_1) -- (ckm_next);
    \draw[->] (ckm_eq_2) -- (ckm_eq_1);
    \draw[->] (cat) -- (ckm_eq_2);
    \draw[->] (ckm_prev) -- (cat);

    \draw[dashed] (dir_prev) -- (dir);
    \draw[dashed] (dir_prev) -- (dir);
    \draw[dashed] (dir) -- (dir_next);
    \draw[->] (dir) -- (cat);
    
    \node[dashed, fit=(ckm_prev)(ckm_next)(dir), draw, inner sep=1.5mm] (box) {};
    \node[above right = 5mm and 3mm of ckm_eq_1] () {\scriptsize{context}};

\end{tikzpicture}}
    \end{minipage}%
    \begin{minipage}{.35\textwidth}
        \centering
        \resizebox{\textwidth}{!}{\begin{tikzpicture}
%nodes
\node [style=deterministic] (A) {$A$};
\node [style=deterministic, below of=A] (mult) {$\times$};
\node [style=deterministic, right of=mult, node distance=22mm] (gaussian) {$\mathcal{N}$};
\node [style=stochastic, above of=gaussian] (N) {$V$};
\node [left of=mult, node distance=20mm] (x) {};
\node [above of=A, node distance=15mm] (theta) {};
\node [right of=gaussian, node distance=15mm] (y) {};
\node [above of=N, node distance=15mm] (gamma) {};

%connections
\draw[->] [>=stealth] (A) -- (mult) node[pos=0.4, left] (Atheta) {\hspace*{-5pt}$\scriptstyle{A(\bm{\theta})}$};
\draw[->] [>=stealth] (theta) -- (A) node[pos=.3, left=2mm] {$\scriptscriptstyle{\overleftarrow{\nu}(\scriptscriptstyle{\bm{\theta})}}$} node[pos=0.3, right=2mm] {$\scriptscriptstyle{\overrightarrow{\nu}(\scriptscriptstyle{\bm{\theta}})}$} node[pos=0.3, left] {$\uparrow$} node[pos=0.3, right] {$\downarrow$};
\draw[->] [>=stealth] (mult) -- (gaussian);
\draw[->] [>=stealth] (gaussian) -- (y) node[pos=0.6, below=2mm] {$\scriptscriptstyle{\overrightarrow{\nu}(\bm{y})}$} node[pos=0.6, above=2mm] {$\scriptscriptstyle{\overleftarrow{\nu}(\bm{y})}$} node[pos=0.6, above] {$\leftarrow{}$} node[pos=0.6, below] {$\rightarrow$};
\draw[->] [>=stealth] (x) -- (mult) node[pos=0.3, below=2mm] {$\scriptscriptstyle{\overleftarrow{\nu}(\bm{x})}$} node[pos=0.3, above=2mm] {$\scriptscriptstyle{\overrightarrow{\nu}(\bm{x})}$} node[pos=0.3, below] {$\leftarrow$} node[pos=0.3, above] {$\rightarrow$};
\draw[->] [>=stealth] (N) -- node[pos=0.4, left] {$\scriptscriptstyle{V(\gamma)}$} (gaussian);
\draw[->] [>=stealth] (gamma) -- node[pos=.3, left=2mm] {$\scriptscriptstyle{\overleftarrow{\nu}(\gamma)}$} node[pos=0.3, right=2mm] {$\scriptscriptstyle{\overrightarrow{\nu}(\gamma)}$} node[pos=0.3, left] {$\uparrow$} node[pos=0.3, right] {$\downarrow$} (N);

\node[dashed, fit=(A)(gaussian)(Atheta), draw, inner sep=1.5mm] (box) {};
\node[below of=box, node distance=15mm](AR){$AR$};
% \draw[->, dotted] (AR) -- (box);

\end{tikzpicture}}
    \end{minipage}
    \caption{(left) A Forney-style factor graph representation of the acoustic source signals model as specified by \eqref{eq:model:acoustic-speech}-\eqref{eq:model:acoustic} at time index $t$. The observation $x_t$ is specified as the sum of a latent speech signal $s_t$ and a latent noise signal $n_t$. The speech signal is modeled by a time-varying auto-regressive process, where its coefficients $\bm{\theta}_t$ are modeled by a Gaussian random walk. The noise signal is a context-dependent auto-regressive process, modeled by Gaussian (GMM) and Gamma mixture models ($\Gamma$MM) for the parameters $\bm{\zeta}_k$ and $\tau_k$, respectively. The selection variable of these mixture models represents the context $\bm{c}_k$. The model for the context dynamics is enclosed by the dashed box. The composite AR factor node represents the auto-regressive transition dynamics specified by \eqref{eq:model:acoustic:AR}. (right) The composite AR node that conceals its internal operations from the rest of the graph \citep{podusenko_message_2021-1}. The arrows show the direction of incoming and outgoing messages. \bdv{This figure misses an edge for the context $c_k$.}}\label{fig:model:coupled-AR}
\end{figure}
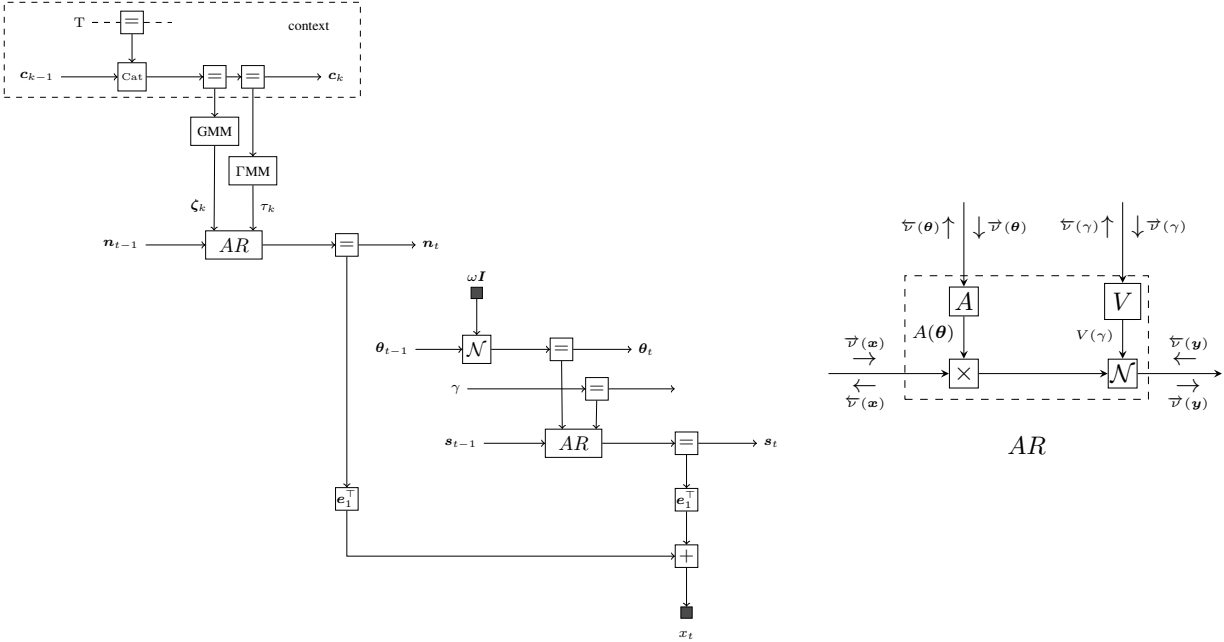

We assume that the observed acoustic signal $\bm{x}$ consists of a speech signal (or more generally, a target signal that the HA client wants to focus on) and an additive noise signal (that the HA client is not interested in), as
\begin{equation}\label{eq:model:acoustic}
    % x_t = \bm{e}_{1}^\intercal(\bm{s}_t + \bm{n}_t)
    x_t = s_t + n_t
\end{equation}
where $x_t \in \mathbb{R}$ represents the observed signal at time $t$, i.e. the input to the HA.
The speech and noise signals are represented by $s_t\in \mathbb{R}$ and $n_t\in \mathbb{R}$, respectively.
At this point, the source dynamics of $s_n$ and $n_t$ need to be further specified.
Here we choose to model the speech signal by a time-varying auto-regressive model and the noise signal by a context-dependent auto-regressive model.
The remainder of this subsection will elaborate on both these source models and will further specify how the hearing aid output is generated.
An FFG visualization of the described acoustic model is depicted in Figure \ref{fig:model:coupled-AR}. 

Historically, Auto-Regressive (AR) models have been widely used to represent speech signals \citep{kakusho_hierarchical_1982, paliwal_speech_1987}. As the dynamics of the vocal tract exhibit non-stationary behavior, speech is usually segmented into individual frames that are assumed to be quasi-stationary. Unfortunately, the signal is often segmented without any prior information about the phonetic structure of the speech signal. Therefore the quasi-stationarity assumption is likely to be violated and time-varying dynamics are more likely to occur in the segmented frames \citep{vermaak_particle_2002}. To address this issue, we can use a time-varying prior for the coefficients of the AR model, leading to a time-varying AR (TVAR) model \citep{rudoy_time-varying_2011}
\begin{subequations}\label{eq:model:acoustic-speech}
\begin{align}
    &\bm{\theta}_t \sim \mathcal{N}\left(\bm{\theta}_{t-1},\ \omega\mathrm{I}_{M}\right) \label{eq:model:acoustic:coefs}\\
    &\bm{s}_t \sim \mathcal{N}\left(A(\bm{\theta}_t)\bm{s}_{t-1},\ V\left(\gamma\right)\right) \label{eq:model:acoustic:AR}
\end{align}
\end{subequations}
where $\bm{\theta}_t = [\theta_{1t}, \theta_{2t}, ..., \theta_{Mt}]^\intercal \in \mathbb{R}^M$, $\bm{s}_t = [s_t, s_{t-1}, ..., s_{t-M+1}]^\intercal \in \mathbb{R}^M$ are the coefficients and states of an $M$-th order TVAR model for speech signal $s_t = \bm{e}_1^\intercal \bm{s}_t$. We use $\mathcal{N}(\mu, \Sigma)$ to denote a Gaussian distribution with mean $\mu$ and covariance matrix $\Sigma$. In this model, the AR coefficients $\bm{\theta}_t$ are represented by a Gaussian random walk with process noise covariance $\omega \mathrm{I}_M$, with $\mathrm{I}_M$ denoting the identity matrix of size $(M\times M)$, scaled by $\omega\in\mathbb{R}_{>0}$. $\gamma\in\mathbb{R}_{>0}$ represents the process noise precision matrix of the AR process. Here, we have adopted the state-space formulation of TVAR models as in \citep{podusenko_online_2020}, where $V(\gamma)=(1/\gamma)\bm{e}_{1} \bm{e}_{1}^\intercal$ creates a covariance matrix with a single non-zero entry. We use $\bm{e}_i$ to denote an appropriately sized Cartesian standard unit vector that represents a column vector of zeros where only the $i^\text{th}$ entry is 1. $A(\bm{\theta})$ denotes the companion matrix of size $(M\times M)$, defined as 
\begin{equation}\label{eq:model:ARmatrix}
 A(\bm{\theta}) = 
    \begin{bmatrix}
         \qquad \bm{\theta}^\intercal  \\
         \mathrm{I}_{M-1} & \bm{0}
    \end{bmatrix}   \,.
\end{equation}
Multiplication of a state vector by this companion matrix, such as $A(\bm{\theta}_t)\bm{s}_{t-1}$, basically performs two operations: an inner product $\bm{\theta}_t^\intercal\bm{s}_{t-1}$ and a shift of $\bm{s}_{t-1}$ by one time step to the past.

The acoustic model also encompasses a model for background noise, such as the sounds at a bar or train station. Many of these background sounds can be well represented by colored noise \citep{popescu_kalman_1998}, which in turn can be modeled by a low-order AR model \citep{gannot_iterative_1998, gibson_filtering_1991}
\begin{equation}\label{eq:model:acoustic-noise}
    \bm{n}_t \sim \mathcal{N}\left(A(\bm{\zeta}_k)\bm{n}_{t-1},\ V\left(\tau_k\right)\right), \qquad\qquad \text{for }t = t^-, t^-+1,\ldots, t^+
\end{equation}
where $\bm{\zeta}_k = [\zeta_{1k}, \zeta_{2k}, ..., \zeta_{Nk}]^\intercal \in \mathbb{R}^N$, $\bm{n}_t = [n_t, n_{t-1}, ..., n_{t-N+1}]^\intercal \in \mathbb{R}^N$ are the coefficients and states of an AR model of order $N \in \mathbb{N}^+$ for noise signal $n_t=\bm{e}_1^\intercal\bm{n}_t$. $\tau_k\in\mathbb{R}_{>0}$ denotes the process noise precision of the AR process.
In contrast to the speech model, we assume the processes $\bm{\zeta}_k$ and $\tau_k$ to be stationary when the user is in a specific acoustic environment or context. 
To make clear that contextual states change much slower that raw acoustic data signals, we index the slower parameters at time index $k$, which is related to index $t$ by
\begin{equation}\label{eq:model:kt}
    k = \left\lceil\frac{t}{W}\right\rceil \,.
\end{equation}
Here, $\lceil\cdot\rceil$ denotes the ceiling function that returns the largest integer smaller or equal than its argument, while $W$ is the window length. The above equation makes sure that $k$ is intuitively aligned with segments of length $W$, i.e. $t\in[1,W]$ corresponds to $k=1$. To denote the start and end indices of the time segment corresponding to context index $k$, we define $t^-=(k-1)W+1$ and $t^+=kW$ as an implicit function of $k$, respectively. The context can be assumed to be stationary within a longer period of time compared to the speech signal.
However, abrupt changes in the dynamics of background noise may occasionally occur. For example, if the user moves from a train station to a bar, the parameters of the AR model that are attributed to the train station will now inadequately describe the background noise of the new environment. To deal with these changing acoustic environments, we introduce context-dependent priors for the background noise, using a Gaussian and Gamma mixture model:
\begin{subequations}\label{eq:model:acoustic-noise-context}
\begin{align}
    \bm{\zeta}_k &\sim \prod_{l=1}^L \mathcal{N}\left(\bm{\mu}_{l}, \Sigma_{l}\right)^{c_{lk}} \label{eq:model:zeta}\\
    \tau_k &\sim \prod_{l=1}^L \Gamma\left(\alpha_{l}, \beta_{l}\right)^{c_{lk}} \label{eq:model:tau}
\end{align}
\end{subequations}
The context at time index $k$, denoted by $\bm{c}_k$, comprises a 1-of-$L$ binary vector with elements $c_{lk}\in \{0, 1\}$, which are constrained by $\sum_l c_{lk} = 1$. $\Gamma(\alpha, \beta)$ represents a Gamma distribution with shape and rate parameters $\alpha$ and $\beta$, respectively.
The hyperparameters $\bm{\mu}_l$, $\Sigma_l$, $\alpha_l$ and $\beta_l$ define the characteristics of the different background noise environments.

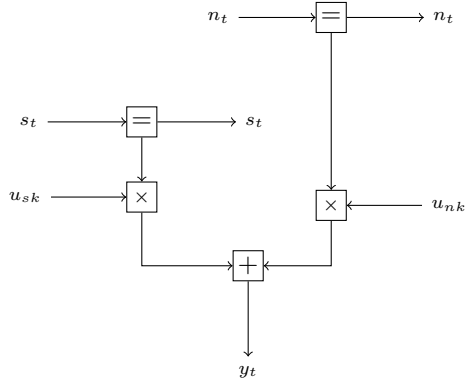
\begin{figure}
    \centering
    \begin{tikzpicture}

    % speech
    \node [] (speech_prev) {$\scriptstyle{s_t}$};
    \node [style=deterministic, right of=speech_prev, node distance=15mm] (constraint_speech) {$=$};
    \node [right of=constraint_speech, node distance=15mm] (speech) {$\scriptstyle{s_t}$};
    \node [right of=speech, node distance=5mm] (next) {};
    \node [style=deterministic, below of=constraint_speech, node distance=10mm] (dot_speech) {$\scriptstyle{\times}$};
    \node [left = 10mm of dot_speech] (dot_speech_u) {$\scriptstyle{u_{sk}}$};

    % noise
    \node [above right= 10mm and -10mm of speech] (noise_prev) {$\scriptstyle{n_t}$};
    \node [style=deterministic, right of=noise_prev, node distance=15mm] (constraint_noise) {$=$};
    \node [right of=constraint_noise, node distance=15mm] (noise) {$\scriptstyle{n_t}$};
    \node [style=deterministic, below of=constraint_noise, node distance=25mm] (dot_noise) {$\scriptstyle{\times}$};
    \node [right = 10mm of dot_noise] (dot_noise_u) {$\scriptstyle{u_{nk}}$};

    % observation
    \node [style=deterministic, below right=5mm and 10mm of dot_speech] (sum) {$+$};
    \node [below=10mm of sum] (gaussian) {$\scriptstyle{y_t}$};
     
    \draw [->] (speech_prev) -- (constraint_speech);
    \draw [->] (constraint_speech) -- (speech);
    \draw [->] (constraint_speech) -- (dot_speech);
    \draw [->] (dot_speech) |- (sum);
    \draw [->] (dot_speech_u) -- (dot_speech);

    \draw [->] (noise_prev) -- (constraint_noise);
    \draw [->] (constraint_noise) -- (noise);
    \draw [->] (constraint_noise) -- (dot_noise);
    \draw [->] (dot_noise) |- (sum);
    \draw [->] (dot_noise_u) -- (dot_noise);

    \draw [->] (sum) -- (gaussian);% node[pos=0.5, left] {$\scriptstyle{y_t}$};

\end{tikzpicture}
    \caption{A Forney-style factor graph representation of the hearing aid output model of \eqref{eq:model:HA-output}. The output of the hearing aid is modeled by reweighing the separated speech and background noise signals.}
    \label{fig:model:HA-output}
\end{figure}

% \subsubsection{Generating the HA output signal}
Now that an acoustic model of the environment has been formally specified, we will extend this model with the goal of obtaining a HA algorithm.
The principal goal of a HA algorithm is to improve audibility and intelligibility of acoustic signals. Audibility can be improved by amplifying the received input signal. Intelligibility can be improved by increasing the Signal-to-Noise Ratio (SNR) of the received signal. Assuming that we can infer the constituent source signals $s_t$ and $n_t$ from received signal $x_t$, the desired HA output signal can be modeled by  
\begin{equation}\label{eq:model:HA-output}
    y_t = u_{sk}s_t + u_{nk}n_t, \qquad\qquad \text{for }t = t^-, t^-+1, \ldots, t^+
\end{equation} where $\bm{u}_k = [u_{sk}, u_{nk}]^\intercal \in [0, 1]^{2}$ 
represents a vector of 2 tuning parameters or source-specific gains for the speech and background noise signal, respectively. In this expression the output of the hearing aid is modeled by a weighted sum of the constituent source signals.
The gains control the amplification of the extracted speech and noise signals individually and thus allows the user to perform source-specific filtering, also known as soundscaping \citep{van_erp_bayesian_2021}. 
Because of imperfections during inference of the source signals (see Section~\ref{sec:inference}), the gains simultaneously reflect a trade-off between residual noise and speech distortion. We show the FFG representation of the HA output in Figure~\ref{fig:model:HA-output}.
%It is assumed that the algorithm designer has no particular information about good values for gains $\bm{u}_k$, as these are user- and context-dependent.
Finding good values for the gains $\bm{u}$ can be a difficult task because the preferred parameter settings may depend on the specific listener and on the acoustic context. 

Next, we describe the acoustic context model that will allow AIDA to make context-dependent parameter proposals. % Then we focus on the "autonomous" agent that takes on the task of assigning "good" values to the tuning parameters, based on this context and user appraisals.

\subsubsection{Model of context dynamics}\label{sec:model:context}

As HA clients move through different acoustic background settings, such as being in a car, doing groceries, watching TV at home, etc.) the preferred parameter settings for HA algorithms tend to vary. The context signal allows to distinguish between these different acoustic environments. %To accurately model the acoustic environment, it is important to model the context transition dynamics to mimic the changes between the acoustic environments.

The hidden context state variable $\bm{c}_k$ at time index $k$ is a 1-of-$L$ encoded binary vector with elements $c_{lk} \in \{0, 1\}$, which are constrained by $\sum_l c_{lk} = 1$.
This context is responsible for the operations of the noise model in \eqref{eq:model:acoustic-noise-context}.
Context transitions are supported by a dynamic model
\begin{equation}\label{eq:model:context-dynamics}
    \bm{c}_k \sim \mathrm{Cat}(\mathrm{T}\bm{c}_{k-1}),
\end{equation}
where the elements of transition matrix $\mathrm{T}$, are defined as $\mathrm{T}_{ij} = p(c_{ik} = 1 \mid c_{j,k-1} = 1)$, which are constrained by $\mathrm{T}_{ij}\in[0,1]$ and $\sum_{j=1}^L \mathrm{T}_{ij} = 1$.
We model the individual columns of $\mathrm{T}$ by a Dirichlet distribution as
\begin{equation}\label{eq:model:context-transition}
    \mathrm{T}_{1:L,j} \sim \mathrm{Dir}(\bm{\alpha}_j),
\end{equation}
where $\bm{\alpha}_j$ denotes the vector of concentration parameters corresponding to the $j^\text{th}$ column of $\mathrm{T}$. The context state is initialized by a categorical distribution as 
\begin{equation}\label{eq:model:context-prior}
    \bm{c}_0 \sim \mathrm{Cat}(\bm{\pi}) = \prod_{l=1}^L \pi_l^{c_{l0}} \quad\text{such that} ~  \sum_{l=1}^L \pi_l=1,
\end{equation}
where the vector $\bm{\pi}=[\pi_1, \pi_2,\ldots,\pi_L]^\intercal$ contains the event probabilities, whose elements can be chosen as $\pi_l = 1/L$ if the initial context is unknown.
An FFG representation of the context dynamics model is shown in the dashed box in Figure~\ref{fig:model:coupled-AR}.

\subsection{AIDA's user response model} \label{sec:model:EFE}

The goal of AIDA is to continually provide the most ``interesting'' settings for the HA tuning parameters $\bm{u}_k$, where interesting has been quantitatively interpreted by minimization of Expected Free Energy. But how does AIDA know what the client wants? In order to learn the client's preferences, she is invited to cast at any time her appraisal $r_k \in \{\varnothing, 0,1\}$ of current HA performance. To keep the user interface very light, we will assume that appraisals are binary, encoded by $r_k=0$ for disapproval and $r_k=1$ indicating a positive experience. If a user does not cast an appraisal, we will just record a missing value, i.e., $r_k=\varnothing$. The subscript $k$ for $r_k$ indicates that we record appraisals at the same rate as the context dynamics. 

If a client submits a negative appraisal $r_k=0$, AIDA interprets this as an expression that the client is not happy with the current HA settings $\bm{u}_k$ in the current acoustic context $\bm{c}_k$ (and vice versa for positive appraisals). To \emph{learn} client preferences from these appraisals, AIDA holds a context-dependent generative model to \emph{predict} user appraisals and updates this model after observing actual appraisals. In this paper, we opt for a Gaussian Process Classifier (GPC) model as the generative model for binary user appraisals. A Gaussian Process (GP) is a very flexible probabilistic model and GPCs have successfully been applied to preference learning in a variety of tasks before \citep{houlsby_bayesian_2011,chu_preference_2005,huszar_gp_2011}. For an in-depth discussion on GPs, we refer the reader to \citep{rasmussen_gaussian_2006}. Specifically, the context-dependent user response model is defined as 
\begin{subequations}\label{eq:user-response-model}
\begin{align}
   v_k(\cdot) &\sim \prod_{l=1}^L \mathrm{GP}(m_l(\cdot),K_l(\cdot,\cdot))^{c_{lk}}  \label{eq:model:GP-1} \\
    r_k &\sim \mathrm{Ber}(\Phi(v_k(\bm{u}_k)))\,. \qquad\qquad\qquad\qquad \text{if }r_k\in\{0,1\} \label{eq:sigmoid-user-response-1}
\end{align}
\end{subequations}
\begin{figure}
    \centering
    \begin{tikzpicture}

    % speech
    \node [style=observation] (u) {};
    \node [above=1mm of u] (u_k) {$\scriptstyle{\bm{u}_k}$};
    \node [style=box, right of=u, node distance=15mm] (GP) {\scriptsize{GPM}};
    \node [above of=GP, node distance=12mm] (switch) {$\scriptstyle{\bm{c}_k}$};
    \node [style=stochastic, right of=GP, node distance=15mm] (Phi) {$\Phi$};
    \node [style=stochastic, right of=Phi, node distance=15mm] (Ber) {$\scriptscriptstyle{\mathrm{Ber}}$};
    \node [right of=Ber, node distance=15mm] (r) {$\scriptstyle{r}_k$};

    \draw [->] (u) -- (GP);
    \draw [->] (GP) -- node[pos=0.5, above] {$\scriptstyle{v}_k$} (Phi);
    \draw [->] (Phi) -- (Ber);
    \draw [->] (Ber) -- (r);
    \draw [->] (switch) -- (GP);

\end{tikzpicture}
    \caption{A Forney-style factor graph representation of the user response model specified by \eqref{eq:user-response-model}. The context state $\bm{c}_k$ is passed to the GP Mixture (GPM) node as a selector variable for its arguments.}
    \label{fig:model:user}
\end{figure}
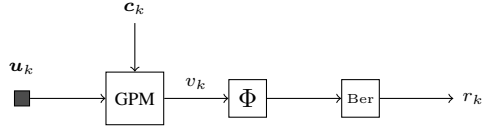

In \eqref{eq:model:GP-1}, $v_k(\cdot)$ is a latent function drawn from a mixture of GPs with mean functions $m_l(\cdot)$ and kernels $K_l(\cdot,\cdot)$. Evaluating $v_k(\cdot)$ at the point $\bm{u}_k$ provides an estimate of user preferences. Without loss of generality, we can set $m_l(\cdot) = 0$. 
Since $\bm{c}_k$ is one-hot encoded, raising to the power $c_{lk}$ serves to select the GP that corresponds to the active context.
$\Phi(\cdot)$ denotes the Gaussian cumulative distribution function, defined as $\Phi(x) = \frac{1}{\sqrt{2\pi}} \int_{-\infty}^x \exp \left( -t^2/2\right) \mathrm{d}t$. This map in \eqref{eq:sigmoid-user-response-1} casts $v_k(\bm{u}_k)$ to a Bernoulli-distributed variable $r_k$. An FFG representation of the user response model is shown in Figure~\ref{fig:model:user}.

\section{Solving tasks by probabilistic inference} \label{sec:inference}
This section elaborates on solving the three tasks of Section~\ref{sec:problem:AIDA}: 1) context classification, 2) trial design and 3) hearing aid algorithm execution.
All tasks can be solved through probabilistic inference in the generative model specified by \eqref{eq:model:acoustic}-\eqref{eq:sigmoid-user-response-1} in Section \ref{sec:model}.
In this section, the inference goals are formally specified based on the previously proposed generative model.

For the realization of the inference tasks we will use variational message passing in a factor graph representation of the generative model. 
Message passing-based inference is highly efficient, modular and scales well to large inference tasks \citep{loeliger_factor_2007, cox_factor_2019}.
With message passing, inference tasks in the generative model reduce to automatable procedures revolving around local computations on the factor graphs.

A thorough discussion on message passing and related topics is omitted here for readability, but made available in Appendix~\ref{sec:appendix:background} to serve as reference.

\subsection{Inference for context classification} \label{sec:inference:context}
% what is the problem

The acoustic context $\bm{c}_k$ describes the dynamics of the background noise model through \eqref{eq:model:acoustic-noise} and \eqref{eq:model:acoustic-noise-context}.
For determining the current environment of the user, the goal is to infer the current context based on the preceding observations.
Technically we are interested in determining the marginal distribution $p(\bm{c}_k\mid \bm{x}_{1:t^+})$, where the index range over $t$ of $\bm{x}$ takes into account the relation between $t$ and $k$ as defined in \eqref{eq:model:kt}. 
In our online setting, we wish to calculate this marginal distribution iteratively by solving
\begin{equation}\label{eq:inference:context}
\begin{split}
    \underbrace{p(\bm{c}_k \mid \bm{x}_{1:t^+})}_\text{posterior} 
    &\propto 
    \int 
    \underbrace{p(z_{t^-:t^+}, \Psi_{k}, \bm{x}_{t^-:t^+} \mid z_{t^--1}, \bm{c}_k)}_\text{observation model}\
    \underbrace{p(\bm{c}_k, \mathrm{T}\mid \bm{c}_{k-1})}_\text{context dynamics} \\
    &\qquad\qquad\qquad \cdot
    \underbrace{p(\bm{c}_{k-1}, z_{t^--1} \mid \bm{x}_{1:t^--1})}_\text{prior}\
    \mathrm{d}z_{t^--1:t^+}\
    \mathrm{d}\Psi_k\
    \mathrm{d}\bm{c}_{k-1}\
    \mathrm{d}\mathrm{T}.
    \end{split}
\end{equation}

The observation model is fully specified by the model specification in Section~\ref{sec:model}, similarly as the context dynamics.
The prior distribution is a joint result of the iterative execution of both \eqref{eq:inference:context} and \eqref{eq:inference:bayes}, where the latter refers to the HA algorithm execution from Section~\ref{sec:inference:acoustic}.
The calculation of this marginal distribution renders intractable and therefore exact inference of the context is not possible.
This is a result of 1) the intractability resulting from the autoregressive model as described in the previous subsection and of 2) the intractability that is a result of performing message passing with mixture models.
In \eqref{eq:model:acoustic-noise-context} the model structure contains a Normal and Gamma mixture model for the AR-coefficients and process noise precision parameter, respectively.
Exact inference with these mixture models quickly leads to intractable inference through message passing, especially when multiple background noise models are involved.
Therefore, we need to resort to a variational approximation where the output messages of these mixture models are constrained to be within the exponential family.

% why don't we just do VMP then?
Although variational inference with the mixture models is feasible \citep{bishop_pattern_2006, van_de_laar_automated_2019, podusenko_message_2021-1}, it is prone to converge to local minima of the Bethe free energy (BFE) for more complicated models.
The variational messages originating from the mixture models are constrained to either Normal or Gamma distributions, possibly losing important multi-modal information, and as a result they can lead to suboptimal inference of the context variable.
Because the context is vital for the above underdetermined source separation stage, we wish to limit the amount of (variational) approximations during context inference.
At the cost of an increased computational complexity, we will remove the variational approximation around the mixture models and instead expand the mixture components into distinct models.
As a result, each distinct model now contains one of the mixture components for a given context and now results in exact messages originating from the priors of $\bm{\xi}_k$ and $\tau_k$.
Therefore we only need to resort to a variational approximation for the auto-regressive node.
By expanding the mixture models into distinct models to reduce the number of variational approximations, calculation of the posterior distribution of the context $p(\bm{c}_k\mid \bm{x}_{1:t^+})$ reduces to an approximate Bayesian model comparison problem, similarly as described in \cite{van_erp_bayesian_2021}.
Appendix~\ref{sec:appendix:realization:context} gives a more in-depth description on how we use Bayesian model comparison for solving the inference task in \eqref{eq:inference:context}.

\subsection{Inference for trial design of HA tuning parameters}\label{sec:inference:trial}

% \subsubsection{Optimal selection of parameter trials} \label{sec:efe}
%
The goal of proposing alternative HA tuning parameter settings (task 3) is to receive positive user responses in the future. Free energy minimization over desired future user responses can be achieved through a procedure called Expected Free Energy (EFE) minimization \citep{friston_active_2015,sajid_active_2021}. 

EFE as a trial selection criterion induces a natural trade-off between explorative (information seeking) and exploitative (reward seeking) behaviour. In the context of situated HA personalization, this is desirable because soliciting user feedback can be burdensome and invasive, as described in Section~\ref{sec:problem:AIDA}. From the agent's point of view, this means that striking a balance between gathering information about user preferences and satisfying learned preferences is vital. The EFE provides a way to tackle this trade-off, inspired by neuro-scientific evidence that brains operate under a similar protocol \citep{friston_active_2015,parr_uncertainty_2017}.
The EFE is defined as \citep{friston_active_2015}
\begin{align}\label{eq:efe}
    G_{\bm{u}}[q]
    &= \mathbb{E}_{q(r,v\mid \bm{u})} \left[ \ln \frac{q(v\mid \bm{u})}{p(r,v\mid \bm{u})} \right]\,,
\end{align}
%which is known as the Expected Free Energy criterion for trials $\bm{u}$. 
where the subscript indicates that the EFE is a function of a trial $\bm{u}$.  The EFE can be decomposed into \citep{friston_active_2015}
\begin{align}
    G_{\bm{u}}[q]
    &\approx -\underbrace{\mathbb{E}_{q(r \mid \bm{u})} \bigg[ \ln p(r) \bigg]}_{\text{Utility drive}}
    -\underbrace{\mathbb{E}_{q(r,v\mid \bm{u})} \bigg[ \ln \frac{q(v \mid \bm{u}, r)}{q(v \mid \bm{u})} \bigg]}_{\text{Information gain}}\,, \label{eq:efe_bound}
\end{align}
which contains an information gain term and a utility-driven term. Minimization of the EFE reduces to maximization of both these terms.
Maximization of the utility drive pushes the agent towards matching predicted user responses $q(r \mid \bm{u})$ with a goal prior over \emph{desired} user responses $p(r)$. This goal prior allows encoding of beliefs about future observations that we wish to observe. Setting the goal prior to match positive user responses then drives the agent towards parameter settings that it believes make the user happy in the future.
The information gain term in \eqref{eq:efe_bound} drives agents that optimize the EFE to seek out responses that are maximally informative about latent states $v$.

To select the next set of gains $\bm{u}$ to propose to the user, we need to find
\begin{align}\label{eq:min_min}
    \bm{u}^* &=  \argmin_{\bm{u}} \left(\min_{q} G_{\bm{u}}[q] \right) \,.
\end{align}
Intuitively, one can think of \eqref{eq:min_min} as a two step procedure with an inner and an outer loop. The inner loop finds the approximate posterior $q$ using (approximate) Bayesian inference, conditioned on a particular action parameter $\bm{u}$. The outer loop evaluates the resulting EFE as a function of $\bm{u}$ and proposes a new set of gains to bring the EFE down.
For our experiments we consider a candidate grid of possible gains. For each candidate we compute the resulting EFE and then select the lowest scoring proposal as the next set of gains to be presented to the user.

The probabilistic model used for AIDA is a mixture GPC. For simplicity we will restrict inference to the GP corresponding to the MAP estimate of $\bm{c}_k$. Between trials, the corresponding GP needs to be updated to adapt to the new data gathered from the user. Specifically, we are interested in finding the posterior over the latent user preference function
\begin{align}
    p(v^* \mid \bm{u}_{1:k}, r_{1:k-1})
    %&= p(v^* \mid \bm{u}_k, \bm{u}_{1:k-1}, r_{1:k-1}) \\
    &= \int p(v^* \mid \bm{u}_{1:k-1},\bm{u}_{k},v) p(v \mid \bm{u}_{1:k-1},r_{1:k-1}) \mathrm{d} v \,.
    %&= \int p(v^* \mid \bm{u}_{1:k-1},\bm{u}_{k},v) p(v | D) \mathrm{d} v \label{eq:joint_gaussian}
\end{align}
where we assume AIDA has access to a dataset consisting of previous queries $\bm{u}_{1:k-1}$ and appraisals $r_{1:k-1}$ and we are querying the model at $\bm{u}_k$.
While this inference task in the GPC is intractable, there exist a number of techniques for approximate inference, such as variational Bayesian methods, Expectation Propagation, and the Laplace approximation \citep{rasmussen_gaussian_2006}. Appendix~\ref{sec:appendix:realization:trial} describes the exact details of the inference realization of the inference tasks of AIDA.

\subsection{Inference for executing the hearing aid algorithm}\label{sec:inference:acoustic}

% \subsubsection{Inference of acoustic source signals}\label{sec:inference:algorithm}
The main goal of the proposed hearing aid algorithm is to improve audibility and intelligibility by re-weighing inferred source signals in the HA output signal. In our model of the observed signal in \eqref{eq:model:acoustic}-\eqref{eq:model:acoustic-noise-context} we are interested in iteratively inferring the marginal distribution over the latent speech and noise signals $p(\bm{s}_t, \bm{n}_t \mid \bm{x}_{1:t})$. This inference task is in literature sometimes referred to as informed source separation \citep{knuth_informed_2013}. Inferring the latent speech and noise signals tries to optimally disentangle these signals from the observed signal based on the sub-models of the speech and noise source.% Clearly, before we can infer suitable gains, we need to infer the latent states of our acoustic environment. 
This requires us to compute the posterior distributions associated with the speech and noise signals. To do so, we perform probabilistic inference by means of message passing in the acoustic model of \eqref{eq:model:acoustic}-\eqref{eq:model:acoustic-noise-context} .
The posterior distributions can be calculated in an online manner using sequential Bayesian updating by solving the Chapman-Kolmogorov equation \citep{sarkka_bayesian_2013}
\begin{equation}\label{eq:inference:bayes}
\begin{split}
    &\underbrace{p(z_t, \Psi_k \mid\bm{x}_{1:t})}_{\text{posterior}} \\
    &\qquad\propto\underbrace{p(x_t\mid z_t)}_{\text{observation}} \int{\underbrace{p(z_t\mid z_{t-1}, \Psi_k)}_{\text{state dynamics}}\underbrace{p(z_{t-1}, \Psi_k \mid  \bm{x}_{1:t-1})}_{\text{prior}}}\d z_{t-1}, \quad \text{for }t=t^-,t^-+1,\ldots,t^+
    \end{split}
\end{equation}
where $z_t$ and $\Psi_k$ denote the sets of dynamic states and static parameters $z_t=\{\bm{\theta}_t, \bm{s}_t, \bm{n}_t\}$ and $\Psi_k=\{\gamma, \tau_k, \bm{\zeta}_k\}$, respectively. Here, the states and parameters correspond to the latent AR and TVAR models of \eqref{eq:model:acoustic-speech} and \eqref{eq:model:acoustic-noise}. Furthermore, we assume that the context does not change, i.e. $k$ is fixed. When the context does change \eqref{eq:inference:bayes} will need to be extended by integrating over the varying parameters.
Unfortunately, the solution of \eqref{eq:inference:bayes} is not analytically tractable. This happens because of 1) the integration over large state spaces, 2) the non-conjugate prior-posterior pairing, and 3) the absence of a closed-form solution for the evidence factor \citep{podusenko_message_2021}.
To circumvent this issue, we resort to a hybrid message passing algorithm that combines structured variational message passing (SVMP) and loopy belief propagation for the minimization of Bethe free energy \citep{senoz_variational_2021}.
Appendix~\ref{sec:appendix:background} describes these concepts in more detail.

For the details of the SVMP and BP algorithms, we refer the reader to Appendix~\ref{sec:appendix:background} and \citep{dauwels_variational_2007, senoz_variational_2021}.
Owing to the modularity of the factor graphs, the message passing update rules can be tabulated and only need to be derived once for each of the included factor nodes. The  derivations of the sum-product update rules for elementary factor nodes can be found in \cite{loeliger_factor_2007} and the derived structured variational rules for the composite AR node can be found in \cite{podusenko_message_2021}. The variational updates in the mixture models can be found in \cite{van_de_laar_automated_2019, podusenko_message_2021-1}. The required approximate marginal distribution of some variable $z$ can be computed by multiplying the incoming and outgoing variational messages on the edges corresponding to the variables of our interest as $q(z) \propto \vec{\nu}(z)\cdot \cev{\nu}(z)$.

Based on the inferred posterior distributions of $s_t$ and $n_t$, these signals can be used for inferring the hearing aid output through \eqref{eq:model:HA-output} to produce a personalized output which compromises between residual noise and speech distortion.

\section{Experimental verification \& validation} \label{sec:experiments}

In this section, we first verify our approach for the three design tasks of Section~\ref{sec:problem:AIDA}.
Specifically, in Section~\ref{sec:experiments:context} we evaluate the context inference approach by reporting the classification performance of correctly classifying the context corresponding to a signal segment.
In Section~\ref{sec:experiments:trial} we evaluate the performance of our intelligent agent that actively proposes hearing aid settings and learns user preferences.
The execution of the hearing aid algorithm is verified in Section~\ref{sec:experiments:HA} by evaluating the source separation performance.
To conclude this section, we present a demonstrator for the entire system in Section~\ref{sec:experiments:demo}.

All algorithms have been implemented in the scientific programming language \texttt{Julia} \citep{bezanson_julia:_2017}. Probabilistic inference in our model is automated using the open source \texttt{Julia} package \texttt{ReactiveMP}\footnote{\texttt{ReactiveMP} \citep{bagaev_reactive_2022} is available at \url{https://github.com/biaslab/ReactiveMP.jl}.}  \citep{bagaev_reactive_2022}. \bdv{Please change Dmitry paper to "submitted.". Also in the references section, the title of the paper is not grammatically separated from the status of the submission process.} All of the experiments presented in this section can be found at our AIDA GitHub repository\footnote{The AIDA GitHub repository with all experiments is available at \url{https://github.com/biaslab/AIDA}.}.

\subsection{Context classification verification}\label{sec:experiments:context}
To verify that the context is appropriately inferred through Bayesian model selection, we generated synthetic data from the following generative model:
\begin{subequations}\label{eq:verification-context-model}
\begin{align}
    \bm{c}_k \sim \mathrm{Cat}(\mathrm{T}\bm{c}_{k-1})
\end{align}
with priors
\end{subequations}
\begin{subequations}
\begin{align}
    \bm{c}_0 &\sim \mathrm{Cat}(\bm{\pi})\\
    \mathrm{T}_{1:L,j} &\sim \mathrm{Dir}(\bm{\alpha}_j),
\end{align}
\end{subequations}
where $\bm{c}_o$ is chosen to have length $L=4$. The event probabilities $\bm{\pi}$ and concentration parameters $\bm{\alpha}_j$ are defined as $\bm{\pi} = [0.25, 0.25, 0.25, 0.25]^\intercal$ and $\bm{\alpha}_j = [1.0, 1.0, 1.0, 1.0]^\intercal$, respectively.
We generated a sequence of 1000 frames, each containing 100 samples, such that we have 100 x 1000 data points. Each frame is associated with one of the 4  different contexts. Each context corresponds to an AR model with the parameters presented in Table~\ref{table:verification-context}.
\begin{table}[t]
\centering
\caption{The parameters of autoregressive processes that are used for generating a time series with simulated context dynamics.}
\label{table:verification-context}
\begin{tabular}{c|llll|c} 
 AR order& $\bm{\zeta}$ & & & & $\tau^{-1}$ \\ 
 \hline &&&&& \\[-1.5ex]
 1 & -0.308 & & & & 1.0 \\ 
 2 & 0.722  & -0.673 & & & 2.0 \\
 3 & -0.081 & 0.079 & -0.362 & & 0.5 \\
 4 & -1.433 & -0.174 & 0.757 & 0.466 & 1.0 \\
\end{tabular}
\end{table}
% Figure \ref{fig:experiments:context_signal} shows a fragment of the generated time series. 
% \begin{figure}
%     \centering
%     % \resizebox{\textwidth}{!}{\input{figures/experiments/context_signals}}
%     \resizebox{\textwidth}{!}{\includesvg{figures/experiments/context/context_signals.svg}}
%     \caption{A fragment of the time series from the generated dataset that is used to verify the context classification procedure. Each color represents a different context that corresponds to one of the autoregressive models from Table \ref{table:verification-context}. Observations are color-coded according to the autoregressive models they are generated from.}
%     \label{fig:experiments:context_signal}
% \end{figure}

\begin{figure}
    \centering
    \resizebox{\textwidth}{!}{\includegraphics{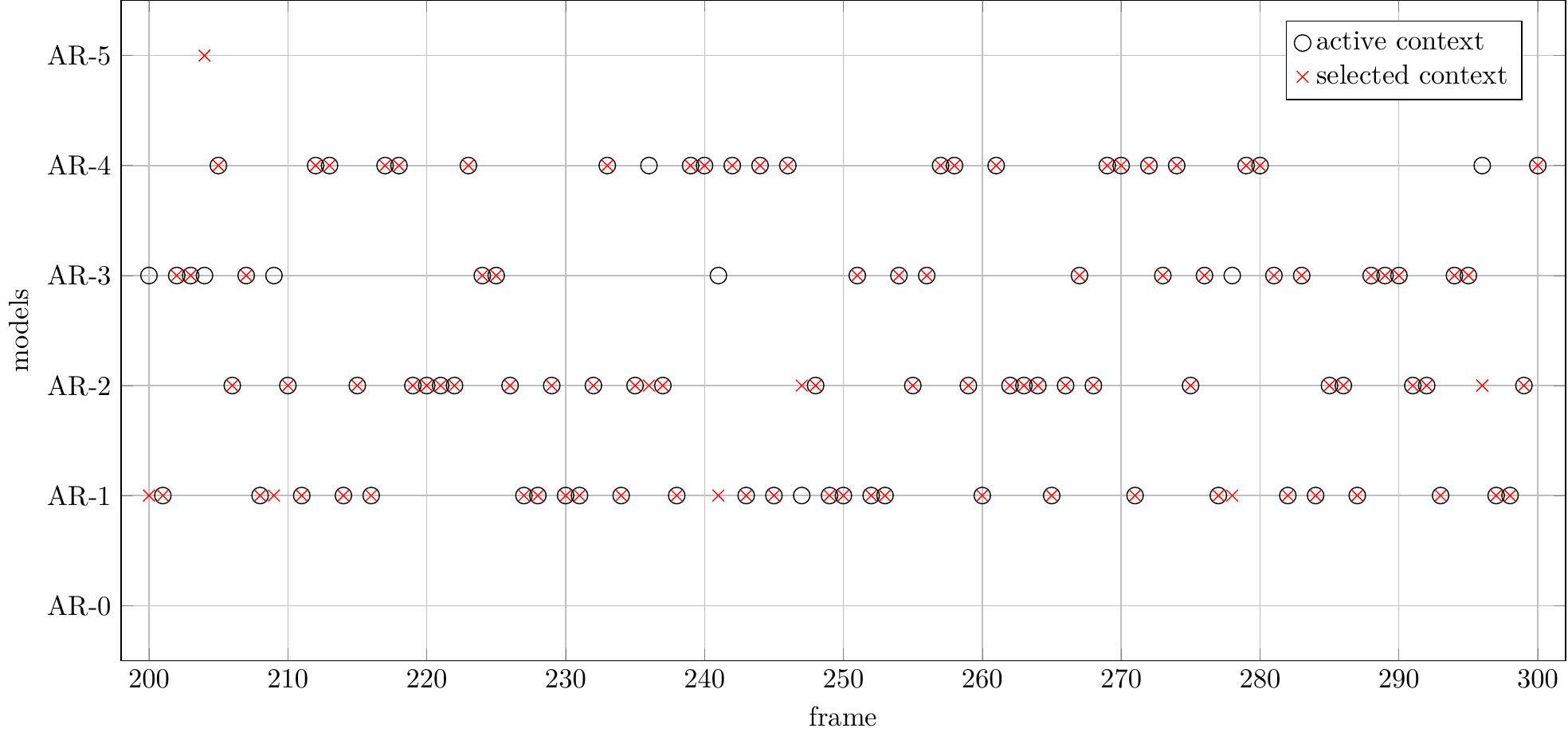}}
    \caption{True and inferred evolution of contexts from frames 200 to 300. Each frame consists of 100 data points. Circles denote the active contexts that were used to generate the frame. Crosses denote the model that achieves the lowest Bethe free energy for a specific frame.}
    \label{fig:experiments:contexts}
\end{figure}

For verification of the context classification procedure, we wish to identify which model best approximates the observed data. To do that, 4 models with the same specifications as were used to generate the dataset were employed. We used informative priors for the coefficients and precision of AR models. Additionally, we extended our set of models with an AR(5) model with weakly informative priors and a Gaussian i.i.d. model that can be viewed as an AR model of zeroth order, i.e. AR(0). The individual frames containing 100 samples each were processed individually and we computed the Bethe free energy for each of the different models. The Bethe free energy is introduced in Appendix~\ref{sec:appendix:background:bfe}. By approximating the true model evidence using the Bethe free energy as described in Appendix~\ref{sec:appendix:realization:context}, we performed approximate Bayesian model selection by selecting the model with the lowest Bethe free energy. This model then corresponds to the most likely context hat we are in. We highlight the obtained inference result in Figure~\ref{fig:experiments:contexts}.

We evaluate the performance of the context classification procedure using approximate Bayesian model selection by computing the categorical accuracy metric defined as
\begin{equation}
    acc = \frac{tp + tn}{N}
\end{equation}
where $tp$, $tn$ are the number of true positive and true negative values, respectively. $N$ corresponds to the number of total observations, which in this experiment is set to $N=1000$. In this context classification experiment, we have achieved a categorical accuracy of $acc = 0.94$.

\subsection{Trial design verification}\label{sec:experiments:trial}
\begin{figure}[t]
    \centering
        \resizebox{0.6\textwidth}{!}{\includegraphics{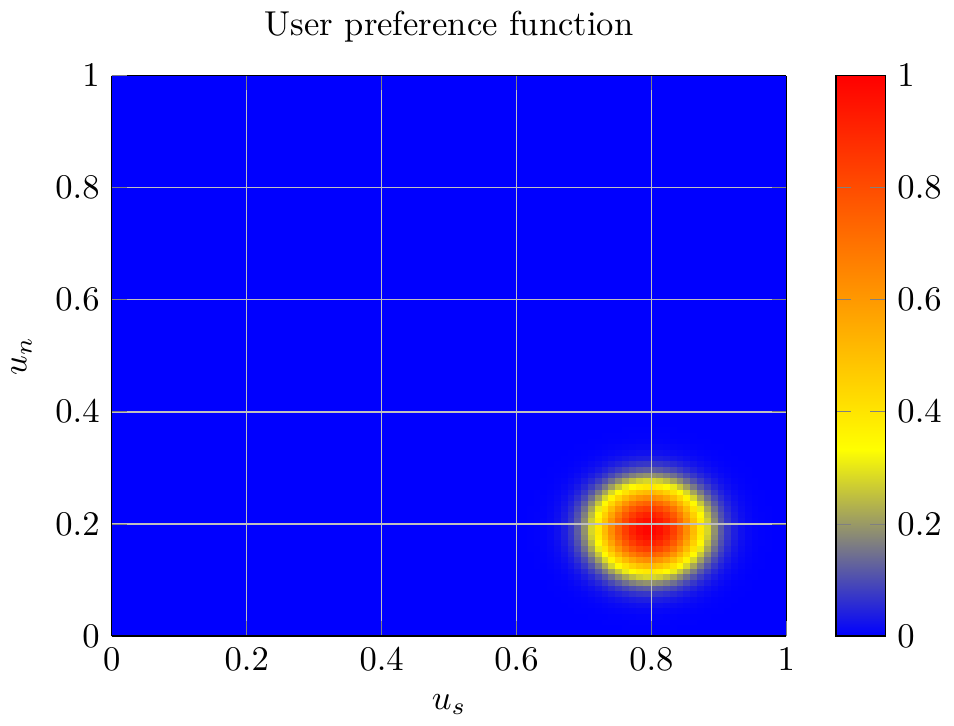}}
    \caption{Simulated user preference function $p(r_k = 1\mid \bm{u}_k)$. The coloring corresponds to the probability of the user giving a positive appraisal for the search space of gains $\bm{u}_k =[u_{sk}, u_{nk}]^\intercal$.}
    \label{fig:user_prefs}
\end{figure}

Evaluating the performance of the intelligent agent is not trivial. Because the agent adaptively trades off exploration and exploitation, accuracy is not an adequate metric. There are reasons for the agent to veer \emph{away} from what it believes is the optimum to obtain more information.
As a verification experiment we can investigate how the agent interacts with a simulated user. Our simulated user samples binary appraisals $r_k$ based on the HA parameters $\bm{u}_k$ as
\begin{align}\label{eq:pref_function}
    r_k &\sim  \mathrm{Ber} \bigg( \frac{2}{1 + \exp \big( (\bm{u}_k - \bm{u}^*)^T \Lambda_{\text{user}} (\bm{u}_k - \bm{u}^*) \big)} \bigg) \,,
\end{align}
where $\bm{u}^*$ denotes the optimal parameter setting, $\bm{u}_k$ is the set of parameters proposed by AIDA at time $k$, $\Lambda_{\text{user}}$ is a diagonal weighing matrix that controls how quickly the probability of positive appraisals decays with the squared distance to $\bm{u}^*$. The constant $2$ ensures that when $\bm{u}_k = \bm{u}^*$, the probability of positive appraisals is $1$ instead of $0.5$. For our experiments, we set $\bm{u}^* = [0.8,0.2]^\intercal$ and the diagonal elements of $\Lambda_{\text{user}}$ to $0.004$. This results in the user preference function $p(r_k=1 \mid \bm{u}_k)$ as shown in Figure~\ref{fig:user_prefs}.
The kernel used for AIDA is a squared exponential kernel, given by 
\begin{align}
    K(\bm{u},\bm{u}^\prime) = \sigma^2 \exp\left\{-\frac{\|\bm{u}-\bm{u}^\prime\|^2_2}{2l^2}\right\},
\end{align}
where $l$ and $\sigma$ are the hyperparameters of this kernel. Intuitively, $\sigma$ is a static noise parameter and $l$ encodes the smoothness of the kernel function. Both hyperparameters were initialized to $\sigma = l = 0.5$, which is uninformative on the scale of the experiment. We let the agent search for $80$ trials and update hyperparameters every $5^\text{th}$ trial using conjugate gradient descent as implemented in \texttt{Optim.jl} \citep{k_mogensen_optim_2018}. We constrain both hyperparameters to the domain $[0.1,1]$ to ensure stability of the optimization. As we will see, for large parts of each experiment AIDA only receives negative appraisals. The generative model of AIDA is fundamentally a classifier and unconstrained optimization can therefore lead to degenerate results when the data set only contains examples of a single class. For all experiments, the first proposal of AIDA was a randomly sampled parameter from the admissible set of parameters, because the AIDA has no prior knowledge about the user preference function. This random initial proposal, lead to distinct behaviour for all simulated agents.

\begin{figure}[h!]
    \centering
        \resizebox{1.0\textwidth}{!}{\includegraphics{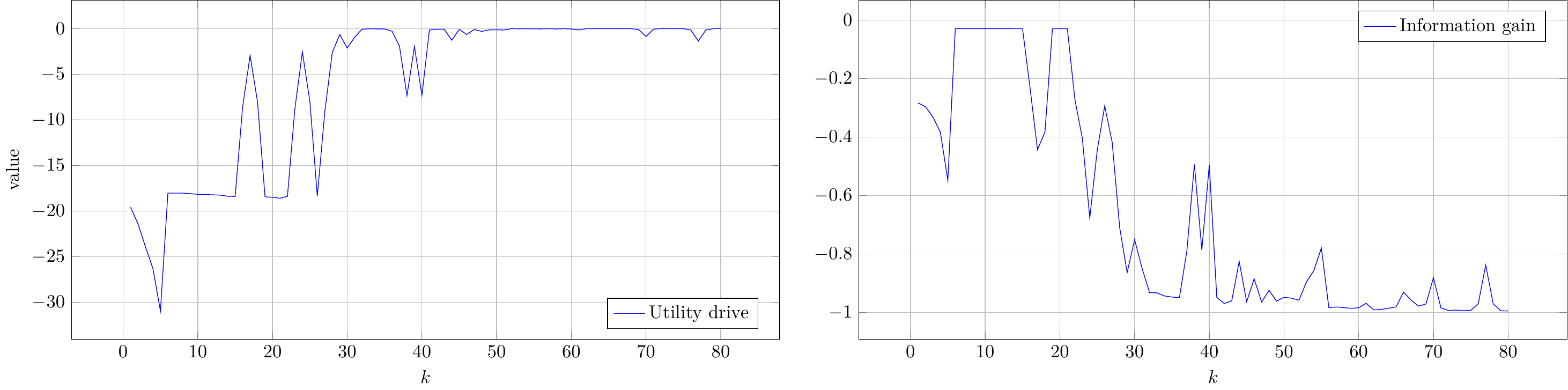}}
    \caption{Evolution of the utility drive and negative information gain after throughout a single experiment.}
    \label{fig:grouped_agent}
\end{figure}

We provide two verification experiments for AIDA. First, we will thoroughly examine a single run in order to investigate how AIDA switches between exploratory and exploitative behaviour. Secondly, we examine the aggregate performance of an ensemble of agents to test the average performance. 
To assess the performance for a single run, we can examine the evolution of the distinct terms in the EFE decomposition of \eqref{eq:efe_bound} over time. We expect that when AIDA is primarily exploring, the utility drive is relatively low while the information gain is relatively high. When AIDA is primarily engaged in exploitation, we expect the opposite pattern. We show these terms separately in Figure~\ref{fig:grouped_agent}.

\begin{figure}[h!]
    \centering
    \begin{minipage}[b]{0.4\textwidth}
        \centering
        \resizebox{1.0\textwidth}{!}{\includegraphics{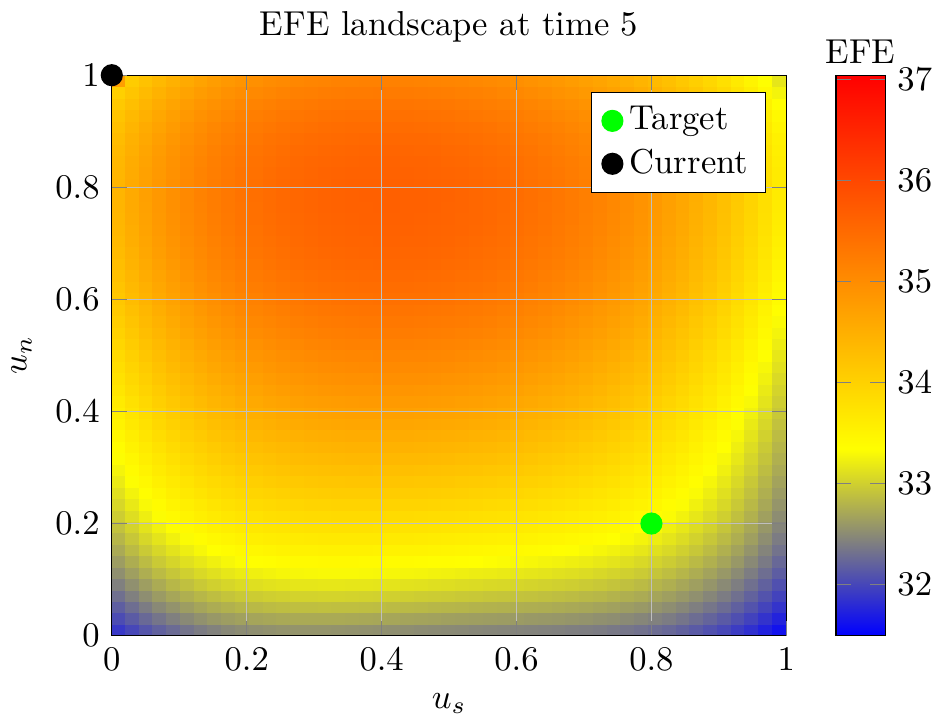}}
    \end{minipage}
    \hspace{2em}
    \begin{minipage}[b]{0.4\textwidth}
        \centering
        \resizebox{1.0\textwidth}{!}{\includegraphics{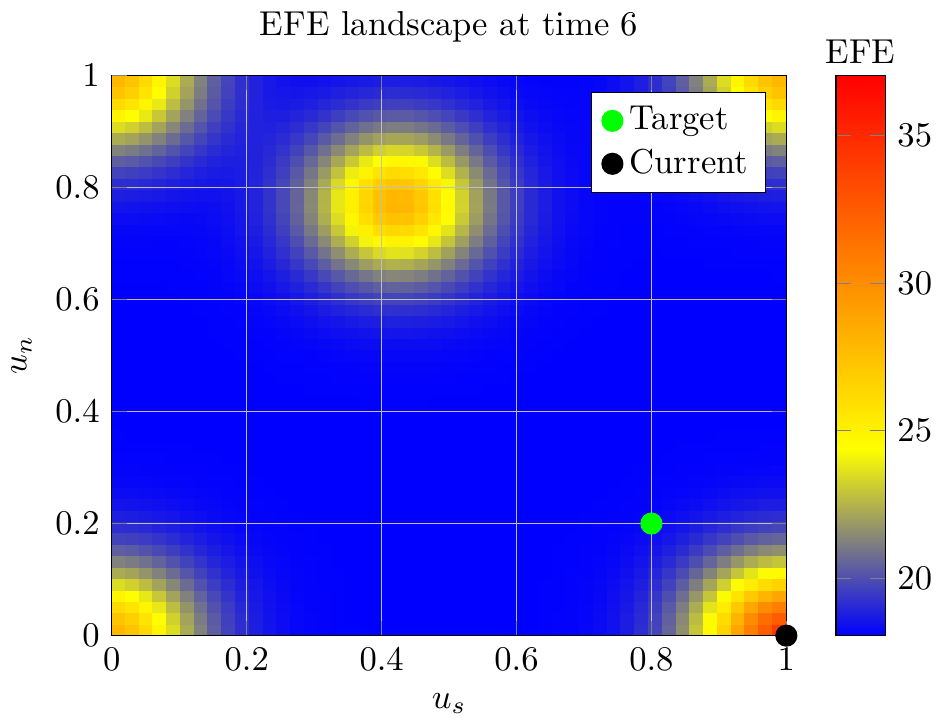}}
    \end{minipage}

    \begin{minipage}[b]{0.4\textwidth}
        \centering
        \resizebox{1.0\textwidth}{!}{\includegraphics{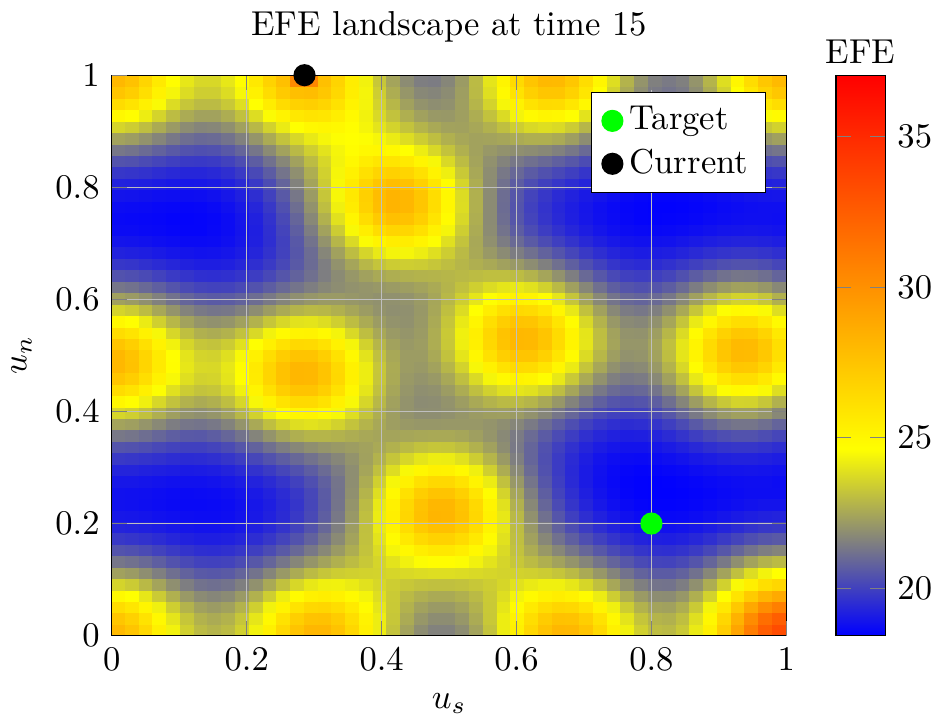}}
    \end{minipage}
    \hspace{2em}
    \begin{minipage}[b]{0.4\textwidth}
        \centering
        \resizebox{1.0\textwidth}{!}{\includegraphics{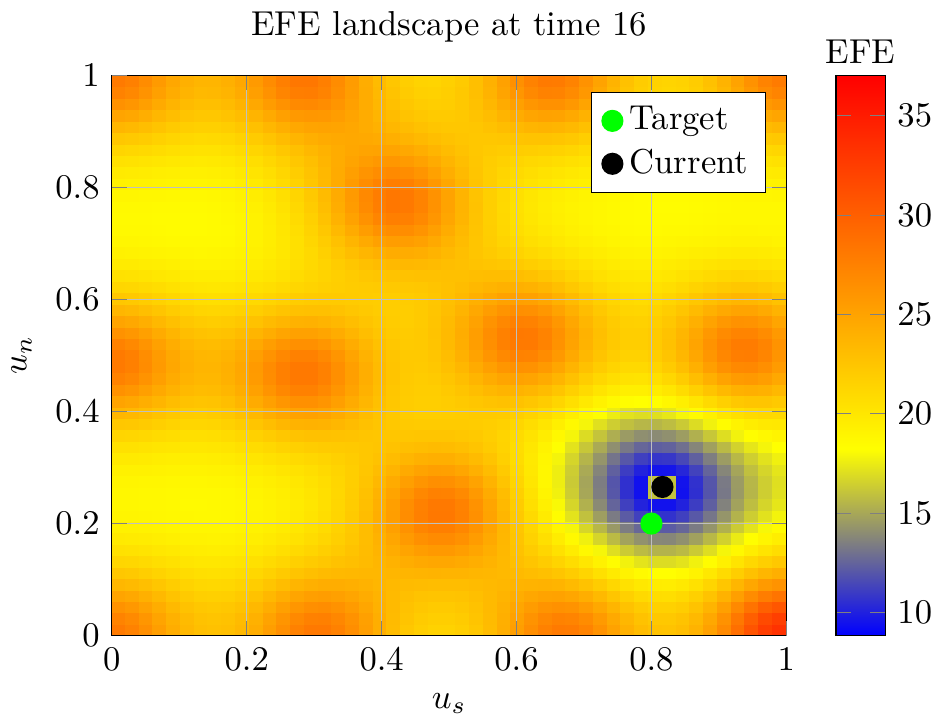}}
    \end{minipage}

    \begin{minipage}[b]{0.4\textwidth}
        \centering
        \resizebox{1.0\textwidth}{!}{\includegraphics{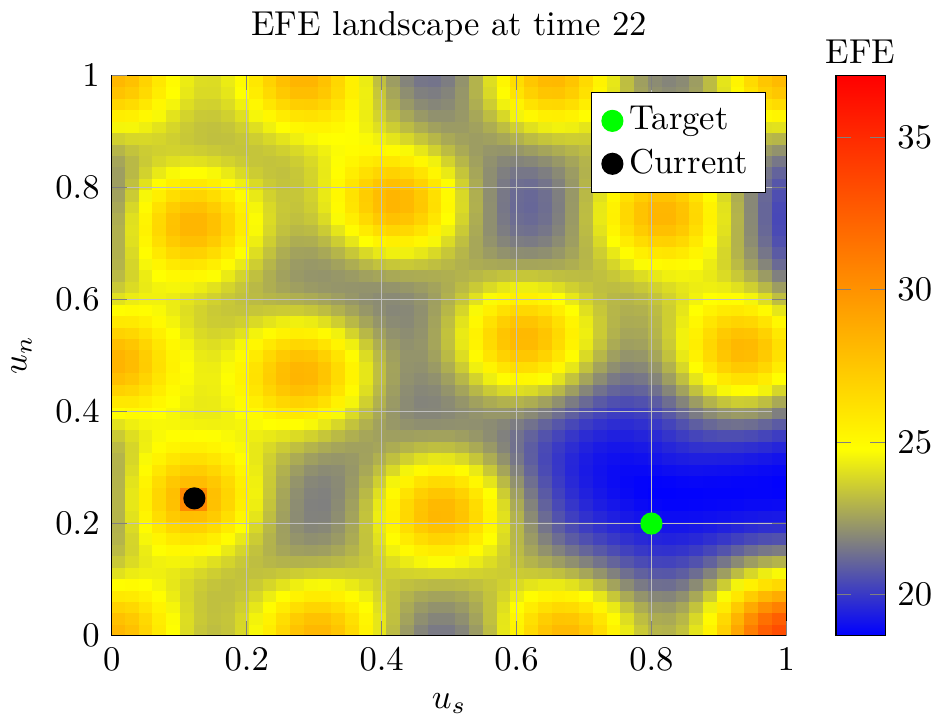}}
    \end{minipage}
    \hspace{2em}
    \begin{minipage}[b]{0.4\textwidth}
        \centering
        \resizebox{1.0\textwidth}{!}{\includegraphics{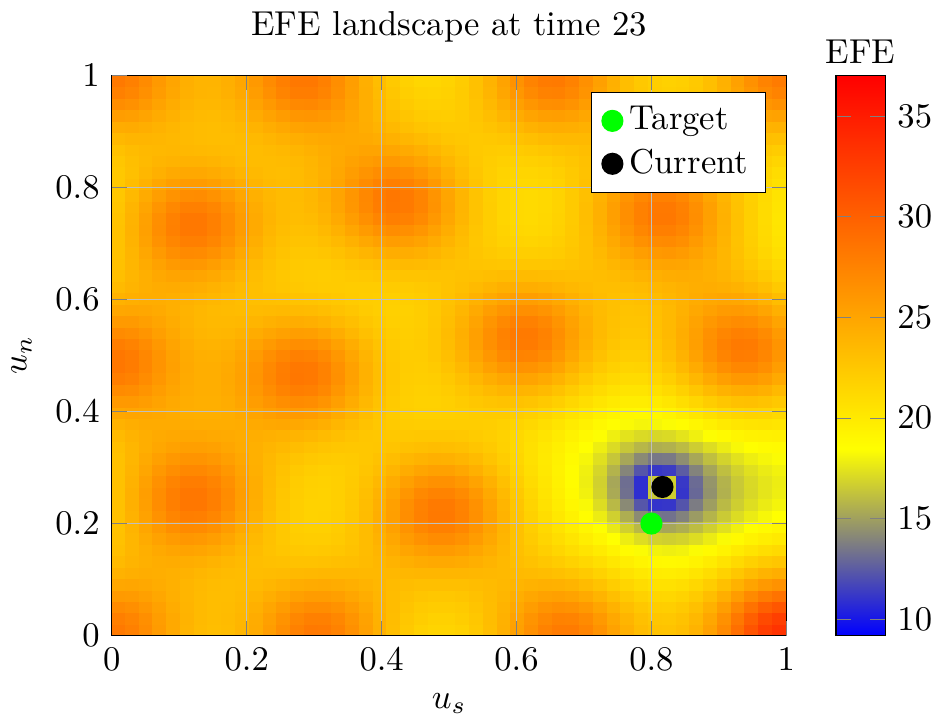}}
    \end{minipage}
    
    \begin{minipage}[b]{0.4\textwidth}
        \centering
        \resizebox{1.0\textwidth}{!}{\includegraphics{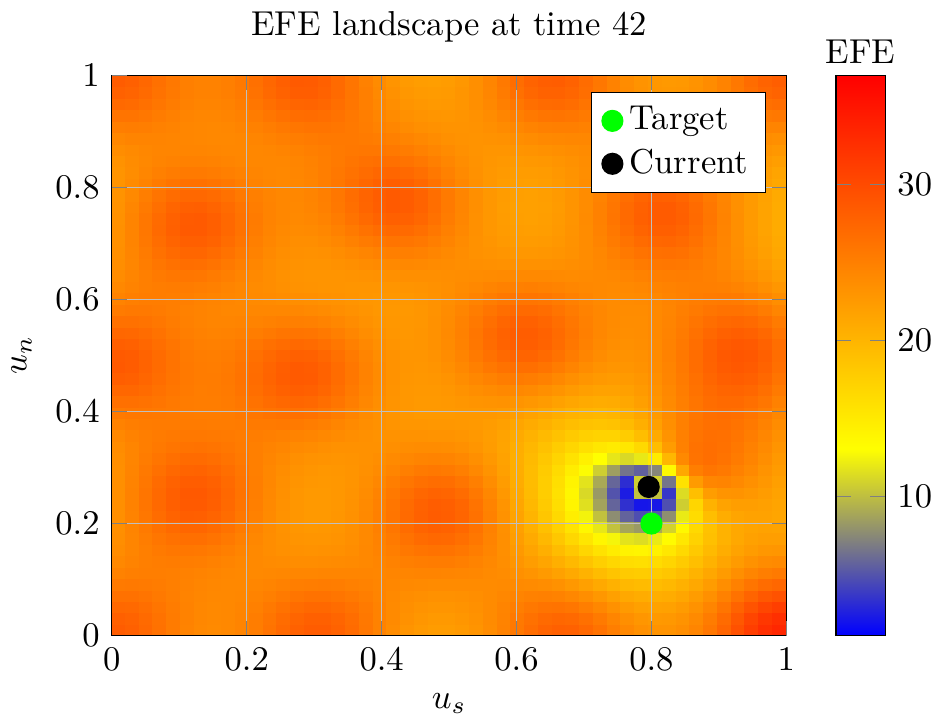}}
    \end{minipage}
    \hspace{2em}
    \begin{minipage}[b]{0.4\textwidth}
        \centering
        \resizebox{1.0\textwidth}{!}{\includegraphics{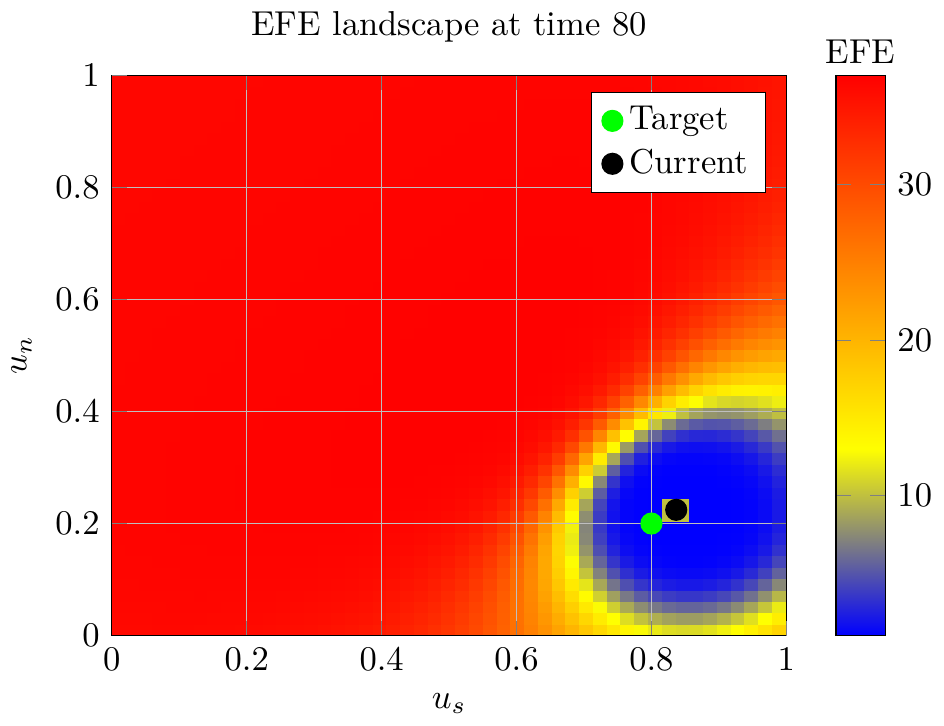}}
    \end{minipage}
    \caption{Snapshot of EFE landscape at different time points as a function of gains $u_{s}$ and $u_{n}$. The black dot denotes the current parameter settings and the green dot denotes $\bm{u}^*$.}
    \label{fig:efe_experiments}
\end{figure}

Figure~\ref{fig:grouped_agent} shows that there are distinct phases to the experiment. In the beginning ($k < 5$) AIDA sees a sharp decrease in utility drive and information gain terms. This indicates a saturation of the search space such that no points present good options. This happens early due to uninformative hyperparameter settings in the GPC. After trial 5, these hyperparameters are optimized and the agent no longer thinks it has saturated the search space, which can be explained by the jumps in Figure~\ref{fig:grouped_agent} from trial $5$ to $6$. From trial $6$ throughout $15$ we observe a relatively high information gain and relatively low utility drive, meaning that the agent is still exploring the search space for parameter settings which yield a positive user appraisal. The agent obtains its first positive appraisal at $k=16$, as denoted by the jump in utility drive and drop in information gain. This first positive appraisal is followed by a period of oscillations in both terms, where the agent is refining its parameters. Finally AIDA settles down to predominantly exploitative behaviour starting from $41^\text{st}$ trial. To examine the first transition, we can visualize the EFE landscape at $k=5$ and $k=6$, the upper row of Figure~\ref{fig:efe_experiments}.

Recall that AIDA is minimizing EFE. Therefore, it is looking for the lowest values corresponding to blue regions and avoiding the high values corresponding to red regions. Between $k=5$ and $k=6$ we perform the first hyperparameter update, which drastically changes the EFE landscape. This indicates that initial parameter settings were not informative, as we did not cover the majority of the search space within the first 5 iterations. The yellow regions at $k=6$ indicates regions corresponding to previous proposals of AIDA that resulted into negative appraisals.
We can visualize snapshots of the exploration phase starting from $k=6$ in a similar manner. The second row of Figure~\ref{fig:efe_experiments} displays the EFE landscape at two different time instances during the exploration phase. It shows that over the course of the experiment, AIDA gradually builds a representation over the search space. In trial $16$ this takes the form of patterns of connected regions that denote areas that AIDA believes are unlikely to results in positive appraisals. 

Once AIDA receives its first positive appraisal at $k=16$, it switches from exploring the search space to focusing only on the local region. If we examine Figure~\ref{fig:grouped_agent}, we see that at this time the information gain term is still reasonably high. This indicates a subtle point: once AIDA receives a positive appraisal, it starts with \emph{local} exploration around where the optimum might be located. However, the agent was located near the boundary of the optimum and next receives a negative appraisal. Therefore in trials $18$ to $22$ AIDA queries points which it deems most informative. At time $23$ the position of AIDA in the search space (black dot in the third row of Figure~\ref{fig:efe_experiments}) returns to the edge of the user preference function in Figure~\ref{fig:user_prefs}. This causes AIDA to receive a mixture of positive and negative appraisals in the following trials, leading to the oscillations seen in Figure~\ref{fig:grouped_agent}. 
Finally, we can examine the landscape after AIDA has confidently located the optimum and switched to purely exploitative behavior.
This happens at $k=42$ where the utility drive goes to $0$ and the information gain concentrates around $-1$. 

The last row of Figure~\ref{fig:efe_experiments} shows that once $\bm{u}^*$ is confidently located, AIDA disregards the remainder of the search space in favour of providing good parameter settings. Finally, if the user continues to supply data to AIDA, it will gradually extend the potential region of samples around the optimum. This indicates that if a user keeps requesting updated parameters, AIDA will once again perform local exploration around the optimum. This further indicates that AIDA accommodates gradual retraining as user's hearing loss profile changes over time.

Having thoroughly examined an example run and investigated the types of behavior produced by AIDA, we can now turn our attention to aggregate performance over an ensemble of agents. To that end we repeat the experiment 80 times with identical hyperparameters, but with different initial proposals. The metric we are most interested in is how quickly AIDA is able to locate the optimum and produce a positive appraisal. 

\begin{figure}[h!]
    \hfill
    \begin{minipage}[b]{0.45\textwidth}
        \centering
        \resizebox{1.0\textwidth}{!}{\includegraphics{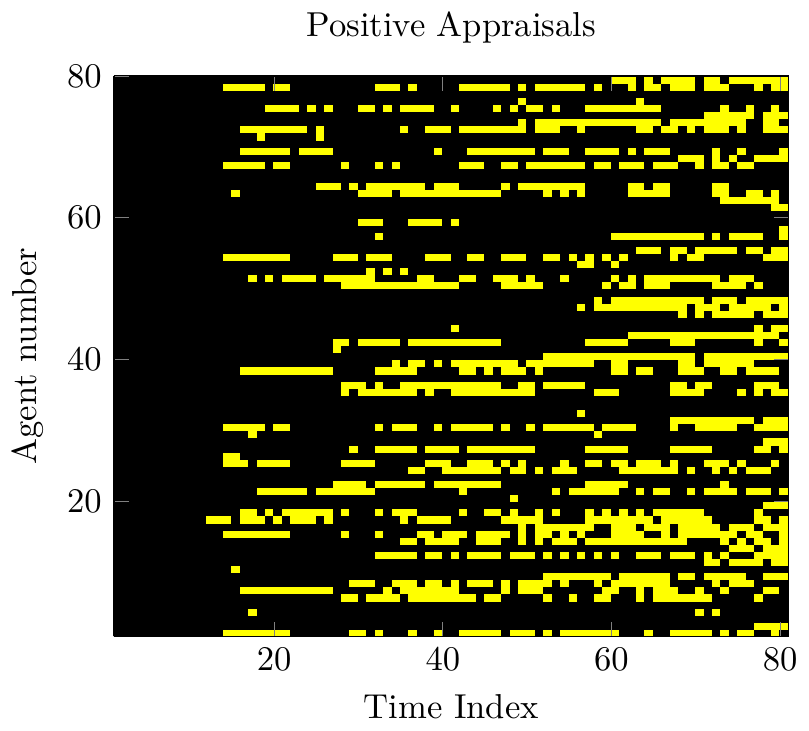}}
    \end{minipage}
    \begin{minipage}[b]{0.45\textwidth}
        \centering
        \resizebox{1.0\textwidth}{!}{\includegraphics{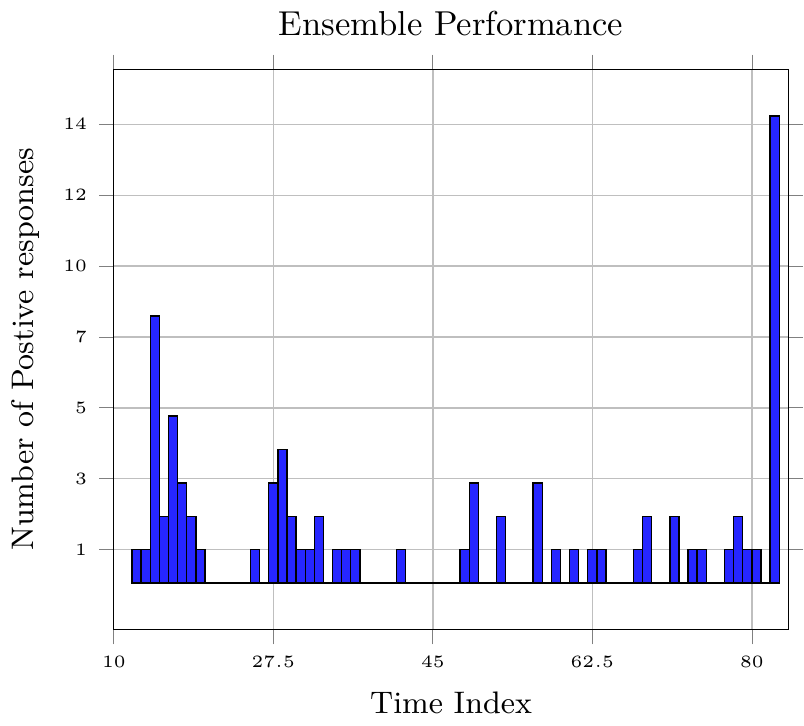}}
    \end{minipage}
    \caption{(Left) Heatmap showing ensemble performance over 80 agents. Positive and negative responses are indicated with yellow and black squares, respectively. (Right) Histogram showing time indices where the agents receive their first positive response. The right most column indicates agents that failed to obtain a positive appraisal. In total, 66/80 agents solve the task, corresponding to a success rate of 82.5\%.}
    \label{fig:ensemble}
\end{figure}

Figure~\ref{fig:ensemble} shows a heatmap of when each agent obtains positive responses. Positive responses are indicated by yellow squares and negative responses by black squares. Each row contains results for a single AIDA-agent and each column indicates a time step of the experiment. Consistent with the results for a single agent, we see that each experiment starts with a period of exploration. A large number of rows also show a yellow square within the first 35 trials, indicating that the optimum was found. Interestingly, no agents receive only positive responses, even after locating the optimum. This follows from AIDA actively trading off exploration and exploitation. When exploring, AIDA can select parameters that are suboptimal with respect to eliciting positive user responses, to gather more information. 
Figure~\ref{fig:ensemble} also shows a histogram indicating when each agent obtains its first positive appraisal. The very right column shows agents that failed to locate the optimum within the designated number of trials. In total, 66/80 agents correctly solve the task, corresponding to a success rate of 82.5\%. Disregarding unsuccessful runs, on average, AIDA obtains a positive response in 37.8 trials with a median of 29.5 trials.

\subsection{Hearing aid algorithm execution verification}\label{sec:experiments:HA}
To verify the proposed inference methodology for the hearing aid algorithm execution, we synthesized data by sampling from the following generative model:
\begin{subequations}\label{eq:verification:acoustic}
\begin{align}
    &\bm{\theta}_t \sim \mathcal{N}\left(\bm{\theta}_{t-1},\ \omega\mathrm{I}_{M}\right) \\
    % &\bm{\zeta}_k \sim \delta\left(\bm{\zeta}_{k-1}\right) \\
    &\bm{s}_t \sim \mathcal{N}\left(A(\bm{\theta}_t)\bm{s}_{t-1},\ V\left(\gamma\right)\right) \label{eq:verification:acoustic-tvar} \\
    &\bm{n}_t \sim \mathcal{N}\left(A(\bm{\zeta})\bm{n}_{t-1},\ V\left(\tau\right)\right) \label{eq:verification:acoustic-ar} \\
    &x_t = s_t + n_t \label{eq:verification:acoustic:observation},
\end{align}
\end{subequations}
with priors
\begin{subequations} \label{eq:verification:acoustic-priors}
\begin{align}
    %p&(M=k) = \prod_{k=4}^{8}\frac{1}{5}^{[M=k]} \label{eq:verification:acoustic-priors-order1} \\
    %p&(L=k) = \prod_{l=1}^{4}\frac{1}{4}^{[L=k]} \label{eq:verification:acoustic-priors-order2} \\
    &\bm{\theta}_0 \sim \mathcal{N}(\bm{0}, \omega\mathrm{I}_M) \\
    &\bm{\zeta} \sim \mathcal{N}(\bm{0}, \mathrm{I}_N) \\
    &\gamma \sim \Gamma(1.0, 1e-4) \\
    &\tau \sim \Gamma(1.0, 1.0) \\
    &\omega = 1e-4
\end{align}
\end{subequations}
where $M$ and $N$ are the orders of TVAR and AR models, respectively, and where $M \geq N$ holds, as we assume that the noise signal can be modeled by a lower AR order in comparison to the speech signal. We use an uninformative prior for the output of the hearing aid $y_t$ as in Figure~\ref{fig:model:coupled-AR} to prevent interactions from that part of the graph. We generated 1000 distinct time series of length 100. For each generated time series, the (TV)AR orders $M$ and $N$ were sampled from the discrete domains $[4,8]$ and $[1,4]$, respectively. We resampled the priors that initially resulted into unstable TVAR and AR processes.

\begin{figure}[h!]
    \hfill
    \begin{minipage}[b]{\textwidth}
        \centering
        \resizebox{\textwidth}{!}{\includegraphics{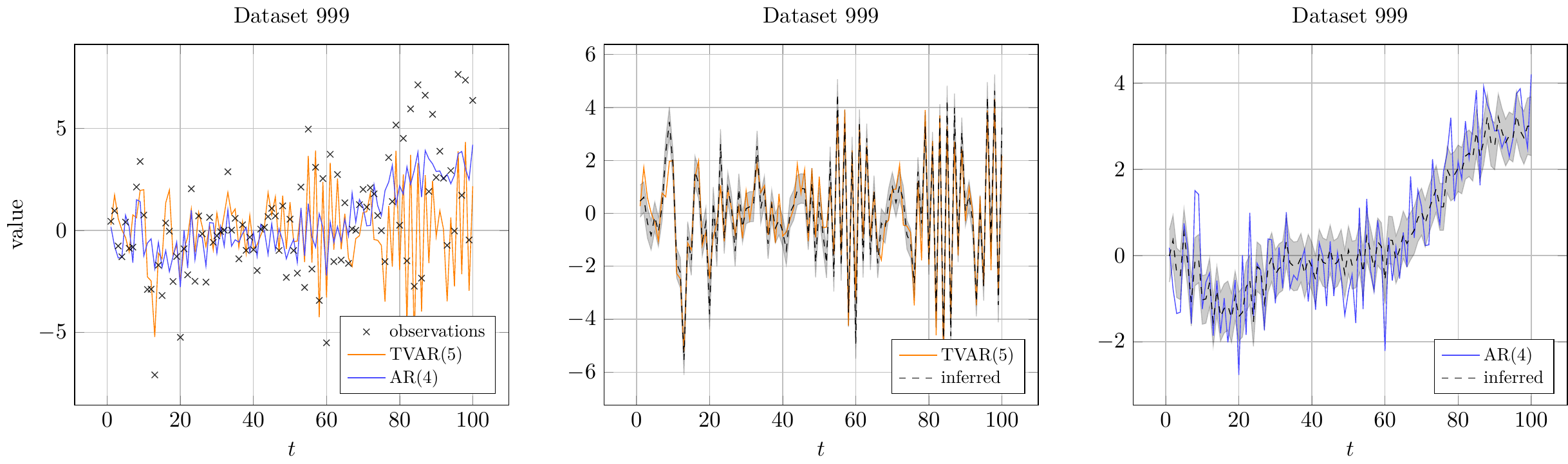}}
    \end{minipage}
    \begin{minipage}[b]{\textwidth}
        \centering
        \resizebox{\textwidth}{!}{\includegraphics{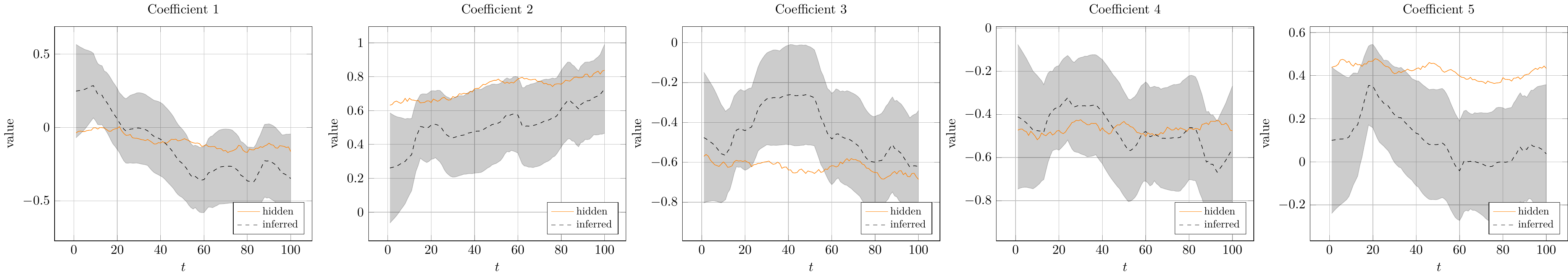}}
    \end{minipage}h
    \caption{(Top) Inference results for the hidden states $s_t$ and $n_t$ of coupled (TV)AR process on dataset $999$. (left) The generated observed signal $x_t$ with underlying generated signals $s_t$ and $n_t$. (center) The latent signal $s_t$ and its corresponding posterior approximation. (right) The latent signal $n_t$ and its corresponding posterior approximation. The dashed lines corresponds to the mean of the posterior estimates. The transparent regions represent the corresponding remaining uncertainty as plus-minus one standard deviation from the mean. (Bottom) Inference results for the coefficients $\bm{\theta}_t$ of dataset 999. The solid lines correspond to the true latent AR coefficients. The dashed lines correspond to the mean of the posterior estimates of the coefficients and the transparent regions correspond to plus-minus one standard deviation from the mean of the estimated coefficients.}
    \label{fig:experiments:coupled-AR}
\end{figure}

\begin{figure}
    \centering
    \resizebox{\textwidth}{!}{\includegraphics{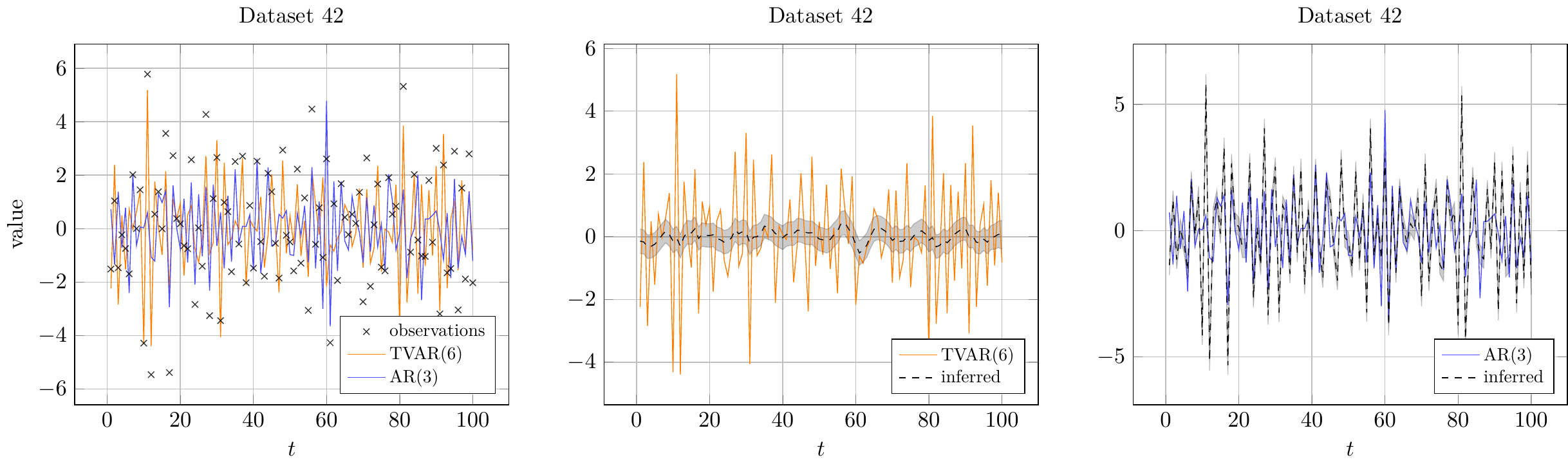}}
    \caption{Inference results for the hidden states $s_t$ and $n_t$ of coupled (TV)AR process on dataset $42$. In this particular case it can be noted that the inferred states are swapped with respect to the true underlying signals. However, the accompanying intelligent agent is able to cope with these kinds of situations, such that the HA clients do not experience any problems as a result. }
    \label{fig:experiments:coupled-AR-bad}
\end{figure}

\begin{figure}
    \centering
    \resizebox{0.7\textwidth}{!}{\includegraphics{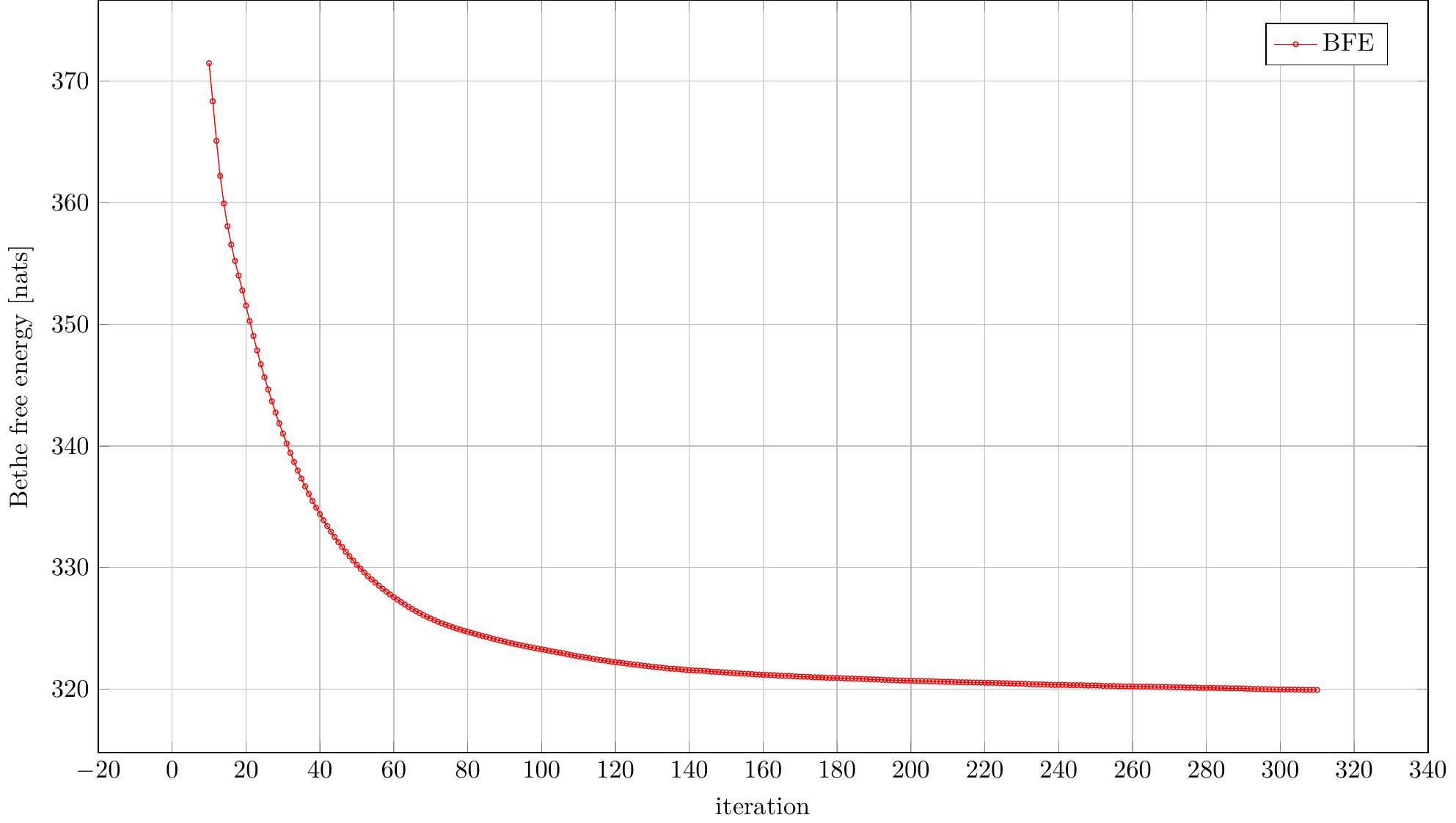}}
    \caption{Evolution of the Bethe free energy for the coupled autoregressive model averaged over all generated time series. The iteration index specifies the number of marginal updates for all edges in the graph.}
    \label{fig:experiments:coupled-AR-fe}
\end{figure}

The generated time series were used in the following experiment. We first created a probabilistic model with the same specifications as the generative model in \eqref{eq:verification:acoustic}. However, we used non-informative priors for the states and parameters of the model that corresponds to the TVAR process in \eqref{eq:verification:acoustic-tvar}. To ensure the identifiability of the separated sources, we used weakly informative priors for the parameters of the AR process in \eqref{eq:verification:acoustic-ar}. Specifically, the mean of the prior for $\bm{\zeta}$ was centered around the real AR coefficients that were used in the data generation process.
The goals of the experiment are 1) to verify that the proposed inference procedure recovers the hidden states $\bm{\theta}_t$, $s_t$ and $n_t$ for each generated dataset and 2) to verify convergence of the BFE as convergence is not guaranteed, because our graph contains loops \citep{murphy_loopy_1999}. For a typical case, the inference results for the hidden states $s_t$ and $n_t$ are shown in the top row of Figure \ref{fig:experiments:coupled-AR}.
The bottom row of Figure \ref{fig:experiments:coupled-AR} shows the tracking of the time-varying coefficients $\bm{\theta}_t$. This plot does not show the correlation between the inferred coefficients, whereas this actually contains vital information for modeling an acoustic signal. Namely, the coefficients together specify a set of poles, which influence the characteristics of the frequency spectrum of the signal.
% \begin{figure}
%     \centering
%     \resizebox{\textwidth}{!}{\includesvg{figures/experiments/coupled_ar/coefs_tvar_999.svg}}
%     \caption{Inference results for the coefficients $\bm{\theta}_t$ of dataset 999. The solid lines correspond to the true latent AR coefficients. The dashed lines correspond to the mean of the posterior estimates of the coefficients and the transparent regions correspond to plus-minus one standard deviation from the mean of the estimated coefficients.}
%     \label{fig:experiments:coefs-tvar}
% \end{figure}
An interesting example is depicted in Figure~\ref{fig:experiments:coupled-AR-bad}. We can see that the inference results for the latent states $s_t$ and $n_t$ are swapped with respect to the true underlying signals. This behaviour is undesirable in standard algorithms when the output of the HA is produced based on hard-coded gains. However, the presence of our intelligent agent can still find the optimal gains for this situation. The automation of the hearing aid algorithm and intelligent agent will relieve this burden on HA clients.
As can be seen from Figure~\ref{fig:experiments:coupled-AR-fe}, the Bethe free energy averaged over all generated time series monotonically decreases. Note that even though the proposed hybrid message passing algorithm results in a stationary solution, it does not provide convergence guarantees.

\subsection{Validation experiments}\label{sec:experiments:demo}
\begin{figure}[!ht]
    %\hfill
    \centering
    \includegraphics[scale=0.85]{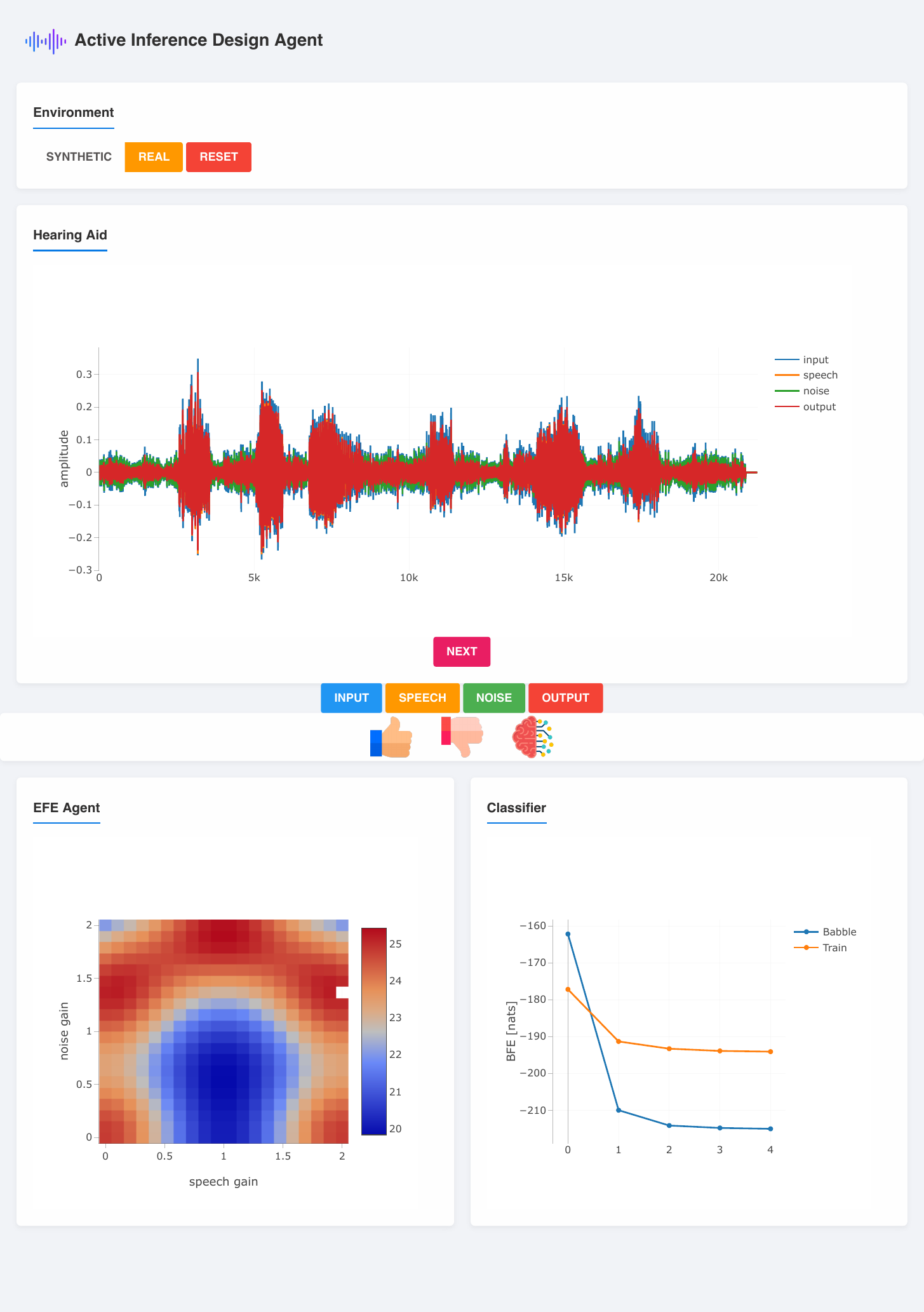}
    \caption{Screenshot of the interactive web application of AIDA. The dashboard consists of four distinct cells. The top cell {\it{Environment}} allows the user to change the interfering noise signal from a generated noise signal (synthetic) to a real noise signal. Furthermore it contains a reset button for resetting the application. The {\it{Hearing Aid}} cell provides an interactive plot of the input, separated speech, separated noise, and generated output waveform signals. Each waveform can be played when the corresponding button is pressed. The {\it{NEXT}} button loads a new audio file for evaluation. The {\it{thumbs-up}} and {\it{thumbs-down}} buttons correspond to providing AIDA with positive and negative appraisals, respectively. The {\it{brain}} button starts optimization of the parameters of GPC. The {\it{EFE Agent}} cell reflects the agent's beliefs about optimal parameters for the user as an EFE heatmap. The {\it{Classifier}} cell shows the Bethe free energy (BFE) score for the different models, corresponding to the different contexts. For the real noise signal, the algorithm automatically determines whether we are surrounded by babble noise, or by noise from a train station.}
    \label{fig:AIDA-GUI}
\end{figure}

For the validation of the proposed HA algorithm and AIDA, we created an interactive web application\footnote{A web application of AIDA is available at \url{https://github.com/biaslab/AIDA-app/}.} to demonstrate the the joint system. Figure~\ref{fig:AIDA-GUI} shows the interface of the demonstrator.

The user listens to the output of the hearing aid algorithm by pressing the "output" button. The buttons "speech" and "noise" correspond to the beliefs of AIDA about the constituent signals of the HA input. Note that in reality the user does not have access to this information and can only listen to HA output.
After listening to the output signal, the user is invited to assess the performance of the current HA setting. The user can send positive and negative appraisals by pressing the thumb up or thumb down buttons respectively. Once the appraisal is sent, AIDA updates its beliefs about the parameters' space and provides new settings for the HA algorithm to make the user happy. As AIDA models user appraisals using a GPC, we provide an additional button that forces AIDA to optimize the parameters of GPC. This could be useful when AIDA has already collected some feedback from the user that contains both positive and negative appraisals.

The demonstrator works in two environments: synthetic and real. The synthetic environment allows the user to listen to a spoken sentence with two artificial noise sources, i.e. either interference from a sinusoidal wave or a drilling machine. In the synthetic environment the hearing aid algorithm exploits the knowledge about acoustic contexts, i.e, it uses informative priors for the AR model that corresponds to noise. The real environment uses the data from NOIZEUS speech corpus\footnote{The NOIZEUS database is available at \url{https://ecs.utdallas.edu/loizou/speech/noizeus/}.}. In particular, the real environment consists of $30$ sentences pronounced in two different noise environments. Here the user is either experiencing surrounding noise at a train station or babble noise. In the real environment, the HA algorithm uses weakly informative priors for the background noise which influences the performance of the HA algorithm.
Both the HA algorithm and AIDA determine the acoustic context based on the Bethe free energy score, which is also shown in the demonstrator. The context with the lower Bethe free energy score corresponds to the selected acoustic context.

\section{Discussion}\label{sec:discussion}
% discussions on AIDA
We have introduced a design agent that is capable of tuning the context-dependent parameters of a hearing aid algorithm by incorporating user feedback. Throughout the paper, we have made several design choices whose implications we shortly review in this section.

% discussion on audio model
The audio model introduced in Section \ref{sec:model:acoustic} describes the dynamics of the speech signal perturbed by colored noise. Despite the fact that the proposed inference algorithm allows for the decomposition of such signals into speech and noise components, there are a few limitations that must be highlighted. First, the identifiability of the coupled AR model depends on the selected priors. Non-informative priors can lead to poor source estimation \citep{hsiao_identification_2008, kleibergen_bayesian_1995}. To tackle the identifiability issue, we use informative context-dependent priors. In other words, for each context, we use a different set of priors that better describe the dynamics of the acoustic signal in that context. Secondly, throughout our experiments we used fixed orders of TVAR and AR models. In reality, we do not have prior information about the actual order of the underlying signals. Therefore, to continuously update our models of the underlying sources we need to perform active order selection, which can be realized using Bayesian model reduction \citep{friston_post_2011, friston_bayesian_2018}. Thirdly, our model assumes that the hearing aid device only has access to a monaural input, which means that the observed signal originates from single microphone. As a result we do not use any spatial information about an acoustic signal that could have been obtained using multiple microphones. This assumption is mostly influenced by our desire to focus on the concept of designing a novel class of hearing aid algorithms rather than building real-world HA engine. Fortunately, the proposed framework allows for the easy substitution of source models with more versatile models that might be better suited for speech. For instance, one can use several microphones, as commonly done in beamforming \citep{ozerov_multichannel_2010}, or use a frequency decomposition for improving the source separation performance \citep{rennie_dynamic_2006, rennie_single-channel_2009, frey_algonquin_2001}. Inevitably, a more complex model will also likely result in a higher computational burden. 
Hence, the implementation of this algorithm on an embedded device remains a challenge.

% Agent in general
The power of the agent comes from the choice of the objective function. Since the objective is independent of the generative model, a straightforward approach to improving the agent is to adapt the generative model. In particular, a GPC is a nonparametric model with very few assumptions on the underlying function. Placing constraints on the preference function, such as was done in \cite{ignatenko_sequential_2021,cox_parametric_2017}, is likely to improve data efficiency of the agent. Arguably a core move of \cite{ignatenko_sequential_2021,cox_parametric_2017} is to acknowledge that user preferences are likely to be peaked around one or a few optima. Even if the true preference function has multiple modes, assuming a single peak for the agent is safe since it only needs to locate one of the modes to provide good parameter settings. Making this assumption allows the authors to work with a parametric model over user preferences. Working with a less flexible model predictably leads to higher data efficiency, which can aid performance of the agent. Given that the target demographic for AIDA consists of HA users, it is of paramount importance that the agent is able to learn an adequate representation of user preferences in as few trials as possible to avoid inconveniencing the user.

During model specification in Section~\ref{sec:model:EFE}, we make some assumptions on the control variable $\bm{u}_k$ and user appraisals $r_k$. First, we set the domain of the elements of control variable $\bm{u}_k$ to $[0, 1]$. Note that this is an arbitrary constraint which we use for illustrative purposes. The domain can be easily rescaled without loss of generality. For example, in our demonstrator, we use the default domain of $\bm{u}_k\in [0, 2]^2$.
Secondly, we opt for binary user appraisals, i.e. $r_k \in \{\varnothing, 0, 1\}$.
This design choice follows from the requirement of allowing users to communicate covertly to AIDA.
Binary user appraisal can more easily be linked to for example covert wrist movements when wearing a smartwatch to update the control variables.
With continuous user appraisals, e.g. $r_k \in [0,1]$, or pairwise comparison tests the convergence of AIDA can be greatly improved as these appraisals yield more information per appraisal.
However, providing AIDA with these appraisals requires more attention, which is undesirable in certain circumstances, for example during business meetings.

% AIDA
Real-world testing of AIDA has not been included in our work as performance evaluation with human HA clients is not straightforward. The performance of AIDA should be evaluated by means of a randomized controlled trial (RCT) where HA clients should be randomly assigned to either an experimental group or a control group. Unfortunately, our implementation is currently not able to achieve real-time performance and hence cannot be tested in the proper RCT setting. Nonetheless, we provide a demo that simulates AIDA and can be tested freely.
\section{Related work} \label{sec:related}
The problem of hearing aid personalization has been explored in various works. In \cite{nielsen_perception-based_2015} \bdv{here use cite rather than citep. Please check the difference. }the HA parameters are tuned according to a pairwise user assessment tests, during which the user's perception is encoded using Gaussian processes. The intractable posterior distribution corresponding to the user's perception is then computed using a Laplace approximation with Expected Improvement as the acquisition function used to select the next set of gains.
Our agent improves upon \cite{nielsen_perception-based_2015} in two concrete ways. Firstly, AIDA places a lower cognitive load on the user by not requiring pairwise comparisons. This means the user does not need to keep in her memory what the HA sounded like at the previous trial but only needs to consider the current HA output. AIDA accomplishes this without requiring more trials for training. In fact, since AIDA does not require pre-training but can be trained fully online under in-situ conditions, AIDA requires less data to locate optimal gains. Secondly, AIDA can be trained and retrained in a continual learning fashion. In case the users preferences change over time, for instance by a change in the hearing loss profile, AIDA can smoothly accommodate the user as long as she continues to provide the agent with feedback. Using EFE as acquisition function means the agent will engage in local exploration once the optimum is located, leading the agent to naturally learn shifts in the users preferences by balancing exploration and exploitation. 
In \cite{alamdari_personalization_2020}, personalization of the hearing aid compression algorithm is framed in terms of deep reinforcement learning. On the contrary, in our work we take inspiration from the active inference framework where agents act to maximize model evidence of their underlying generative model. Importantly, this does not require us to explicitly specify a loss function that drives exploitative and epistemic behaviour. In the recent work of \cite{ignatenko_sequential_2021}, the hearing aid preference learning algorithm is implemented through sequential Bayesian optimization with pairwise comparisons. Their hearing aid system comprises two subsystems representing a user with their preferences and the agent that guides the learning process. However, \cite{ignatenko_sequential_2021} focus only on exploration through maximising information gain with a parametric model. The EFE additionally adds a goal directed term that ensures the agent will stay near the optimum once located, even if other parameter settings provide more information. Extending the model of \cite{ignatenko_sequential_2021} to employ the full EFE is an exciting potential direction for future work. Finally neither \cite{nielsen_perception-based_2015} nor \cite{ignatenko_sequential_2021} takes context dependence into account.

\cite{friston_active_2021} introduces Active Listening (AL), which performs speech recognition based on the principles of active inference. In \cite{friston_active_2021}, they regard listening as an active process that is largely influenced by lexical, speaker and prosodic information. 
\cite{friston_active_2021} distinguishes itself from conventional audio processing algorithms, because it explicitly includes the process of word boundary selection before word classification and recognition, and that they regard this as an active process.
Word boundaries are selected from a group of candidate word boundaries, based on Bayesian model selection, by choosing the word boundary that optimizes the VFE during classification.
In the future, we see the potential of incorporating the AL approach into AIDA. Active inference is successfully applied in the work \cite{holmes_active_2021} that studies to model selective attention in a cocktail party listening setup. 

The audio processing components of AIDA essentially perform informed source separation \citep{knuth_informed_2013}, where sources are separated based on prior knowledge. Even though blind source separation approaches \citep{laufer_bayesian_2021, xie_time-frequency_2012} always use some degree of prior information, we do not focus on this direction and instead we actively try to model the underlying sources based on variations of auto-regressive processes.
For audio processing applications source separation has often been performed in the log-power domain \citep{rennie_dynamic_2006, rennie_single-channel_2009, frey_algonquin_2001}.
However, the interaction of the signals in this domain is no longer linear.
The intractability that results from performing exact inference in this model is often resolved by simplifying the interaction function \citep{hershey_signal_2010, radfar_nonlinear_2006}.
Although this approach has shown to be successful in the past, its performance is limited because of the negligence of phase information.
\section{Conclusions} \label{sec:conclusion}

This paper has presented AIDA, an active inference design agent for novel situation-aware personalized hearing aid algorithms.
AIDA and the corresponding hearing aid algorithm are based on probabilistic generative models that model the user and the underlying speech and context-dependent background noise signals of the observed acoustic signal, respectively.
Through probabilistic inference by means of message passing, we perform informed source separation in this model and use the separated signals to perform source-specific filtering.
AIDA then learns personalized source-specific gains through user interaction, depending on the environment that the user is in.
Users can give a binary appraisal after which the agent will make an improved proposal, based on expected free energy minimization for encouraging both exploitative and epistemic behaviour.
AIDA's operations are context-dependent and uses the context from the hearing aid algorithm, which is based on Bayesian model selection.
Experimental results show that hybrid message passing is capable of finding the hidden states of the coupled AR model that are associated with the speech and noise components. Moreover, Bayesian model selection has been tested for the context inference problem where each source is modelled by AR process. The experiments on preference learning showed the potential of applying expected free energy minization for finding the optimal settings of the hearing aid algorithm.
Although real-world implementations still present challenges, this novel class of audio processing algorithms has the potential to change the leading approach to hearing aid algorithm design. 
Future plans encompass developing AIDA towards real-time applications.

\section*{Acknowledgments}
This work was partly financed by GN Advanced Science, which is the research department of GN Hearing A/S, and by research programs ZERO and EDL with project numbers P15-06 and P16-25, respectively, which are (partly) financed by the Netherlands Organisation for Scientific Research (NWO). 
The authors would also like to thank the \href{https://biaslab.github.io/}{BIASlab} team members for insightful discussions on various topics related to this work. 

%Bibliography
\bibliographystyle{unsrt}  
\bibliography{main}

\begin{thebibliography}{10}

\bibitem{kates_multichannel_2005}
James Kates and Kathryn Arehart.
\newblock Multichannel {Dynamic}-{Range} {Compression} {Using} {Digital}
  {Frequency} {Warping}.
\newblock {\em EURASIP Journal on Applied Signal Processing}, 18:3003--3014,
  2005.

\bibitem{van_de_laar_probabilistic_2016}
Thijs van~de Laar and Bert de~Vries.
\newblock A {Probabilistic} {Modeling} {Approach} to {Hearing} {Loss}
  {Compensation}.
\newblock {\em IEEE/ACM Transactions on Audio, Speech, and Language
  Processing}, 24(11):2200--2213, November 2016.

\bibitem{nielsen_perception-based_2015}
J.B.B. Nielsen, J.~Nielsen, and J.~Larsen.
\newblock Perception-{Based} {Personalization} of {Hearing} {Aids} {Using}
  {Gaussian} {Processes} and {Active} {Learning}.
\newblock {\em IEEE/ACM Transactions on Audio, Speech, and Language
  Processing}, 23(1):162--173, January 2015.

\bibitem{alamdari_personalization_2020}
Nasim Alamdari, Edward Lobarinas, and Nasser Kehtarnavaz.
\newblock Personalization of {Hearing} {Aid} {Compression} by
  {Human}-in-the-{Loop} {Deep} {Reinforcement} {Learning}.
\newblock {\em IEEE Access}, 8:203503--203515, 2020.

\bibitem{reddy_individualized_2017}
C.~Karadagur~Ananda Reddy, N.~Shankar, G.~Shreedhar Bhat, R.~Charan, and
  I.~Panahi.
\newblock An {Individualized} {Super}-{Gaussian} {Single} {Microphone} {Speech}
  {Enhancement} for {Hearing} {Aid} {Users} {With} {Smartphone} as an
  {Assistive} {Device}.
\newblock {\em IEEE Signal Processing Letters}, 24(11):1601--1605, November
  2017.

\bibitem{friston_free_2006}
Karl Friston, James Kilner, and Lee Harrison.
\newblock A free energy principle for the brain.
\newblock {\em Journal of Physiology, Paris}, 100(1-3):70--87, September 2006.

\bibitem{van_de_laar_simulating_2019}
Thijs van~de Laar and Bert de~Vries.
\newblock Simulating {Active} {Inference} {Processes} by {Message} {Passing}.
\newblock {\em Frontiers in Robotics and AI}, 6:20, 2019.

\bibitem{van_de_laar_application_2019}
Thijs van~de Laar, Ay{\c c}a {\"O}z{\c c}elikkale, and Henk Wymeersch.
\newblock Application of the {Free} {Energy} {Principle} to {Estimation} and
  {Control}.
\newblock {\em arXiv preprint arXiv:1910.09823}, 2019.

\bibitem{millidge_deep_2019}
Beren Millidge.
\newblock Deep {Active} {Inference} as {Variational} {Policy} {Gradients}.
\newblock {\em arXiv:1907.03876 [cs]}, July 2019.
\newblock arXiv: 1907.03876.

\bibitem{tschantz_scaling_2020}
Alexander Tschantz, Manuel Baltieri, Anil~K. Seth, and Christopher~L. Buckley.
\newblock Scaling active inference.
\newblock In {\em 2020 {International} {Joint} {Conference} on {Neural}
  {Networks} ({IJCNN})}, pages 1--8. IEEE, 2020.

\bibitem{friston_active_2015}
Karl Friston, Francesco Rigoli, Dimitri Ognibene, Christoph Mathys, Thomas
  Fitzgerald, and Giovanni Pezzulo.
\newblock Active inference and epistemic value.
\newblock {\em Cognitive Neuroscience}, 6(4):187--214, March 2015.

\bibitem{da_costa_active_2020}
Lancelot Da~Costa, Thomas Parr, Noor Sajid, Sebastijan Veselic, Victorita
  Neacsu, and Karl Friston.
\newblock Active inference on discrete state-spaces: a synthesis.
\newblock {\em arXiv:2001.07203 [q-bio]}, January 2020.
\newblock arXiv: 2001.07203.

\bibitem{friston_sophisticated_2021}
Karl Friston, Lancelot Da~Costa, Danijar Hafner, Casper Hesp, and Thomas Parr.
\newblock Sophisticated {Inference}.
\newblock {\em Neural Computation}, 33(3):713--763, March 2021.

\bibitem{rix_perceptual_2001}
Antony~W. Rix, John~G. Beerends, Michael~P. Hollier, and Andries~P. Hekstra.
\newblock Perceptual evaluation of speech quality ({PESQ})-a new method for
  speech quality assessment of telephone networks and codecs.
\newblock In {\em 2001 {IEEE} {International} {Conference} on {Acoustics},
  {Speech}, and {Signal} {Processing}, 2001. {Proceedings}.}, volume~2, pages
  749--752. IEEE, 2001.

\bibitem{kates_hearing-aid_2010}
James~M. Kates and Kathryn~H. Arehart.
\newblock The hearing-aid speech quality index ({HASQI}).
\newblock {\em Journal of the Audio Engineering Society}, 58(5):363--381, 2010.

\bibitem{taal_algorithm_2011}
Cees~H. Taal, Richard~C. Hendriks, Richard Heusdens, and Jesper Jensen.
\newblock An {Algorithm} for {Intelligibility} {Prediction} of
  {Time}{\textendash}{Frequency} {Weighted} {Noisy} {Speech}.
\newblock {\em IEEE Transactions on Audio, Speech, and Language Processing},
  19(7):2125--2136, September 2011.

\bibitem{beerends_perceptual_2013}
John~G. Beerends, Christian Schmidmer, Jens Berger, Matthias Obermann, Raphael
  Ullmann, Joachim Pomy, and Michael Keyhl.
\newblock Perceptual {Objective} {Listening} {Quality} {Assessment} ({POLQA}),
  {The} {Third} {Generation} {ITU}-{T} {Standard} for {End}-to-{End} {Speech}
  {Quality} {Measurement} {Part} {I}{\textemdash}{Temporal} {Alignment}.
\newblock {\em Journal of the Audio Engineering Society}, 61(6):366--384, July
  2013.
\newblock Publisher: Audio Engineering Society.

\bibitem{hines_visqol_2015}
Andrew Hines, Jan Skoglund, Anil~C Kokaram, and Naomi Harte.
\newblock {ViSQOL}: an objective speech quality model.
\newblock {\em EURASIP Journal on Audio, Speech, and Music Processing},
  2015(1):13, December 2015.

\bibitem{chinen_visqol_2020}
Michael Chinen, Felicia S.~C. Lim, Jan Skoglund, Nikita Gureev, Feargus
  O'Gorman, and Andrew Hines.
\newblock {ViSQOL} v3: {An} {Open} {Source} {Production} {Ready} {Objective}
  {Speech} and {Audio} {Metric}.
\newblock {\em arXiv:2004.09584 [cs, eess]}, April 2020.

\bibitem{forney_codes_2001}
G.David Forney.
\newblock Codes on graphs: normal realizations.
\newblock {\em IEEE Transactions on Information Theory}, 47(2):520--548,
  February 2001.

\bibitem{loeliger_introduction_2004}
Hans-Andrea Loeliger.
\newblock An introduction to factor graphs.
\newblock {\em Signal Processing Magazine, IEEE}, 21(1):28--41, January 2004.

\bibitem{kakusho_hierarchical_1982}
O.~Kakusho and M.~Yanagida.
\newblock Hierarchical {AR} model for time varying speech signals.
\newblock In {\em {ICASSP} '82. {IEEE} {International} {Conference} on
  {Acoustics}, {Speech}, and {Signal} {Processing}}, volume~7, pages
  1295--1298, Paris, France, May 1982.

\bibitem{paliwal_speech_1987}
K.~Paliwal and A.~Basu.
\newblock A speech enhancement method based on {Kalman} filtering.
\newblock In {\em {ICASSP} '87. {IEEE} {International} {Conference} on
  {Acoustics}, {Speech}, and {Signal} {Processing}}, volume~12, pages 177--180,
  Dallas, TX, USA, April 1987.

\bibitem{vermaak_particle_2002}
J.~Vermaak, C.~Andrieu, A.~Doucet, and S.J. Godsill.
\newblock Particle methods for {Bayesian} modeling and enhancement of speech
  signals.
\newblock {\em IEEE Transactions on Speech and Audio Processing},
  10(3):173--185, March 2002.

\bibitem{rudoy_time-varying_2011}
Daniel Rudoy, Thomas~F. Quatieri, and Patrick~J. Wolfe.
\newblock Time-{Varying} {Autoregressions} in {Speech}: {Detection} {Theory}
  and {Applications}.
\newblock {\em IEEE Transactions on Audio, Speech, and Language Processing},
  19(4):977--989, May 2011.

\bibitem{podusenko_online_2020}
Albert Podusenko, Wouter~M. Kouw, and Bert de~Vries.
\newblock Online {Variational} {Message} {Passing} in {Hierarchical}
  {Autoregressive} {Models}.
\newblock In {\em 2020 {IEEE} {International} {Symposium} on {Information}
  {Theory} ({ISIT})}, pages 1337--1342, Los Angeles, CA, USA, June 2020.
\newblock ISSN: 2157-8117.

\bibitem{popescu_kalman_1998}
D.C. Popescu and I.~Zeljkovic.
\newblock Kalman filtering of colored noise for speech enhancement.
\newblock In {\em Proceedings of the 1998 {IEEE} {International} {Conference}
  on {Acoustics}, {Speech} and {Signal} {Processing}, {ICASSP} '98}, volume~2,
  pages 997--1000, Seattle, WA, USA, May 1998.
\newblock ISSN: 1520-6149.

\bibitem{gannot_iterative_1998}
S.~Gannot, D.~Burshtein, and E.~Weinstein.
\newblock Iterative and sequential {Kalman} filter-based speech enhancement
  algorithms.
\newblock {\em IEEE Transactions on Speech and Audio Processing},
  6(4):373--385, July 1998.

\bibitem{gibson_filtering_1991}
J.D. Gibson, B.~Koo, and S.D. Gray.
\newblock Filtering of colored noise for speech enhancement and coding.
\newblock {\em IEEE Transactions on Signal Processing}, 39(8):1732--1742,
  August 1991.

\bibitem{podusenko_message_2021-1}
Albert Podusenko, Bart van Erp, Dmitry Bagaev, Ismail Senoz, and Bert de~Vries.
\newblock Message {Passing}-{Based} {Inference} in the {Gamma} {Mixture}
  {Model}.
\newblock In {\em 2021 {IEEE} 31st {International} {Workshop} on {Machine}
  {Learning} for {Signal} {Processing} ({MLSP})}, pages 1--6, Gold Coast,
  Australia, October 2021. IEEE.

\bibitem{van_erp_bayesian_2021}
Bart van Erp, Albert Podusenko, Tanya Ignatenko, and Bert de~Vries.
\newblock A {Bayesian} {Modeling} {Approach} to {Situated} {Design} of
  {Personalized} {Soundscaping} {Algorithms}.
\newblock {\em Applied Sciences}, 11(20):9535, October 2021.
\newblock Number: 20 Publisher: Multidisciplinary Digital Publishing Institute.

\bibitem{houlsby_bayesian_2011}
Neil Houlsby, Ferenc Husz{\'a}r, Zoubin Ghahramani, and M{\'a}t{\'e} Lengyel.
\newblock Bayesian {Active} {Learning} for {Classification} and {Preference}
  {Learning}.
\newblock {\em arXiv:1112.5745 [cs, stat]}, December 2011.

\bibitem{chu_preference_2005}
Wei Chu and Zoubin Ghahramani.
\newblock Preference learning with {Gaussian} processes.
\newblock In {\em Proceedings of the 22nd international conference on {Machine}
  learning}, {ICML} '05, pages 137--144, New York, NY, USA, August 2005.
  Association for Computing Machinery.

\bibitem{huszar_gp_2011}
Ferenc Huszar.
\newblock A {GP} classification approach to preference learning.
\newblock In {\em {NIPS} {Workshop} on {Choice} {Models} and {Preference}
  {Learning}}, page~4, Sierra Nevada, Spain, 2011.

\bibitem{rasmussen_gaussian_2006}
Carl~Edward Rasmussen and Christopher K.~I Williams.
\newblock {\em Gaussian {Processes} for {Machine} {Learning}}.
\newblock MIT Press, 2006.

\bibitem{loeliger_factor_2007}
Hans-Andrea Loeliger, Justin Dauwels, Junli Hu, Sascha Korl, Li~Ping, and
  Frank~R. Kschischang.
\newblock The {Factor} {Graph} {Approach} to {Model}-{Based} {Signal}
  {Processing}.
\newblock {\em Proceedings of the IEEE}, 95(6):1295--1322, June 2007.

\bibitem{cox_factor_2019}
Marco Cox, Thijs van~de Laar, and Bert de~Vries.
\newblock A factor graph approach to automated design of {Bayesian} signal
  processing algorithms.
\newblock {\em International Journal of Approximate Reasoning}, 104:185--204,
  January 2019.

\bibitem{bishop_pattern_2006}
Christopher~M. Bishop.
\newblock {\em Pattern {Recognition} and {Machine} {Learning}}.
\newblock Springer-Verlag New York, Inc., 2006.

\bibitem{van_de_laar_automated_2019}
Thijs van~de Laar.
\newblock {\em Automated {Design} of {Bayesian} {Signal} {Processing}
  {Algorithms}}.
\newblock PhD thesis, Eindhoven University of Technology, Eindhoven, The
  Netherlands, 2019.

\bibitem{sajid_active_2021}
Noor Sajid, Philip~J. Ball, Thomas Parr, and Karl~J. Friston.
\newblock Active {Inference}: {Demystified} and {Compared}.
\newblock {\em Neural Computation}, 33(3):674--712, March 2021.

\bibitem{parr_uncertainty_2017}
Thomas Parr and Karl~J. Friston.
\newblock Uncertainty, epistemics and active inference.
\newblock {\em Journal of The Royal Society Interface}, 14(136):20170376,
  November 2017.

\bibitem{knuth_informed_2013}
Kevin~H. Knuth.
\newblock Informed {Source} {Separation}: {A} {Bayesian} {Tutorial}.
\newblock {\em arXiv:1311.3001 [cs, stat]}, November 2013.
\newblock arXiv: 1311.3001.

\bibitem{sarkka_bayesian_2013}
Simo S{\"a}rkk{\"a}.
\newblock {\em Bayesian {Filtering} and {Smoothing}}.
\newblock Cambridge University Press, London ; New York, October 2013.

\bibitem{podusenko_message_2021}
Albert Podusenko, Wouter~M. Kouw, and Bert de~Vries.
\newblock Message {Passing}-{Based} {Inference} for {Time}-{Varying}
  {Autoregressive} {Models}.
\newblock {\em Entropy}, 23(6):683, June 2021.
\newblock Number: 6 Publisher: Multidisciplinary Digital Publishing Institute.

\bibitem{senoz_variational_2021}
{\.I}smail {\c S}en{\"o}z, Thijs van~de Laar, Dmitry Bagaev, and Bert de~Vries.
\newblock Variational {Message} {Passing} and {Local} {Constraint}
  {Manipulation} in {Factor} {Graphs}.
\newblock {\em Entropy}, 23(7):807, 2021.
\newblock Publisher: Multidisciplinary Digital Publishing Institute.

\bibitem{dauwels_variational_2007}
Justin Dauwels.
\newblock On {Variational} {Message} {Passing} on {Factor} {Graphs}.
\newblock In {\em {IEEE} {International} {Symposium} on {Information}
  {Theory}}, pages 2546--2550, Nice, France, June 2007.

\bibitem{bezanson_julia:_2017}
J.~Bezanson, A.~Edelman, S.~Karpinski, and V.~Shah.
\newblock Julia: {A} {Fresh} {Approach} to {Numerical} {Computing}.
\newblock {\em SIAM Review}, 59(1):65--98, January 2017.

\bibitem{bagaev_reactive_2022}
Dmitry Bagaev and Bert de~Vries.
\newblock Reactive {Message} {Passing} for {Scalable} {Bayesian} {Inference}.
\newblock 2022.
\newblock Submitted to the Journal of Machine Learning Research.

\bibitem{k_mogensen_optim_2018}
Patrick K~Mogensen and Asbj{\o}rn N~Riseth.
\newblock Optim: {A} mathematical optimization package for {Julia}.
\newblock {\em Journal of Open Source Software}, 3(24):615, April 2018.

\bibitem{murphy_loopy_1999}
Kevin~P. Murphy, Yair Weiss, and Michael~I. Jordan.
\newblock Loopy belief propagation for approximate inference: {An} empirical
  study.
\newblock In {\em Proceedings of the {Fifteenth} conference on {Uncertainty} in
  artificial intelligence}, pages 467--475. Morgan Kaufmann Publishers Inc.,
  1999.

\bibitem{hsiao_identification_2008}
Tesheng Hsiao.
\newblock Identification of {Time}-{Varying} {Autoregressive} {Systems} {Using}
  {Maximum} a {Posteriori} {Estimation}.
\newblock {\em IEEE Transactions on Signal Processing}, 56(8):3497--3509,
  August 2008.

\bibitem{kleibergen_bayesian_1995}
Frank Kleibergen and Henk Hoek.
\newblock Bayesian {Analysis} of {ARMA} models using {Noninformative} {Priors}.
\newblock {\em CentER Discussion Paper}, 1995-116:24, 1995.

\bibitem{friston_post_2011}
Karl Friston and Will Penny.
\newblock Post hoc {Bayesian} model selection.
\newblock {\em Neuroimage}, 56(4-2):2089--2099, June 2011.

\bibitem{friston_bayesian_2018}
Karl Friston, Thomas Parr, and Peter Zeidman.
\newblock Bayesian model reduction.
\newblock {\em arXiv:1805.07092 [stat]}, May 2018.
\newblock arXiv: 1805.07092.

\bibitem{ozerov_multichannel_2010}
A.~Ozerov and C.~Fevotte.
\newblock Multichannel {Nonnegative} {Matrix} {Factorization} in {Convolutive}
  {Mixtures} for {Audio} {Source} {Separation}.
\newblock {\em IEEE Transactions on Audio, Speech, and Language Processing},
  18(3):550--563, March 2010.

\bibitem{rennie_dynamic_2006}
Steven Rennie, Trausti Kristjansson, Peder Olsen, and Ramesh Gopinath.
\newblock Dynamic noise adaptation.
\newblock In {\em 2006 {IEEE} {International} {Conference} on {Acoustics}
  {Speech} and {Signal} {Processing} {Proceedings}}, volume~1, pages 1--4,
  Toulouse, France, 2006. IEEE.

\bibitem{rennie_single-channel_2009}
S.J. Rennie, J.R. Hershey, and P.A. Olsen.
\newblock Single-channel speech separation and recognition using loopy belief
  propagation.
\newblock In {\em {IEEE} {International} {Conference} on {Acoustics}, {Speech}
  and {Signal} {Processing}, 2009. {ICASSP} 2009}, pages 3845--3848, Taipei,
  Taiwan, April 2009.

\bibitem{frey_algonquin_2001}
Brendan~J Frey, Li~Deng, Alex Acero, and Trausti Kristjansson.
\newblock {ALGONQUIN}: {Iterating} {Laplace}'s {Method} to {Remove} {Multiple}
  {Types} of {Acoustic} {Distortion} for {Robust} {Speech} {Recognition}.
\newblock In {\em Proceedings of the {Eurospeech} {Conference}}, pages
  901--904, Aalborg, Denmark, September 2001.

\bibitem{ignatenko_sequential_2021}
Tanya Ignatenko, Kirill Kondrashov, Marco Cox, and Bert de~Vries.
\newblock On {Sequential} {Bayesian} {Optimization} with {Pairwise}
  {Comparison}.
\newblock {\em arXiv:2103.13192 [cs, math, stat]}, March 2021.
\newblock arXiv: 2103.13192.

\bibitem{cox_parametric_2017}
Marco Cox and Bert de~Vries.
\newblock A parametric approach to {Bayesian} optimization with pairwise
  comparisons.
\newblock In {\em {NIPS} {Workshop} on {Bayesian} {Optimization} ({BayesOpt}
  2017)}, pages 1--5, Long Beach, USA, December 2017.

\bibitem{friston_active_2021}
Karl~J. Friston, Noor Sajid, David~Ricardo Quiroga-Martinez, Thomas Parr,
  Cathy~J. Price, and Emma Holmes.
\newblock Active listening.
\newblock {\em Hearing Research}, 399(Stimulus-specific adaptation, MMN and
  predicting coding):107998, January 2021.

\bibitem{holmes_active_2021}
Emma Holmes, Thomas Parr, Timothy~D. Griffiths, and Karl~J. Friston.
\newblock Active inference, selective attention, and the cocktail party
  problem.
\newblock {\em Neuroscience and Biobehavioral Reviews}, 131:1288--1304, October
  2021.

\bibitem{laufer_bayesian_2021}
Y.~Laufer and S.~Gannot.
\newblock A {Bayesian} {Hierarchical} {Model} for {Blind} {Audio} {Source}
  {Separation}.
\newblock In {\em 2020 28th {European} {Signal} {Processing} {Conference}
  ({EUSIPCO})}, pages 276--280, January 2021.
\newblock ISSN: 2076-1465.

\bibitem{xie_time-frequency_2012}
S.~Xie, L.~Yang, J.~Yang, G.~Zhou, and Y.~Xiang.
\newblock Time-{Frequency} {Approach} to {Underdetermined} {Blind} {Source}
  {Separation}.
\newblock {\em IEEE Transactions on Neural Networks and Learning Systems},
  23(2):306--316, February 2012.

\bibitem{hershey_signal_2010}
John~R Hershey, Peder Olsen, and Steven~J Rennie.
\newblock Signal {Interaction} and the {Devil} {Function}.
\newblock In {\em Proceedings of the {Interspeech} 2010}, pages 334--337,
  Makuhari, Chiba, Japan, 2010.

\bibitem{radfar_nonlinear_2006}
M.H. Radfar, A.H. Banihashemi, R.M. Dansereau, and A.~Sayadiyan.
\newblock Nonlinear minimum mean square error estimator for
  mixture-maximisation approximation.
\newblock {\em Electronics Letters}, 42(12):724--725, June 2006.

\bibitem{pearl_reverend_1982}
Judea Pearl.
\newblock Reverend {Bayes} on {Inference} {Engines}: {A} {Distributed}
  {Hierarchical} {Approach}.
\newblock In {\em Proceedings of the {Second} {AAAI} {Conference} on
  {Artificial} {Intelligence}}, {AAAI}'82, pages 133--136, Pittsburgh,
  Pennsylvania, 1982. AAAI Press.

\bibitem{kschischang_factor_2001}
Frank~R. Kschischang, Brendan~J. Frey, and H.-A. Loeliger.
\newblock Factor graphs and the sum-product algorithm.
\newblock {\em IEEE Transactions on information theory}, 47(2):498--519, 2001.

\bibitem{yedidia_bethe_2001}
Jonathan~S Yedidia, William~T Freeman, and Yair Weiss.
\newblock Bethe free energy, {Kikuchi} approximations, and belief propagation
  algorithms.
\newblock {\em Advances in neural information processing systems}, 13:24, 2001.

\bibitem{minka_divergence_2005}
Thomas Minka.
\newblock Divergence {Measures} and {Message} {Passing}.
\newblock Technical report, Microsoft Research, 2005.

\end{thebibliography}

\appendix
\newpage
\section{Factor graphs, free energy and message passing-based inference} \label{sec:appendix:background}

% why this section + short overview
This appendix introduces the basics of the probabilistic modeling approach, which underlies all computations in this paper.
First, in Section~\ref{sec:appendix:background:ffg} we describe factor graphs as a useful tool for visualizing probabilistic models.
In Section~\ref{sec:appendix:background:sp} we describe how exact probabilistic inference can be performed through message passing.
We then introduce the variational free energy in Section~\ref{sec:appendix:background:vfe} and the Bethe free energy in Section~\ref{sec:appendix:background:bfe} for quantifying the performance of the probabilistic model when exact inference is not possible.
By adding constraints to the approximate posterior distributions, we end up with hybrid message passing algorithms, such as variational message passing, as will be described in Section~\ref{sec:appendix:background:vmphybrid}.
% Finally, Section~\ref{sec:appendix:background:efe} describes the expected free energy, a quantity that enforces both extrinsic and epistemic behaviour in intelligent agents.
% Appendix~\ref{appendix:symbols}, and specifically Table~\ref{tab:symbols-background}, gives an overview of the symbols used throughout this section.

\subsection{Forney-style factor graphs} \label{sec:appendix:background:ffg}
For the remainder of this section consider the factorized function
\begin{equation} \label{eq:appendix:background:factorized}
    f(\bm{z}) = \prod_{a \in \mathcal{V}} f_a(\bm{z}_a),
\end{equation}
where $\left\{f_a \mid a \in \mathcal{V} \right\}$ denotes the set of factors $f_a(\bm{z}_a)$ indexed by $a\in\mathcal{V}$ from the set of vertices $\mathcal{V}$ and $\bm{z}$ denotes the set of variables $\bm{z}=\{z_i \mid i\in \mathcal{E}\}$, which are indexed by $i\in\mathcal{E}$ from the set of edges $\mathcal{E}$.
The variable $\bm{z}_a$ represents the set of all neighbouring variables of factor $f_a$.

Factor graphs are a specific type of probabilistic graphical models.
In this paper we will focus on Forney-style factor graphs (FFG) as introduced in \cite{forney_codes_2001} with notational conventions adopted from \cite{loeliger_introduction_2004}.
FFGs visualize factorized functions as undirected graphs, whose nodes represent the individual factors of the global function.
The nodes are interconnected by edges representing the mutual arguments of the factors.
In FFGs, a node can be connected to an arbitrary number of edges, but edges are constrained to have a maximum degree of two.
As an example, consider the factorized function
\begin{equation} \label{eq:appendix:background:examplefunction}
    f(z_1, z_2, z_3, z_4, z_5) = f_a(z_1) f_b(z_1, z_2, z_3) f_c(z_3, z_4, z_5) f_d(z_4).
\end{equation}
The FFG representation of this function is visualized in Figure \ref{fig:appendix:background:example_messages}.

% \begin{figure}[b]
%     \centering
%     \input{figures/background/exampleFFG}
%     \caption{A Forney-style factor graph representation of the factorized function in \eqref{eq:appendix:background:examplefunction}.}
%     \label{fig:appendix:background:example_ffg}
% \end{figure}

When a variable occurs in more than three factors, this constraint can be satisfied by introducing equality factors, defined as $f_{=}(z, z^{\prime}, z^{\prime\prime}) = \delta(z - z^{\prime})\delta(z - z^{\prime\prime})$.
Here $z^{\prime}$ and $z^{\prime\prime}$ are variable copies of $z$, whose posterior distributions are constrained to be identical as a result of the equality factor.
For a more extensive overview of factor graphs, we refer the interested reader to \citep{loeliger_introduction_2004, loeliger_factor_2007}.

\subsection{Sum-product message passing} \label{sec:appendix:background:sp}
Probabilistic inference concerns the calculation of the posterior distribution in our model.
The posterior distribution of the factorized function in \eqref{eq:appendix:background:factorized} is defined as 
\begin{equation} \label{eq:appendix:background:posterior}
    p(\bm{z}) = \frac{f(\bm{z})}{Z},
\end{equation}
where $Z = \int f(\bm{z})\ \mathrm{d}\bm{z}$ is the normalization constant.
Here and throughout the rest of this section, we assume to be dealing with continuous variables for generality.
For discrete variables, this integration simply reduces to a summation.
Furthermore we implicitly assume that the factors $f_a$ are appropriate (possibly unnormalized) probability density functions, meaning that their mapping is specified as $f_a: \mathbb{R}^{|\bm{z}_a|} \rightarrow \mathbb{R}_{\geq 0}$, where $|\bm{z}_a|$ denotes the cardinality of set $\bm{z}_a$.

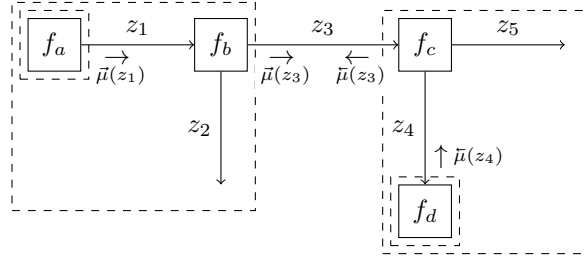
\begin{figure}
    \centering
    \begin{tikzpicture}
  
    % nodes
    \node[box] (fa) {$f_a$};
    \node[box, right = 15mm of fa] (fb) {$f_b$};
    \node[box, right = 20mm of fb] (fc) {$f_c$};
    \node[box, below = 15mm of fc] (fd) {$f_d$};
    \node[below = 15mm of fb] (s2) {};
    \node[right = 15mm of fc] (s5) {};
    
    % dashed boxes
    \node [fit = (fa), draw, inner sep = 1mm, dashed] (fa-box) {};
    \node [fit = (fa-box) (fb) (s2), draw, inner sep = 1mm, dashed] (fb-box) {};
    \node [fit = (fd), draw, inner sep = 1mm, dashed] (fd-box) {};
    \node [fit = (fd-box) (fc) (s5), draw, inner sep = 1mm, dashed] (fc-box) {};
    
    % edges
    \draw[->] (fa) -- (fb) 
        node[pos=0.5, above] {$z_1$}
        node[pos=0.35, below=2mm, fill=white, inner sep=2pt] {\scriptsize $\vec{\mu}(z_1)$}
        node[pos=0.35, below] {$\rightarrow{}$};
    \draw[->] (fb) -- (fc) 
        node[pos=0.5, above] {$z_3$}
        node[pos=0.25, below=2mm, fill=white, inner sep=2pt] {\scriptsize $\vec{\mu}(z_3)$}
        node[pos=0.25, below] {$\rightarrow{}$}
        node[pos=0.75, below=2mm, fill=white, inner sep=2pt] {\scriptsize $\cev{\mu}(z_3)$}
        node[pos=0.75, below] {$\leftarrow{}$};
    \draw[->] (fc) -- (fd) 
        node[pos=0.5, left] {$z_4$}    node[pos=0.75, right=3mm, fill=white, inner sep=2pt] {\scriptsize $\cev{\mu}(z_4)$}
        node[pos=0.75, right] {$\uparrow{}$};
    \draw[->] (fb) -- (s2) 
        node[pos=0.5, left] {$z_2$};
    \draw[->] (fc) -- (s5) 
        node[pos=0.5, above] {$z_5$};
    
\end{tikzpicture}
    \caption{A Forney-style factor graph representation of the factorized function in \eqref{eq:appendix:background:examplefunction}. The edges are arbitrarily directed to distinguish between forward and backward messages on the graph. The drawn messages can be regarded as summaries of the dashed boxes, used for solving \eqref{eq:appendix:background:examplemarginal}.}
    \label{fig:appendix:background:example_messages}
\end{figure}

% formal SP update rule
Computing the marginal distributions of this posterior requires integration over all nuisance variables.
Because of the conditional (in)dependencies in a factorized model, this computationally complex global integration can be performed through a set of smaller local computations.
This approach is better known as the sum-product algorithm or belief propagation \citep{pearl_reverend_1982, kschischang_factor_2001}.
The results of the local computations are termed messages $\mu$ and they propagate over the edges of the corresponding FFG.
In order to distinguish between the messages in the forward and backward direction, the edges in the corresponding FFG are made arbitrarily directed as shown in Figure~\ref{fig:appendix:background:example_messages}.
Now the messages $\vec{\mu}$ and $\cev{\mu}$ are specified to propagate in and against the direction of the edge, respectively.
The sum-product message $\vec{\mu}(z_i)$ \citep{kschischang_factor_2001} flowing out of some node $f_a$ with $z_i \in \bm{z}_a$ is defined as
\begin{equation} \label{eq:appendix:background:sumproductupdaterule}
    \vec{\mu}(z_i) \propto \int f_a(\bm{z}_a) \prod_{z_j \in \bm{z}_{a\backslash i}} \vec{\mu}(z_j) \mathrm{d}\bm{z}_{a\backslash i},
\end{equation}
where the notation $\bm{z}_{a \backslash i}$ refers to the set $\bm{z}_a$ excluding the element $z_i$, formally defined as $\bm{z}_{a\backslash i} = \{z \in \bm{z}_a \mid z \notin \{z_i\}\}$.
The messages $\vec{\mu}(z_j)$ are the incoming messages to the factor node.
Propagating these messages throughout the graph allows us to determine the marginal distributions of some variable $z_i$ as the product of the messages propagating on that respective edge as
\begin{equation} \label{eq:appendix:background:sumproductmarginal}
    p(z_i) \propto \vec{\mu}(z_i) \cdot \cev{\mu}(z_i).
\end{equation}

% example
To clarify the above points, suppose that we are interested in calculating the marginal distribution of $s_3$ from the model in \eqref{eq:appendix:background:examplefunction}.
This distribution can be calculated (up to a scaling constant) as
\begin{equation} \label{eq:appendix:background:examplemarginal}
    \begin{split}
        p(z_3)
        &= \int p(\bm{z})\ \mathrm{d}\bm{z}_{\backslash 3} \\
        % &\propto \int f(\bm{s})\ \mathrm{d}\bm{s}_{\backslash 3} \\
        &\propto \underbrace{\iint f_a(z_1) f_b(z_1, z_2, z_3)\ \mathrm{d}z_1\ \mathrm{d}z_2}_{\vec{\mu}(z_3)} \cdot \underbrace{\iint f_d(z_4) f_c(z_3, z_4, z_5) \ \mathrm{d}z_4\ \mathrm{d}z_5}_{\cev{\mu}(z_3)}.
    \end{split}
\end{equation}
From this derivation the marginal distribution of $z_3$ can be calculated as the product of two terms that each summarize a different part of the model.
Figure \ref{fig:appendix:background:example_messages} visualizes the example model of \eqref{eq:appendix:background:examplefunction}, now with directed edges, and visualizes the corresponding messages as summaries of the dashed parts of the graph.
In the example the prior distributions over $z_1$ and $z_4$ can be regarded as messages themselves, meaning that $\vec{\mu}(z_1) = f_a(z_1)$ and $\cev{\mu}(z_4) = f_d(z_4)$.
Furthermore, the edges corresponding to $z_2$ and $z_5$ are dangling, meaning that they only receive information from one side.
The messages in the reverse direction are defined to be uninformative as $\cev{\mu}(z_2) = \cev{\mu}(z_5) = 1$, because then the posterior distribution is fully derived by the messages in the forward direction as e.g. $p(z_2) \propto \vec{\mu}(z_2)\cev{\mu}(z_2) = \vec{\mu}(z_2)$.

\subsection{Variational free energy} \label{sec:appendix:background:vfe}
The calculation of the normalization constant $Z$ in \eqref{eq:appendix:background:posterior} is often difficult or even intractable.
Similarly, also the marginal distributions might be unobtainable in closed form.
To resolve this, we will approximate the posterior distribution $p(\bm{z})$ of \eqref{eq:appendix:background:posterior} with an approximate distribution $q(\bm{z})$.
Using this approximation we can define the variational free energy (VFE) function as
\begin{equation}\label{eq:appendix:background:vfe}
    F[q, f] = \mathbb{E}_{q(\bm{z})} \left[ \ln \frac{q(\bm{z})}{f(\bm{z})} \right] = \mathrm{KL}\left[ q(\bm{z})\ \| \ p(\bm{z}) \right] - \ln Z \geq - \ln Z,
\end{equation}
which provides an upper-bound on the negative log-normalization constant and is used for approximating the model performance in intractable models.
This bound is attained at the minimum of the VFE when the approximate posterior distribution equals the exact posterior distributions $q(\bm{z}) = p(\bm{z})$.
However, to allow for tractable inference, $q(\bm{z})$ often has to be constrained to some family of distributions $\mathcal{Q}$ as $q(\bm{z})\in\mathcal{Q}$ under some factorization.

\subsection{Bethe free energy} \label{sec:appendix:background:bfe}

The Bethe assumption  
\begin{equation}
    q(\bm{z}) = \prod_{a \in \mathcal{V}} q_a(\bm{z}_a) \prod_{i \in \mathcal{E}} q_i(z_i)^{-1}.
\end{equation}
is a useful constraint on the approximate posterior $q(\bm{z})$,   \citep{yedidia_bethe_2001}. 

Here we made use of the fact that all edges in the FFG have a maximum degree of two, which can be strictly enforced by adding uninformative priors $p(z_i)=1$ to dangling edges.
Under the Bethe assumption, the VFE reduces to the Bethe free energy (BFE)
\begin{equation}\label{eq:appendix:background:bfe}
    F_B[q, f] = - \sum_{a \in \mathcal{V}} \mathbb{E}_{q(\bm{z}_a)}\left[\ln f_a(\bm{z}_a)\right] - \sum_{a \in \mathcal{V}} \mathrm{H}[q_a(\bm{z}_a)] + \sum_{i \in \mathcal{E}} \mathrm{H}[q_i(z_i)],
\end{equation}
which equals the VFE for acyclic graphs (i.e. trees).
The BFE decomposes the VFE into a sum of node-local free energies contributions and edge-specific entropies $\mathrm{H}$.

\subsection{Variational and hybrid message passing} \label{sec:appendix:background:vmphybrid}
Under the variational approximation we can employ variational inference in the model, which iteratively finds stationary points on the BFE by fixing all approximate posterior distributions besides the one that is being optimized.
This inference procedure can be cast to a message passing paradigm and is called variational message passing \citep{dauwels_variational_2007}.
Here the exact message update rule of \eqref{eq:appendix:background:sumproductupdaterule} then reduces to the variational message update rule \citep{dauwels_variational_2007}
\begin{equation} \label{eq:background:vmpupdaterule}
    \vec{\nu}(z_i) \propto \exp \left\{ \mathbb{E}_{q(\bm{z}_{a\backslash i})} \left[\ln f_a(\bm{z}_a)\right]\right\}
\end{equation}
where $\vec{\nu}(z_i)$ denotes the outgoing variational message on edge $z_i$.
The approximate marginal distributions are then iteratively updated as 
\begin{equation}\label{eq:appendix:background:approximateposterior}
    q_i(z_i) \propto \vec{\nu}(z_i) \cdot  \cev{\nu}(z_i).
\end{equation}
The calculations of variational messages and approximate marginal distributions are then iteratively repeated until convergence of the VFE is reached.

In addition to the structure imposed by the Bethe approximation, additional constraints can be enforced.
Depending on these local constraints different inference algorithms naturally follow \citep{senoz_variational_2021}.
\citep{senoz_variational_2021} shows that amongst others the sum-product algorithm \citep{pearl_reverend_1982, kschischang_factor_2001}, variational message passing \citep{dauwels_variational_2007} and expectation propagation \citep{minka_divergence_2005} can be recovered.
By combining different local constraints we can achieve hybrid message passing-based inference in the probabilistic model.
We highly recommend the interested reader the work of \citep{senoz_variational_2021} for an extensive overview of hybrid message passing schemes.

\newpage
\section{Probabilistic model overview}\label{appendix:model}
This appendix gives a concise overview of the generative model of the acoustic model and AIDA. The prior distributions are uninformative unless stated otherwise in Section~\ref{sec:experiments}.
\subsection{Acoustic model}
\begin{subequations}
    \begin{align*}
        \intertext{The observed signal $x_t$ is the sum of a speech and noise signal as}
        x_t &= s_t + n_t
        \intertext{The speech signal $s_t = \bm{e}_1^\intercal\bm{s}_t$ is modeled by a time-varying auto-regressive process as}
        \bm{s}_t &\sim \mathcal{N}\left(A(\bm{\theta}_t)\bm{s}_{t-1},\ V\left(\gamma\right)\right)
        \intertext{The auto-regressive coefficients of the speech signal are time-varying as}
        \bm{\theta}_t &\sim \mathcal{N}\left(\bm{\theta}_{t-1},\ \omega\mathrm{I}_{M}\right)
        \intertext{The noise signal $n_t = \bm{e}_1^\intercal\bm{n}_t$ is also modeled by an auto-regressive process}
        \bm{n}_t &\sim \mathcal{N}\left(A(\bm{\zeta}_k)\bm{n}_{t-1},\ V\left(\tau_k\right)\right) \qquad\qquad \text{for }t = t^-, t^-+1, \ldots, t^+
        \intertext{The parameters of the noise model depend on the context}
        \bm{\zeta}_k &\sim \prod_{l=1}^L \mathcal{N}\left(\bm{\mu}_{l}, \Sigma_{l}\right)^{c_{lk}} \\
        \tau_k &\sim \prod_{l=1}^L \Gamma\left(\alpha_{l}, \beta_{l}\right)^{c_{lk}} \\
        \intertext{The context $\bm{c}_k$ evolves over a different time scale indexed by $k$ as}
        \bm{c}_k &\sim \mathrm{Cat}(\mathrm{T}\bm{c}_{k-1})
        \intertext{The transition matrix of the context is modeled as}
        \mathrm{T}_{1:L,j} &\sim \mathrm{Dir}(\bm{\alpha}_j)
        \intertext{Finally, the output of the hearing aid algorithm $y_t$ is formed as the weighted sum of the speech and noise signals as}
        y_t &= u_{sk}s_t + u_{nk}n_t \qquad\qquad \text{for }t = t^-, t^-+1, \ldots, t^+
    \end{align*}
\end{subequations}

\subsection{AIDA's user response model}
\begin{subequations}
    \begin{align*}
        \intertext{The user responses are modeled by a Bernoulli distribution containing a Gaussian cumulative probability distribution that enforces the output $v_k(\bm{u}_k$) to the allowed domain for the argument of the Bernoulli distribution}
        r_k &\sim \mathrm{Ber}(\Phi(v_k(\bm{u}_k))) \qquad\qquad \text{if } r_k \in\{0,1\}
        \intertext{$v_k(\bm{u}_k)$ encodes our beliefs about the user response function (evaluated at $\bm{u}_k$), modeled by a mixture of Gaussian processes as}
        v_k &\sim \prod_{l=1}^L\mathrm{GP}(m_l(\cdot),K_l(\cdot,\cdot))^{c_{lk}}
        \intertext{whose kernel function is defined as}
        K(\bm{u},\bm{u}^\prime) &= \sigma^2 \exp\left\{-\frac{\|\bm{u}-\bm{u}^\prime\|^2_2}{2l^2}\right\}
        \intertext{where $\sigma$ denotes noise and $l$ the length scale of the kernel.}
        %K_l(a,b) &= \sigma_l^2 \exp \left( -\frac{(a - b)^2}{2l_l^2} \right)
    \end{align*}
\end{subequations}
%The agent model is a standard Gaussian Process Classifier (GPC) with a squared exponential kernel. The GPC is given by 
%
%\begin{subequations}
%\begin{align*}
   
%   v &\sim \mathrm{GP}(m(\cdot),K(\cdot,\cdot)) \label{eq:model:GP}
% \end{align*}
% \end{subequations}
%
% where $\Phi(\cdot)$ denotes the Gaussian CDF. Without loss of generality, we can set the mean function of the GP to 0 \citep{rasmussen_gaussian_2006}. The kernel function $K(\cdot,\cdot)$ is given by
% \begin{align*}
%     K(a,b) &= \sigma^2 \exp \bigg( -\frac{(a - b)^2}{2l^2} \bigg)
% \end{align*}

%\bart{Below details should go in the experimental validation section}
%\mtk{Updated}

%For our experiments we initialised $\sigma = l = 0.5$ and performed updates using the LBFGS implementation of Optim.jl, v1.5.0 at every $5^th$ iteration. 
%User responses are sampled from a Bernoulli distribution with parameter $\theta_{user}$ computed as
%\begin{align*}
%    \theta_{user} &= \frac{2}{1 + \exp( \beta \sum_k  \frac{(\bm{u}_k - \bm{u}_k^*)^2}{\sigma_k} )}
%\end{align*}
%For our experiments we set 
%\begin{align*}
%    \bm{u}^* &= [0.8,0.2] \\
%    \sigma &= [1,1] \\
%    \beta &= 25 
%\end{align*}
\begin{table}[ht]
    \centering
    \caption{Summary of the notational conventions used throughout Section~\ref{sec:model} in this paper.}
    \label{tab:symbols-model}
    \begin{tabular}{c|c|p{13cm}}
        Notation & Definition & Explanation \\ \hline \hline
        $x_t$ & \eqref{eq:model:acoustic} & Observed signal at time index $t$. \\
        $y_t$ & \eqref{eq:model:HA-output} & Output of the hearing aid at time index $t$. \\
        $s_t$ & & Latent speech signal at time index $t$.\\
        $n_t$ & & Latent noise signal at time index $t$. \\
        $\bm{s}_t$ & \eqref{eq:model:acoustic:AR} & Vector with hidden states of the auto regressive model of the speech signal at time index $t$. \\
        $\bm{n}_t$ & \eqref{eq:model:acoustic-noise} & Vector with hidden states of the auto regressive model of the noise signal at time index $t$. \\
        $\bm{\theta}_t$ & \eqref{eq:model:acoustic:coefs} & Vector of auto regressive coefficients of the speech signal at time index $t$. \\
        $\omega$ & & Covariance matrix scaling of the dynamics of the auto regressive coefficients of the speech signal. \\
        $\mathrm{I}_M$ & & Identity matrix of size $(M\times M)$. \\
        $M$ & & Auto regressive order of the speech signal. \\
        $N$ & & Auto regressive order of the noise signal. \\
        $W$ & & Sampling ratio between $t$ and $k$. \\
        $A(\theta)$ & \eqref{eq:model:ARmatrix} & State space transition matrix of an auto regressive process with coefficients $\bm{\theta}$. \\
        $\gamma$ & & Process noise precision of the auto regressive model for the speech signal. \\
        $V(\gamma)$ & & Covariance matrix of the auto regressive model for the speech signal, containing only zeros except for the first element, which is $1/\gamma$. \\
        $\bm{e}_i$ & & Appropriately sized cartesian standard unit vector containing only zeros, except for the $i^\text{th}$ entry, which is 1. \\
        $\bm{0}$ & & Appropriately sized vector of zeros. \\
        $\mathcal{N}(\mu, \Sigma)$ & & Normal distribution with mean $\mu$ and covariance matrix $\Sigma$. \\
        $\Gamma(\alpha, \beta)$ & & Gamma distribution with shape and rate parameters $\alpha$ and $\beta$, respectively. \\
        $\bm{\zeta}_k$ & \eqref{eq:model:zeta} & Vector of auto-regressive coefficients of the noise signal at context time scale index $k$.\\
        $\tau_k$ & \eqref{eq:model:tau} & Process noise precision of the auto regressive model of the noise signal at context time scale index $k$. \\
        $L$ & & Number of contexts. \\
        $\bm{c}_k$ & \eqref{eq:model:context-dynamics} & The context at context time scale $k$, containing a 1-of-$L$ binary vector with elements $c_{lk}\in\{0,1\}$. \\
        $\bm{\pi}$ & & Vector of event probabilities. \\
        $\mathrm{T}$ & \eqref{eq:model:context-transition} & Transition matrix representing the discrete context dynamics. \\
        $\bm{\alpha}_j$ & & Concentration parameters for the prior over the $j^\text{th}$ column of $T$. \\
        $u_{sk}$ & & Gain of the speech signal at context time index $k$. \\
        $u_{nk}$ & & Gain of the noise signal at context time index $k$. \\
        $\bm{u}_k$ & & Vector of tuning parameters at context time index $k$ defined as $[u_{sk}, u_{nk}]^\intercal$. \\
        $v_k(\cdot)$ & \eqref{eq:model:GP-1} & Latent function drawn from a mixture of GPs. \\
        $r_k$ & \eqref{eq:sigmoid-user-response-1}& Binary user response at time index $k$. \\
        $\Phi(\cdot)$ & & Standard Gaussian cumulative distribution function. \\
        $m_l(\cdot)$ & & Mean function of the $l^\text{th}$ GP. \\
        $K_l(\cdot, \cdot)$ & & Kernel function of the $l^\text{th}$ GP.
    \end{tabular}
\end{table}
\clearpage
\section{Inference realization}\label{sec:appendix:realization}
This appendix describes in detail how the inference tasks of Sections~\ref{sec:inference:context} and \ref{sec:inference:trial} are realized.
The inference task of Section~\ref{sec:inference:acoustic} is performed by automated message passing using the update rules of \citep{podusenko_message_2021}.

\subsection{Realization of inference for context classification} \label{sec:appendix:realization:context}
The inference task for context classification of \eqref{eq:inference:context} renders intractable as discussed in Section~\eqref{sec:inference:context}.
To circumvent this problem, we will solve this task as a Bayesian model comparison task.

% how do we solve it (in theory)
In a Bayesian model comparison task, we are interested in calculating the posterior probability $p(m_l \mid \bm{x})$ of some model $m_l$ after observing data $\bm{x}$.
% The most probable model $m_l^*$ that best defines the observed data is then selected as the true underlying model as 
% \begin{equation}
%     m_l^* = \argmax_{m_l} p(m_l \mid \bm{x}).
% \end{equation}
The posterior model probability $p(m_l \mid \bm{x})$ can be calculated using Bayes' rule as 
\begin{equation}\label{eq:inference:BMC}
    p(m_{l} \mid \bm{x}) = \frac{p(\bm{x} \mid m_{l})p(m_{l})}{\sum_{j} p(\bm{x}\mid m_{j})p(m_{j})},
\end{equation}
where the denominator represents the weighted model evidence $p(\bm{x})$, i.e. the model evidence obtained for the individual models $p(\bm{x}\mid m_l)$, weighted by their priors $p(m_l)$.

% how do we solve it (in practice)
To formulate our inference task as a Bayesian model comparison task, the distinct models $m_l$ first have to be specified.
In order to do so, we first note that we obtain the priors of $\bm{c}_{k-1}$ and $z_{t^--1}$ in \eqref{eq:inference:context} separately, and therefore we implicitly assume a factorization of our prior $p(\bm{c}_{k-1}, z_{t^--1}\mid \bm{x}_{1:t^--1})$ as 
\begin{equation}
    p(\bm{c}_{k-1}, z_{t^--1} \mid \bm{x}_{1:t^--1}) = p(\bm{c}_{k-1} \mid \bm{x}_{t^--1}) \ p(z_{t^--1} \mid \bm{x}_{1:t^--1}).
\end{equation}
As a result \eqref{eq:inference:context} can be rewritten as 
\begin{equation}\label{eq:inference:contextdecomposition}
\begin{split}
    p(\bm{c}_k \mid \bm{x}_{1:t^+})
    \propto
    &\underbrace{\int p(\bm{c}_k, \mathrm{T} \mid \bm{c}_{k-1})\ p(\bm{c}_{k-1} \mid \bm{x}_{1:t^--1})\ \mathrm{d}\mathrm{T}\ \mathrm{d}\bm{c}_{k-1}}_{\vec{\mu}(\bm{c}_k)} \\
    \cdot
    &\underbrace{\int p(z_{t^-:t^+}, \Psi_k, \bm{x}_{t^-:t^+} \mid z_{t^--1}, \bm{c})\ p(z_{t^--1} \mid \bm{x}_{1:t^--1}) \ \mathrm{d}z_{t^--1:t^+}\ \mathrm{d}\Psi_k}_{p(\bm{x}_{t^-:t^+} \mid \bm{x}_{1:t^--1}, \bm{c}_k)}.
    \end{split}
\end{equation}
The first term $\vec{\mu}(\bm{c}_k)$ can be regarded as the forward message towards the context $\bm{c}_k$ originating from the previous context.
It gives us an estimate of the new context solely based on the context dynamics as stipulated by the transition matrix $\mathrm{T}$.
The second term $p(\bm{x}_{t^-:t^+} \mid \bm{x}_{1:t^--1}, \bm{c}_k)$ can be regarded as the incremental model evidence under some given context $\bm{c}_k$.
Comparison of \eqref{eq:inference:contextdecomposition} and \eqref{eq:inference:BMC} allows us to formulate our inference problem in \eqref{eq:inference:context} into a Bayesian model comparison problem by defining
\begin{subequations}
\begin{align}
    p(m_l) &= \vec{\mu}(\bm{c}_{k} = \bm{e}_l), \label{eq:inference:BMC_prior}\\
    p(\bm{x} \mid m_l) &= p(\bm{x}_{t^-:t^+} \mid \bm{x}_{1:t^--1}, \bm{c}_k = \bm{e}_l).\label{eq:inference:BMC_evidence}
\end{align}
\end{subequations}
We can therefore define a model $m_l$ by clamping the context variable in generative model as $\bm{c}_k = \bm{e}_l$.
This means that each model only has one active component for both the Gaussian and Gamma mixture nodes and therefore the messages originating from these nodes are exact and do not require a variational approximation.

%Consider the probabilistic model of Section \ref{sec:model} for a single context time scale index, \bdv{What does that mean? A single context time scale index?}such that the context transitions only possibly occur at the boundaries of the segment. \bdv{Which segment? If you do this, then please write down the model assumptions.}
%This model will be split into distinct probabilistic models, based on the mixture parameters corresponding to the different contexts.
%The mixture models are expanded such that each model $m_l$ only contains a single Normal or Gamma prior for the parameters in \eqref{eq:model:acoustic-noise}, corresponding to a specific context.
%The model index $l$ simultaneously reflects the non-zero element in the context variable $\bm{c}_k$ over the entire time segment. 
% Here we are  temporal dynamics of the context within the segment and by fixing its value to one of the possible contexts, i.e., each model performs an inference for one specific context.

% how do we calculate these terms?
Despite the expansion of the mixture models, the incremental model evidence $p(\bm{x}_{t^-:t^+} \mid \bm{x}_{1:t^--1}, \bm{c}_k=\bm{e}_l)$ cannot be computed exactly as the auto-regressive source models lead to intractable inference.
As a result we approximate the model evidence in \eqref{eq:inference:BMC_evidence} using the Bethe free energy, as defined in \eqref{eq:appendix:background:bfe} in Appendix~\ref{sec:appendix:background}, as 
\begin{equation}\label{eq:inference:evidenceapprox}
    p(\bm{x}\mid m_l) \approx \exp \{-F_B[q, m_l]\},
\end{equation}
where $F_B[q, m_l]$ denotes the Bethe free energy observed after convergence of the inference algorithm for model $m_l$.
Similarly the calculation of \eqref{eq:inference:BMC_prior} is intractable.
Therefore we will approximate the model prior with the variational message towards $\bm{c}_k$ instead as
\begin{equation}
    p(m_{l}) \approx \vec{\nu}(\bm{c}_{k}=\bm{e}_l).
\end{equation}

\subsection{Realization of inference for trial design}\label{sec:appendix:realization:trial}
Probabilistic inference in AIDA encompasses 2 tasks: 1) optimal proposal selection and 2) updating of the Gaussian process classifier (GPC). Here we specify how these inference tasks are executed in more detail.

\subsubsection*{Optimal proposal selection}
A closed-form expression of the EFE decomposition in \eqref{eq:efe_bound} can be obtained for the GPC as shown in \citep{houlsby_bayesian_2011}.

The first term in the decomposition, the negative utility drive, resembles the cross-entropy loss between our goal prior and posterior marginal. Since user responses are binary, we can evaluate this binary cross-entropy term as \citep{houlsby_bayesian_2011}
\begin{align}\label{eq:bin_cross}
   -\mathbb{E}_{q(r \mid \bm{u})} \left[ \ln p(r) \right] &=
    \Phi \left(\frac{\mu_{\bm{u},D}}{\sqrt{\sigma^2_{\bm{u},D} +1}} \right)  \ln \mathbb{E}_{p(r)}[r] + 
    \left(1-\Phi \left(\frac{\mu_{\bm{u},D}}{\sqrt{\sigma^2_{\bm{u},D} +1}} \right) \right) \ln \left(1-\mathbb{E}_{p(r)}[r]\right),
\end{align}
where $\mu_{\bm{u},D}$ and $\sigma^2_{\bm{u},D}$ denote the posterior mean and variance returned by the GPC when queried at the point $\bm{u}$ given some data set $D=\{\bm{u}_{1:k-1}, r_{1:k-1}\}$, respectively.
More concretely, the GPC returns a Gaussian distribution from which the posterior mean and variance are extracted as $v(\bm{u}) = \mathcal{N}(\mu_{\bm{u},D}, \sigma^2_{\bm{u},D})$.
$\Phi(\cdot)$ denotes the standard Gaussian cumulative distribution function. $p(r)$ denotes the Bernoulli goal prior over desired user feedback.
$h$ is the binary entropy function and $C = \sqrt{\frac{\pi \ln 2}{2}}$. For brevity, we denote the data set of parameters and matching user responses collected so far as $D$.

% Subtracting \eqref{eq:BALD} from \eqref{eq:bin_cross} provides our approximation to \eqref{eq:efe_bound}.

% AIDA uses the EFE as an acquisition function in order to adaptively trade off exploration (learning latent user preferences) and exploitation (finding the best parameter settings). 
% To that end, we work out how to compute the information gain and the utility drive of \eqref{eq:efe_bound}. The end result is that we can approximate the information gain term of \eqref{eq:efe_bound} for a particular choice of $\bm{u}$, given the dataset $D$ collected so far as 
%
The second term in the decomposition, the (negative) information gain, describes how much information we gain by observing a new user appraisal. This information gain term ($\mathrm{IG}$) can be expressed in a GPC as \citep{houlsby_bayesian_2011}
\begin{equation}\label{eq:BALD}
    % \mathrm{IG}[r,v|D, \bm{u} ] \approx 
    % \underbrace{h\left( \Phi \left(\frac{\mu_{\bm{u},D}}{\sqrt{\sigma^2_{\bm{u},D} +1}} \right) \right)}_{\mathrm{H}[r | \bm{u},D]} -
    % \underbrace{\frac{C}{\sqrt{\sigma^2_{\bm{u},D} + C^2}} \exp \left(- \frac{\mu^2_{\bm{u},D}}{2\left( \sigma^2_{\bm{u},D} + C^2 \right)} \right)}_{\mathbb{E}_{q(v |\bm{u}, D)} \mathrm{H}[r | v, D, \bm{u}]} \,.
    \mathrm{IG}[r,v\mid D, \bm{u} ] \approx 
    h\left( \Phi \left(\frac{\mu_{\bm{u},D}}{\sqrt{\sigma^2_{\bm{u},D} +1}} \right) \right) -
    \frac{C}{\sqrt{\sigma^2_{\bm{u},D} + C^2}} \exp \left(- \frac{\mu^2_{\bm{u},D}}{2\left( \sigma^2_{\bm{u},D} + C^2 \right)} \right) \,,
\end{equation}
where the constant $C$ is defined as $C = \sqrt{\frac{\pi \ln 2}{2}}$ and where $h(\cdot)$ is defined as $h(p) = -p\ln(p) - (1-p)\ln(1-p)$.

\subsubsection*{Inference in the Gaussian process classifier}
For our experiments we use Laplace approximation as described in \citep[Chapter 3.4]{rasmussen_gaussian_2006} for performing inference in the GPC. The Laplace approximation is a two-step procedure, where we approximate the posterior distribution by a Gaussian distribution.
We first find the mode of the exact posterior, which resembles the mean of the approximated Gaussian distribution. Then we approximate the corresponding precision as the negative Hessian around the mode. Finding the exact posterior $p(v \mid D)$ amounts to calculating
\begin{subequations}
\begin{align}
    p(v \mid D) 
    &= \frac{p(r_{1:k-1} \mid v) p(v \mid \bm{u}_{1:k-1})}{p(r_{1:k-1} \mid \bm{u}_{1:k-1})} \\
    &\propto p(r_{1:k-1} \mid v) p( v \mid \bm{u}_{1:k-1}) \label{eq:gp_bayes} \,.
\end{align}
\end{subequations}
Taking the logarithm of \eqref{eq:gp_bayes} and differentiating twice with respect to $v$ gives
\begin{subequations}
\begin{align}
    \nabla_v \ln p(v \mid D) &=  \nabla_v \ln p(r_{1:k-1} \mid v) - K^{-1}v \label{eq:nabla_logp} \\
    \nabla_v \nabla_v \ln p(v \mid D) &=  \nabla_v \nabla_v \ln p(r_{1:k-1} \mid v) - K^{-1}
    = -W - K^{-1} \label{eq:2nabla_logp}
\end{align}
\end{subequations}
where $\nabla_v$ denotes the gradient with respect to $v$, $K = K(\bm{u}_{1:k-1},\bm{u}_{1:k-1})$ is the kernel matrix over the queries $\bm{u}_{1:k-1}$ and $W = - \nabla_v \nabla_v \ln p(r_{1:k-1} \mid v)$ is a diagonal matrix since the likelihood factorizes over independent observations. At the mode $\hat{v}$ \eqref{eq:nabla_logp} equals zero which implies
\begin{align}
    \hat{v} &= K \nabla_v \ln p(r_{1:k-1} | \hat{v}) \label{eq:hatv} \,.
\end{align}
Directly solving \eqref{eq:hatv} is intractable, because of the recursive non-linear relationship. Instead we can estimate $\hat{v}$ using Newton's method, where we perform iterations with an adaptive step size. We omit the computational and implementation details here and instead refer to \citep[Algorithm 3.1]{rasmussen_gaussian_2006}. We determine the step size using a line search as implemented in \texttt{Optim.jl} \citep{k_mogensen_optim_2018}.
Having found the mode $\hat{v}$, we can construct our posterior approximation as 
\begin{align}
    p(v \mid D) &\approx \mathcal{N}\left(\hat{v}, \left(K^{-1} + W \right)^{-1}\right), \label{eq:q_laplace} 
\end{align}
where $W$ is evaluated at $v=\hat{v}$.
If we now  recall that evaluating a GP at any finite number of points results in a Gaussian, we see that under the Laplace approximation the solution can be obtained using standard results for marginalization of jointly Gaussian variables. We define the shorthand $K(\bm{u}_k, \bm{u}_{1:k-1}) = K_{1:k}$ and $K(\bm{u}_k, \bm{u}_{k}) = K_k$ and find the posterior mean $\mu_{\bm{u}}$ as \citep[p. 44]{rasmussen_gaussian_2006}
\begin{align}
    \mu_{\bm{u},D} &= K_{1:k}^\intercal K^{-1} \hat{v} = K_{1:k}^\intercal \nabla \ln p(r_{1:k-1} \mid \hat{v}) \,.
\end{align}
The posterior covariance $\sigma^2_{\bm{u},D}$ is given by \citep[p. 44]{rasmussen_gaussian_2006}
\begin{align}
    \sigma^2_{\bm{u},D} &= K_k - K_{1:k}^\intercal \left( K + W^{-1} \right)^{-1} K_{1:k} \,.
\end{align}
%
% \subsubsection{Optimal proposal selection}
% \albert{Suggestion. Move most of the contents to appendix and keep only equation \eqref{eq:min_min}}

\end{document}